# Observations of a Possible Transient Magnetically Arrested Accretion State in a Nearby Quasar: OQ208

Brian Punsly,[1,2] Cormac Reynolds,[3] Paola Marziani,[4,5] Gary J. Hill,[6,7] Alexander B. Pushkarev,[8] Carlo Stanghellini,[9] Frank Schinzel,[10] Christopher O'Dea,[11] Gregory R. Zeimann,[12,6] Andrew Biggs,[13] Jian-Min Wang, Pu Du,[14] Alberto Floris,[15,16,4] Mauro D'Onofrio,[17,4] and Levi Malmstrom[18]

[1] ICRANet, Piazza della Repubblica 10 Pescara 65100, Italy
[2] ICRA, Physics Department, University La Sapienza, Roma, Italy
[3] CSIRO Space and Astronomy, PO Box 1130, Bentley WA 6102, Australia
[4] INAF, Osservatorio Astronomico di Padova, Italy
[5] Instituto de Astrofísica de Andalucía (IAA-CSIC), Glorieta de Astronomía s/n, ES 18008, Granada, Spain
[6] McDonald Observatory, University of Texas at Austin, 2515 Speedway, Austin, TX 78712, USA
[7] Department of Astronomy, University of Texas at Austin, 2515 Speedway, Austin, TX 78712, USA
[8] Crimean Astrophysical Observatory, Nauchny 298409, Crimea, Russia
[9] INAF-Istituto di Radioastronomia, Via Gobetti 101, I-40129 Bologna, Italy
[10] National Radio Astronomy Observatory, Socorro, NM 87801, USA
[11] Department of Physics and Astronomy, University of Manitoba, Winnipeg, Manitoba, Canada
[12] Hobby-Eberly Telescope, 32 Mt. Locke Rd., McDonald Observatory, TX 79734, USA
[13] UK Astronomy Technology Centre, Royal Observatory, Blackford Hill, Edinburgh EH9 3HJ, UK
[14] Key Laboratory for Particle Astrophysics, Institute of High Energy Physics, CAS, China
[15] Department of Physics University of Crete, Voutes University Campus, 70013 Heraklion, Greece
[16] Institute of Astrophysics, FORTH, N.Plastira 100, Vassilika Vouton,70013 Heraklion, Greece
[17] Dipartimento di Fisica & Astronomia, University of Padua, Italy
[18] Department of Physics, University of Maryland Baltimore County, 1000 Hilltop Circle, Baltimore, MD 21250, USA

## ABSTRACT

OQ208 is a nearby, partially obscured quasar (z=0.077) that is a young, bright, parsec scale radio source. We assemble archival and new high frequency VLBA and VLA observations and optical spectra to form a data-set spanning 39 years. Radio light curves covering 58 years were also compiled. We utilize new spectrophotmetry to calibrate previous spectroscopy using forbidden narrow lines that are expected to be stable on much longer time scales. VLBA and VLA observations of a light-year scale bright nuclear flare at 15.4 GHz and 22 GHz reveal a rise (fade) beginning in mid-1996 (early-2000). Quasi-contemporaneously, from 2/7/1997-6/3/2000, the H$\alpha$ broad line equivalent widths (EWs) and fluxes dropped dramatically. In the context of the tendency of radio loud quasars to have a depressed extreme ultraviolet (EUV) continuum (the main source of ionizing flux for H$\alpha$) relative to radio quiet quasars at matched UV luminosity (the EUV deficit of radio loud quasars), this may not be a coincidence. Analytic models previously developed to explain the relationship between jet power and the EUV deficit are consistent with (but not direct observational proof of) the small EWs being a consequence of transient magnetically arrested accretion states from $\sim 1997-2001$. The 22 GHz VLBA nucleus gradually fades, in 2023 the flux density is $< 5\%$ of its value in 2000. The environs of the nucleus also fade at 22 GHz, but in a time delayed fashion.

Corresponding author: Brian Punsly
brian.punsly@cox.net

# 1. INTRODUCTION

The low redshift (z=0.077) broadline radio galaxy, OQ208, was discovered in the mid 1960's as one of the strongest radio sources missed in the 4C surveys. It is a partially obscured quasar, the $12\mu$ luminosity measured by Spitzer is $7.32 \times 10^{44}$ergs/sec and the 1516 Å luminosity (near the typical peak of a quasar spectral energy distribution) is only $8.91 \times 10^{42}$ergs/sec (Willett et al. 2010; Afanasiev et al. 2019). The broad lines were quantitatively evaluated for the first time in 1979 and were found to be shifted relative to the reference frame of the narrow lines by $\sim 2800$ km/sec to the red (Osterbrock and Cohen 1979). This led to further exploration of the source in the optical and the discovery of evolving double peaked broad lines (Marziani et al. 1993; Gezari et al. 2007). The broad lines have changed dramatically, on time scales of years, during the last 45 years. From a radio perspective, this is a very compact source, unresolved with the Very Large Array (VLA) at all frequencies. It is considered a high frequency peaked ($\sim 6$ GHz), compact steep spectrum radio source (Dallacasa et al. 2000; O'Dea and Saikia 2021). It was very bright, $\sim 2.5$ Jy ($\sim 1.0$ Jy) at 10.5 GHz (22.5 GHz) in the mid-1970's (McCutcheon and Gregory 1978). Like the broad lines, the radio nucleus also changes dramatically on the time scales of years (Wu et al. 2013). Due to its brightness, it is an excellent candidate for high frequency Very Long Baseline Interferometry (VLBI). The low redshift, for a quasar, translates to high spatial resolution of the evolving nucleus with the Very Long Baseline Array (VLBA) at 15.4 GHz and 22.5 GHz. The evolution on time scales of years makes this an excellent laboratory for studying the interaction of the jet launching region and the neighboring gas (like the accretion flow and molecular torus) within one human lifetime. We present a compilation and analysis of radio observations and optical spectroscopy spanning nearly six decades, including new high-quality spectroscopy from the Hobby-Eberly Telescope.

Our primary result is the depiction of a strong 22.5 GHz flare of a parsec scale nucleus. During the radio flare there is an apparent anti-correlation between the radio luminosity and the Balmer broadline equivalent width (EW). Thus, we are motivated to explore physical explanations of this provocative circumstance. Magnetically arrested accretion has become one of the primary tenents of the theory of radio loud active galactic nuclei (AGN) (Igumenshchev 2008; EHT Collaboration et al. 2021; Zamaninasab et al. 2014; Nemmen and Tchekhovskoy 2015). It is widely believed that powerful jets in AGN are produced in magnetically arrested accretion states (Tchekhovskoy et al. 2011; Zamaninasab et al. 2014; EHT Collaboration et al. 2021). Thus, we explore the possibility of magnetically arrested accretion during the 22.5 GHz flare. Previous studies indicate that magnetically arrested accretion (if it exists in an AGN) would disrupt the innermost accretion flow and therefore the gas that radiates the ionizing continuum in a quasar (Igumenshchev 2008; Punsly 2015; Punsly et al. 2016). In the case of OQ208, this effect would then be imprinted in the broad lines $\sim 2$ months later (Afanasiev et al. 2019).

The paper is organized as follows. Section 2 develops a 58 year light curve based primarily on 22.5 GHz total flux density. In Section 3, a study of the Balmer broad line strength from 1985 to 2024 (26 spectra) is undertaken with special emphasis on flux calibration. From the first two sections, we can already see the abrupt decrease in broad line EWs during a 22.5 GHz flare that peaks in 1998. We determine the location of the jet central engine and the fine scale structure of the flaring nucleus in Sections 4 and 5. In Sections 6 and 7, the primary flaring nuclear sub-components are identified and the temporal connection between these flaring sub-components and the depressed broad line EW is made more precise and evident. We describe a possible straightforward connection to magnetically arrested accretion in Sections 8 and 9. In Section 10, we find that the morphology of the nuclear region changes dramatically after the flare. The nucleus, the optical continuum and the broad lines all fade dramatically from 2015 to 2023. Throughout this paper, we adopt the following cosmological parameters: $H_0$=69.6 km s$^{-1}$Mpc$^{-1}$, $\Omega_\Lambda = 0.714$ and $\Omega_m = 0.286$ and use Ned Wright's Javascript Cosmology Calculator [1] (Wright 2006). Using the flat Universe assumption this corresponds to a spatial scaling relation of 1.467 pc/mas for OQ208.

# 2. A 58 YEAR RADIO HISTORY OF OQ208

OQ208 was one of the brightest radio sources that was missed in early 178 MHz Fourth Cambridge Radio Catalogue since it is peaked at ∼5 GHz. At this frequency, the source was ∼3 Jy and very compact, so it became a VLA calibrator when it was commissioned (Stanghellini et al. 1997). It has been extensively observed, but it has faded and is no longer a useful calibrator for flux density. The emission is very depolarized, making it still relevant as standard, so it is still a standard polarization leakage calibrator, especially if it re-brightens in the future. This section describes our efforts to track the history of the nuclear flux density. As such, we try to utilize the highest possible radio frequency in construction of the light curve.

[1] http://www.astro.ucla.edu/ wright/CosmoCalc.html

In our efforts to create a long term light curve, it became apparent that there are significant gaps in the temporal analysis at every frequency. A radio frequency of $\sim$ 20 GHz is preferable because our primary interest is the cental engine and the radio core. Observations at 22 GHz have a long history for this object (see Figure 1). However, there are the aforementioned gaps and coverage at 10.7 GHz begins 10 years earlier. We piece together the light curves at various frequencies $\geq$ 10.5 GHz in Figure 1 with linear scaling relationships. The validity of the scaling relationships (effected by the offset and graticule spacings of the two vertical axes) are verified by the approximate agreement in the overlap regions (within measurement uncertainty). The method is not exact, but it should trace gradual and major changes in the light curve. It also provides a consistency check on extreme behavior detected by only one telescope. For example, the abrupt flare that peaks on 12/30/1984 at 14.95 GHz with the University of Michigan survey is not supported by the 22.4 GHz VLA observation on the same date in which the flux calibrator 3C 286 was observed a few minutes before OQ208. Therefore, we consider this flare as not real. Another date of interest is the brief large flare in 1981 from the University of Michigan Survey that is somewhat supported by the slightly elevated 14.9 GHz VLA observations on 5/10/1981 (perhaps the beginning of the flare) and 6/30/1982 (perhaps the end of the flare). There is an apparent shoulder in the VLA light curve in $\sim$1981. This is a brief event ($\sim$1 year) and could be considered a minor flare. The large, extremely narrow University of Michigan peak is possible, but not verified. Finally, the very low 22 GHz Metsahovi detection in 1989 with large error bars, Wiren et al. (1992), is not supported by the 14.9 GHz VLA data (after linear re-scaling) from 1988 and 1989 in Figure 1. This must have been at the limit of detection and the 14.9 GHz VLA data points suggest that the top of the error bars of the Metshaovi flux density is likely the best value.

The plot does clearly shows two large flares in $\sim$1998 and $\sim$1976. However, due to sparse time sampling, the larger flare between 1966 and 1976 could be 2 or more individual flares. The diminished flux density at the present time is striking. The data for Figure 1 is compiled in Table 1. .

## 3. BROAD LINE ANALYSIS

The broad Balmer lines of OQ208 have been observed extensively due to the double-peaked or disk-like nature. Most conspicuously, the H$\beta$ broad emission line (BEL) is shifted from the H$\beta$ narrow component (NC) by $\sim$2000 km/s redward when the BEL is bright (Marziani et al. 1993). The double peaked Balmer lines naturally lend themselves to a 3 component Gaussian fit (Brotherton et al 1994; Marziani et al. 1996; Sulentic et al. 2000). First, there is a red broad component (red BC, sometimes just called the BC in objects in which it is not highly red-shifted) that is shifted to the red by varying amounts. There is a very broad component (VBC) with a FWHM $\sim$8,000km/sec -10,000 km/sec that is red-shifted, typically less than the red BC. There is also a blue-shifted component, the blue BC. To further complicate the issues, there are semi-broad components (SBCs) in OQ208 that are associated with most of the NCs. The NCs have FWHM$\sim$600 km/s, while the SBCs typically have a FWHM$\sim$ 1800-2500 km/s. An example of this decomposition is given in Figure 2. What seems to be depicted is a complex emission region with a wide range of velocity that is being modeled with mathematical modules (Gaussian functions) that are not necessarily fundamental physical elements. The 3 BCs appear with varying prominence over the 40 years of our data. The exact decomposition is not unique. The least uncertain quantities are probably the total BEL flux and the red/blue asymmetry. Thus, in this study, we do not attempt to classify the changes in individual components over time. We accept that a method based on the Gaussian decomposition or any simple mathematical depiction is prone to uncertainty in any individual epoch. Thus, this effort focuses on multi-epoch trends of the most reliable measurable quantity, the total BEL flux.

### 3.1. *Broad Emission Line Flux Calibration*

In order to compare BEL fluxes over many decades requires utilizing the most accurate flux calibrations possible, to place data of varying quality on a common flux density scale. Spectroscopic data from a range of archival sources are compiled in Table 2. The new data in this paper were obtained with the 10 m Hobby-Eberly Telescope (HET, Ramsey et al. 1998; Hill et al. 2021) as discussed below. We have two reliable standard star calibrations, SDSS in 2006 and HET in 2024, with flux uncertainties $\sim$5%-10%. Typically observations obtained with slit spectrographs can have difficulty accounting for slit losses or differential atmospheric refraction (DAR, Filippenko 1982) that influence the accuracy of the system response and absolute flux density calibration.

To remedy this uncertainty in flux calibration, narrow lines (NLs) are traditionally used as "standard candles"(Gezari et al. 2007) to calibrate the spectrophotometry. The narrow line region is far from the quasar and

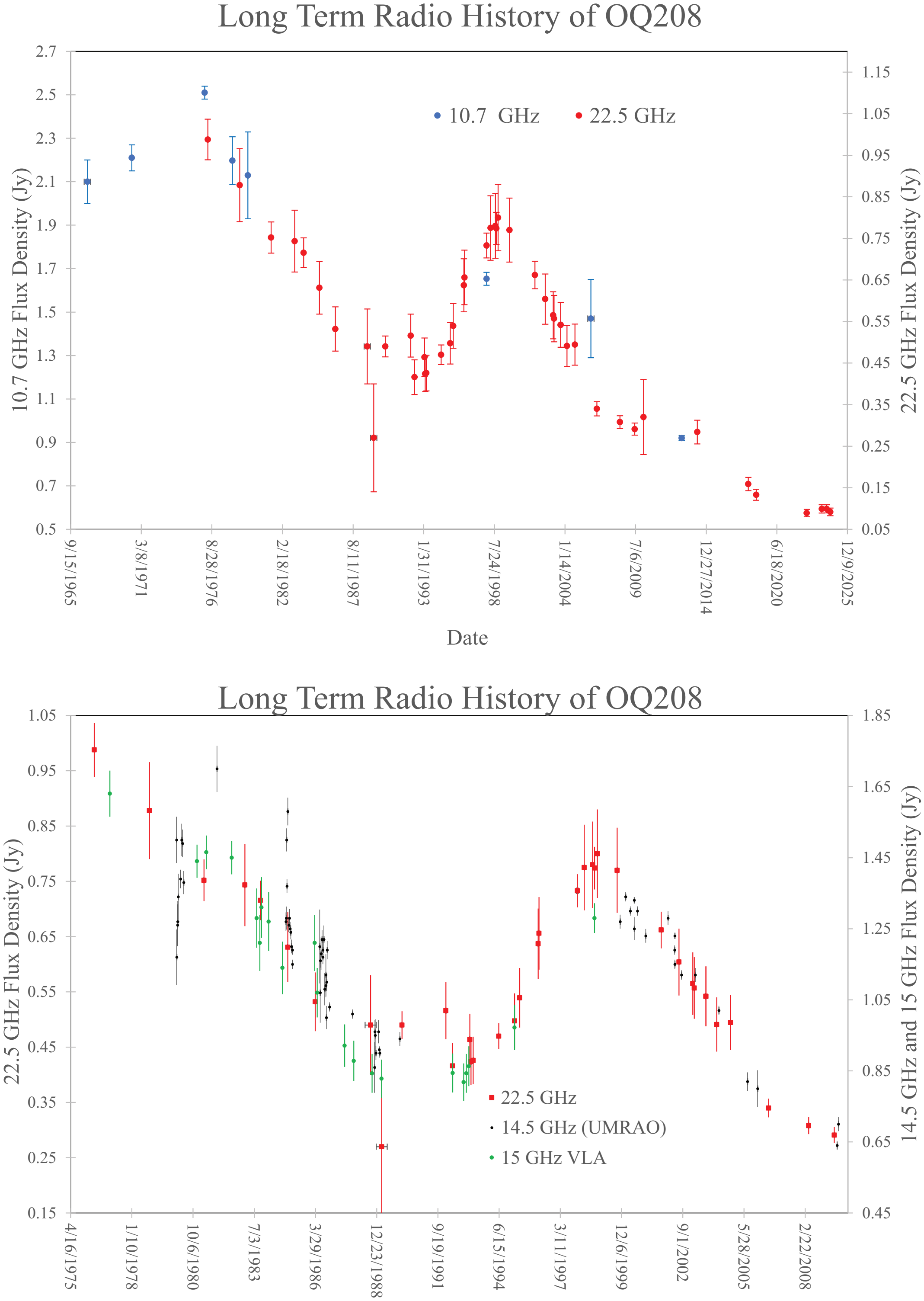

**Figure 1.** The ∼ 58 year history of the radio source OQ208. High radio frequency observations provide the most direct tool for tracking the evolution of the jet central engine. The total OQ208 flux density is plotted at multiple frequencies in order to improve the temporal coverage. The top plot highlights the elevated flux from 1967 until the start of 22 GHz monitoring. The bottom plot uses 15 GHz data to help fill in the gaps in the 22 GHz data. From these plot we identified an elevated flux from 1966-1970, a large flare with a peak ∼ 1975, that declines to a trough in 1989. Then there is another flare that peaks in 1998, the main subject of this study. A possible small, short flare occurs in 1981. The plots are not intended to provide spectral information. For that information Table 1 can be useful.

**Table 1.** 58 Year Radio History of OQ208: Total Flux Density

| Observed Frequency (GHz) | Date | Flux Density (Jy) | Telescope | Project Code | References |
|---|---|---|---|---|---|
| 10.7 | $\sim$1/1/1967[a] | $2.1 \pm 0.1$ | 46m Algonquin Park | N/A | Krauss and Scheer (1967); Krauss et al. (1968) |
| 10.7 | $\sim$6/1/1970-6/30/1970 | $2.21 \pm 0.06$ | NRAO 140 foot | N/A | Kellermann and Pauliny-Toth (1973) |
| 10.52 | 2/6/1976 | $2.51 \pm 0.03$ | 46m Algonquin Park | N/A | McCutcheon and Gregory (1978) |
| 10.7 | 4/2/1978 | $2.20 \pm 0.11$ | 91m Green Bank | N/A | Bicay et al. (1995) |
| 10.7 | $\approx$6/15/1979 | $2.13 \pm 0.12$ | OVRO 40m | N/A | Israel et al. (1988) |
| 10.7 | 12/16/1997 | $1.65 \pm 0.03$ | RATAN-600 | N/A | Kovalev et al. (1999) |
| 10.45 | $\sim$1/16/2006[a] | $1.47 \pm 0.18$ | 100m Effelsberg | N/A | Vollmer et al. (2008) |
| 10.45 | $\approx$1/30/2013[a] | $0.92 \pm 0.01$ | 100m Effelsberg | N/A | Pasetto et al. (2016) |
| 15.1 | 1/16/1977 | $1.63 \pm 0.1$ | 11m NRAO | N/A | Owen et al. (1978) |
| 14.9 | 12/19/1980 | $1.44 \pm 0.07$ | VLA | WROB | This paper |
| 14.9 | 5/10/1981 | $1.46 \pm 0.07$ | VLA | RUDN | This paper |
| 14.9 | 6/30/1982 | $1.45 \pm 0.07$ | VLA | BURN | This paper |
| 14.9 | 8/12/1983 | $1.28 \pm 0.13$ | VLA | AR98 | This paper[b] |
| 14.9 | 10/1/1983 | $1.21 \pm 0.12$ | VLA | AH129 | This paper[b] |
| 14.9 | 10/29/1983 | $1.31 \pm 0.13$ | VLA | AB287 | This paper[b] |
| 14.9 | 2/22/1984 | $1.27 \pm 0.13$ | VLA | AM67 | This paper[b] |
| 14.9 | 10/4/1984 | $1.14 \pm 0.11$ | VLA | AR115 | This paper[b] |
| 14.9 | 3/15/1986 | $1.21 \pm 0.12$ | VLA | AF113 | This paper[b] |
| 14.9 | 4/27/1986 | $1.07 \pm 0.11$ | VLA | AL119 | This paper[b] |
| 14.9 | 7/18/1987 | $0.92 \pm 0.09$ | VLA | AM205 | This paper[b] |
| 14.6 | 12/14/1987 | $0.88 \pm 0.09$ | VLA | AM224 | This paper[b] |
| 14.9 | 10/8/1988 | $0.84 \pm 0.08$ | VLA | AM275B | This paper[b] |
| 14.9 | 3/13/1989 | $0.83 \pm 0.08$ | VLA | AC234 | This paper[b] |
| 14.9 | 5/19/1992 | $0.84 \pm 0.08$ | VLA | AD286 | This paper[b] |
| 14.9 | 11/13/1992 | $0.82 \pm 0.08$ | VLA | AK312 | This paper[b] |
| 14.9 | 12//27/1992 | $0.84 \pm 0.08$ | VLA | A357 | This paper[b] |
| 14.9 | 2/4/1993 | $0.86 \pm 0.09$ | VLA | AW230 | This paper[b] |
| 14.9 | 2/21/1995 | $0.97 \pm 0.10$ | VLA | AH593 | This paper[b] |
| 14.9 | 9/21/1998 | $1.28 \pm 0.06$ | VLA | AS645 | This paper |
| 22.5 | 5/4/1976 | $0.998 \pm 0.049$ | 46m Algonquin Park | N/A | McCutcheon and Gregory (1978) |
| 22.5 | 10/24/1978 | $0.878 \pm 0.088$ | VLA | SHAF | This paper |
| 22.5 | 4/2/1981 | $0.752 \pm 0.038$ | VLA | WALK | This paper |
| 22.5 | 1/30/1983 | $0.743 \pm 0.074$ | VLA | AW87 | This paper[b] |
| 22.5 | 10/10/1983 | $0.716 \pm 0.072$ | VLA | AW87 | This paper |
| 22.4 | 12/30/1984 | $0.631 \pm 0.63$ | VLA | AO53 | This paper |
| 22.4 | 3/25/1986 | $0.532 \pm 0.53$ | VLA | AM1 | This paper |
| 22.2 | $\approx$9/15/1988 | $0.49 \pm 0.09$ | 14m Metsahovi | N/A | Wiren et al. (1992) |
| 22.2 | $\approx$3/15/1989 | $0.27 \pm 0.13$ | 14m Metsahovi | N/A | Wiren et al. (1992) |
| 22.4 | 2/9/1990 | $0.490 \pm 0.025$ | VLA | | Stanghellini et al. (1998) |
| 22.4 | 2/27/1992 | $0.516 \pm 0.052$ | VLA | TEST | This paper[b] |
| 22.2 | 5/14/1992 | $0.416 \pm 0.042$ | VLA | BY1Y | This paper[b] |
| 22.4 | 2/21/1993 | $0.464 \pm 0.046$ | VLA | AL292 | This paper[b] |
| 22.4 | 3/14/1993 | $0.424 \pm 0.042$ | VLA | AA149 | This paper[b] |
| 22.4 | 4/17/1993 | $0.454 \pm 0.045$ | VLA | AL288 | This paper[b] |
| 22.4 | 6/8/1994 | $0.470 \pm 0.024$ | VLA | BL7B | This paper |
| 22.4 | 2/21/1995 | $0.498 \pm 0.050$ | VLA | AD353 | This paper[b] |
| 22.4 | 5/17/1995 | $0.539 \pm 0.054$ | VLA | AM474 | This paper |
| 22.4 | 3/13/1996 | $0.637 \pm 0.064$ | VLA | AB766 | This paper[b] |
| 22.4 | 3/29/1996 | $0.656 \pm 0.066$ | VLA | BM49 | This paper[b] |
| 22 | 12/16/1997 | $0.733 \pm 0.03$ | RATAN-600 | N/A | Kovalev et al. (1999) |
| 22.5 | 4/6/1998 | $0.775 \pm 0.078$ | VLA | AB859 | This paper |
| 22.4 | 8/22/1998 | $0.780 \pm 0.078$ | VLA | AS637 | This paper[b] |
| 22.5 | 9/21/1998 | $0.774 \pm 0.038$ | VLA | AS645 | This paper |
| 22.5 | 11/7/1998 | $0.800 \pm 0.080$ | VLA | AS645 | Dallacasa et al. (2000) |
| 22.5 | 9/25/1999 | $0.771 \pm 0.077$ | VLA | AD428 | This paper |
| 22.5 | 9/14/2001 | $0.661 \pm 0.033$ | VLA | AG617 | This paper |
| 22.5 | 7/3/2002 | $0.604 \pm 0.060$ | VLA | AT275 | Tinti et al. (2005) |
| 22.4 | 3/10/2003 | $0.557 \pm 0.056$ | VLA | AP450 | This paper[b] |
| 22.2 | 9/15/2003 | $0.542 \pm 0.054$ | VLA | | Orienti et al. (2007) |
| 22.4 | 3/9/2004 | $0.491 \pm 0.049$ | VLA | TESTGM | This paper[b] |
| 22.4 | 10/22/2004 | $0.495 \pm 0.050$ | VLA | STUDEN | This paper[b] |
| 22.4 | 7/4/2006 | $0.340 \pm 0.017$ | VLA | AR603 | This paper |
| 22.4 | 4/19/2008 | $0.308 \pm 0.015$ | VLA | AR642 | This paper |
| 22.4 | 6/11/2009 | $0.291 \pm 0.015$ | VLA | AM991 | This paper |
| 21.7 | $\approx$2/15/2010 | $0.32 \pm 0.09$ | KVN Yonsei | N/A | Lee et al. (2017) |
| 22.5 | 4/16/2014 | $0.284 \pm 0.028$ | VLA | 14A-460 | This paper |
| 22.0 | 4/2/2018 | $0.159 \pm 0.016$ | VLA | AM1560 | This paper |
| 22.5 | 11/11/2018 | $0.133 \pm 0.013$ | VLA | AC1369 | Deanne Coppejans for this paper |
| 22.5 | 10/13/2022 | $0.089 \pm 0.009$ | VLA | AL1247 | Tanmoy Laskar for this paper |
| 22.5 | 12/14/2023 | $0.099 \pm 0.010$ | VLA | AP851 | Eli Pattie for this paper |
| 22.0 | 4/27/2024 | $0.099 \pm 0.010$ | VLA | AM1763 | Joe Michail for this paper |
| 22.0 | 4/27/2024 | $0.092 \pm 0.009$ | VLA | AL1310 | Tanmoy Laskar for this paper |
| 43.3 | 3/9/1999 | $0.205 \pm 0.021$ | VLA | AB902 | This paper |
| 43.3 | 9/15/2000 | $0.178 \pm 0.036$ | VLA | AG598 | This paper |
| 150 | $\approx$6/1/1998 | $0.041 \pm 0.010$ | SCUBA/JCMT | N/A | Anton et al. (2004) |

[a]The exact date was not recorded. Accurate to within 2-3 months.

[b]NRAO Image Archive https://www.vla.nrao.edu/astro/archive/pipeline/position/J140700.3+282714

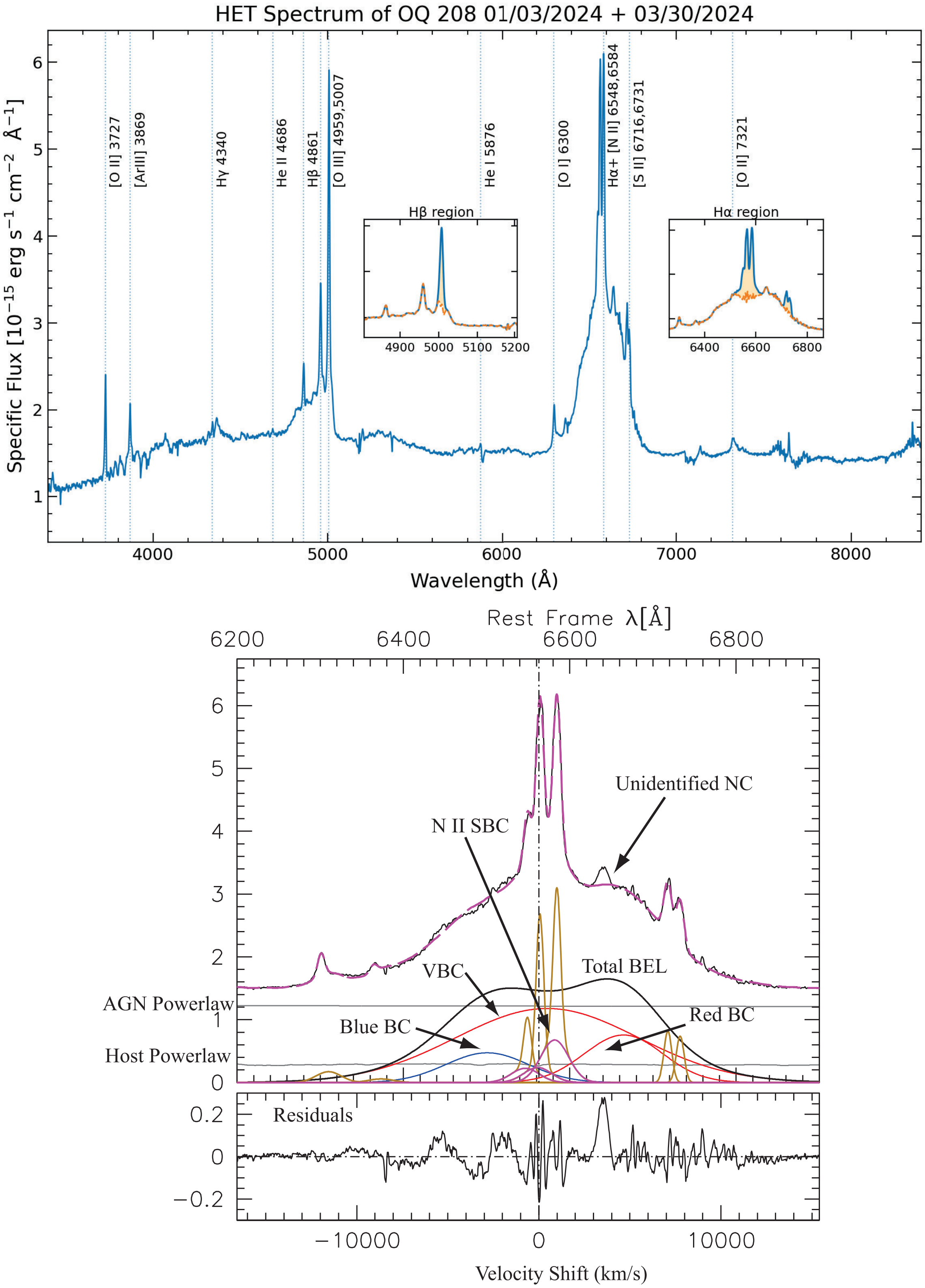

**Figure 2.** The Hobby-Eberly Telescope LRS2 2024 observation is shown in the top panel. Abscissa is rest-frame wavelength in Å, and ordinate is rest frame flux density. This observation is used to set the absolute flux of the narrow lines, as shown by the expanded view of the H$\beta$ and H$\alpha$ spectral ranges in the insets. These narrow component fluxes are used as the calibration standard for other epochs. The deep HET observation during a low BEL flux state is the most accurate narrow component flux standard that we can obtain. The bottom panel is an example of the elements involved in the fitting of the H$\alpha$ lines for the Hobby-Eberly Telescope observation on March 30, 2024. The proportion of flux in the various broad components changes over time. The unidentified narrow component also appears (less prominently) in the February 20, 2023 spectrum and is considered real.

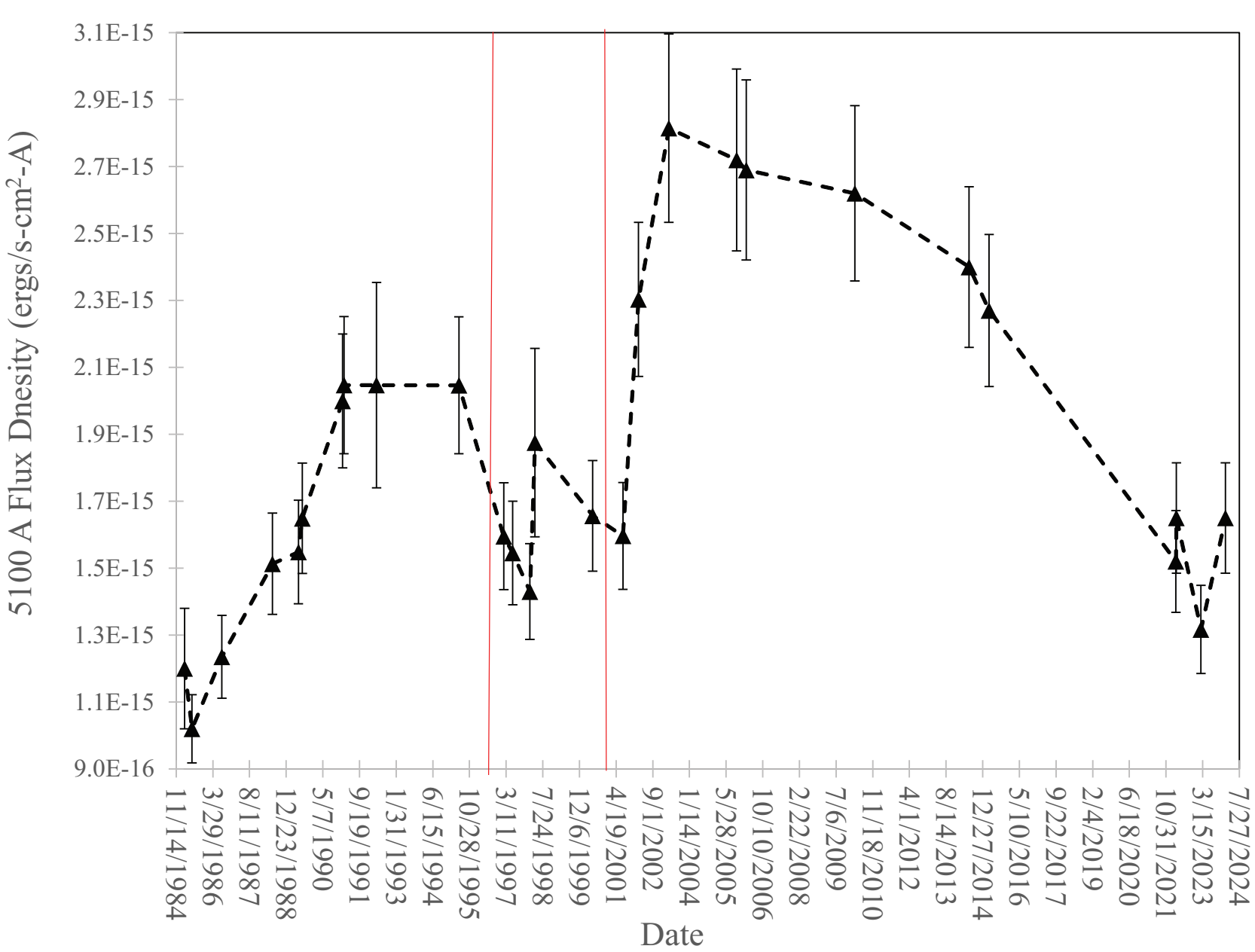

**Figure 3.** The 5100Å light curve from Table 2. Significant changes in the specific flux occur with smooth gradients, consistent with a linear rise or fall. The red lines demarcate the time frame where weak H$\alpha$ broad lines were found in Gezari et al. (2007), see our Figure 4. Furthermore, notice that the optical flux is relatively low in this period. However, similar or lower flux is seen between 1985 and 1989 and between 2022 and 2024.

should be relatively stable for decades. We are interested in three quantities in particular, the continuum flux density at 5100 Å in the rest frame (a standard indicator of accretion disk luminosity), and the H$\alpha$ and H$\beta$ BEL fluxes (Table 2). The H$\alpha$ broad line has historically been the most studied in this source, since it has much higher signal-to-noise (S/N) ratio than H$\beta$ and not blended with the difficult to characterize strong FeII blends (Eracleous and Halpern 1994; Gezari et al. 2007). With this program in mind, we proceed to characterize the best available narrow lines for calibration.

The strength of this method relies on having an accurate measurement of the narrow lines that can be applied to data of varying resolving power and S/N ratio.This is best accomplished with a large telescope (high S/N) and a spectrograph that can be calibrated against simultaneous photometry and which does not suffer from significant slit losses. Furthermore, the observation should occur in a state with weak broad lines, so the contrast of the narrow lines is more pronounced. This reduces the confusion produced by overlapping lines. Thus, we observed OQ208 with 10 m Hobby-Eberly Telescope (HET) during a low state in early 2024, utilizing the integral field Low Resolution Spectrograph 2 (LRS2, Chonis et al. 2016) and calibrating against guide-star photometry obtained simultaneously with the data. Both units of LRS2 were employed in separate observations. The source was viewed with the blue unit of the integral field spectrograph, LRS2-B, on 1/03/2024 and with the red unit, LRS2-R, on 3/30/2024. Each unit is fed by an integral field unit with 6x12 square arcsec field of view, 0.6$''$ spatial elements, and full fill-factor, and has two spectral channels. LRS2-B has two channels, UV (3700 Å - 4700 Å) and Orange (4600 Å - 7000 Å), observed simultaneously. LRS2-R also has two channels, Red (6500 Å - 8470 Å) and Far-red (8230 Å - 10500 Å). Hence there is overlap between 6500 Å and 7000 Å with which to cross-check the spectrophotometry. The system responses (relative response as a function of wavelength) of the four channels of LRS2 are very well known and stable over long periods (Zeimann et al, 2025, in prep.), and were calibrated with standard star observations the same night. Conditions were good with transparency in the g band averaging about 80% and 1.5 to 1.6 arcsec image quality, measured from the guide cameras. The January 2024 observation was obtained in dark time while the March observation had some moonlight $\sim$0.5 magnitudes brighter sky in the SDSS g' bandpass. The raw LRS2 data were processed using the **Panacea** pipeline (Zeimann et al. 2025, in prep.)[2], which performs detector-level calibrations including overscan and bias subtraction, flat-fielding, fiber tracing and extraction, and wavelength calibration using arc-lamp exposures. Panacea also derives an initial relative flux calibration from the instrument's default response curves and estimates of the telescope illumination and exposure throughput, as determined from guide-camera photometry of field stars. The simultaneous flux calibration against $g'$-band imaging ensures a system response stable to better than 5% and an absolute calibration accurate to approximately 10%. To combine data from multiple channels and exposures, we applied additional modeling using **LRS2Multi**[3], which operates on the un-sky-subtracted, flux-calibrated fiber spectra. For each exposure, we modeled the sky background using an annulus surrounding the target. The resulting model was subtracted prior to optimally extracting the target spectrum. We emphasize that the use of an integral field spectrograph eliminates slit- and aperture-related flux losses, and that DAR effects are corrected as part of the reconstruction process, resulting in negligible wavelength-dependent flux loss For each night, the LRS2-B and LRS2-R spectra were internally scaled to match in their overlap region. The spectra from the two nights were then cross-normalized to ensure consistent flux calibration between epochs. The inter-night scaling factors were 0.98 and 1.02, respectively, as determined from the mean flux density in the same overlap region, demonstrating the stability of the relative calibration. The final combined spectra were then telluric-corrected and shifted to the rest frame for presentation in Figure 2.

We consider the following NLs, [OIII]$\lambda$5007, [OI]$\lambda$6300, H$\alpha\lambda$6563, [NII]$\lambda$6583 and [SII]$\lambda\lambda$6716,6731. [OIII]$\lambda$5007 is high signal to noise. However, its extraction is complicated by an additional semi-broad component (SBC) line region in this object with FWHM typically $\sim$ 1800 km/sec -2200 km/sec. Its existence along with a complicated "double-peaked," strongly evolving BEL shape makes the definition of the [OIII]$\lambda$5007 NL nontrivial. [OIII] will be used to calibrate the continuum flux density at 5100 Å and the H$\beta$/[OIII]/FeII complex whenever it is available. For, example the flux of [OIII]$\lambda$5007 from our HET observation is $3.49 \times 10^{-14}$ergs/s/cm$^2$. The [OIII] correction or calibration factor for the visible band is,

$$\mathrm{[OIII]\,cal} = \mathrm{F([OIII]}\lambda 5007)/3.49 \times 10^{-14}\mathrm{ergs/s-cm^2}\,, \tag{1}$$

where F([OIII]$\lambda$5007) is the fitted flux of NL [OIII]$\lambda$5007 in the observed spectrum.

In the red region, [OI]$\lambda$6300 is a relatively weak line making it sensitive to the S/N. Furthermore, it is mixed with an SBC making for large uncertainties in the fit. This is not useful for these purposes. H$\alpha\lambda$6563 and [NII]$\lambda$6583 are both strong lines. However, they also have a strong SBC component and all of this makes for a complex of overlapping lines, the H$\alpha$ complex. Since the region is so complicated, empirically it was determined that there was less scatter if we combined the fitted fluxes of the two adjacent narrow lines. Thus, a more robust calibration standard is F(H$\alpha\lambda$6563)+F([NII]$\lambda$6583). We add an additional constraint of, F(H$\alpha\lambda$6563)/F([NII]$\lambda$6583) = 0.9-1.1 which is less restrictive than the constraint, F(H$\alpha\lambda$6563)/F([NII]$\lambda$6583) = 0.97 used in (Gezari et al. 2007). The HET value of F(H$\alpha\lambda$6563)+F([NII]$\lambda$6583) is $5.4 \times 10^{-14}$ergs/s-cm$^2$. The H$\alpha\lambda$6563 + [NII]$\lambda$6583 correction or calibration factor is

$$\mathrm{H\alpha + [NII]\,cal} = [\mathrm{F(H\alpha\lambda 6563) + F([NII]\lambda 6583)}]/5.4 \times 10^{-14}\mathrm{ergs/s-cm^2}\,. \tag{2}$$

On the other hand, [SII]$\lambda\lambda$6716,6731 is in the tail of the H$\alpha$ broad line and does not have a measurable SBC. These are not strong lines and their extraction is somewhat susceptible to S/N and spectral resolution. As such, it is much more robust to consider the total flux from the two lines than each line separately for calibration purposes. The HET fitted value of F([SII]$\lambda\lambda$6716,6731)is $1.86 \times 10^{-14}$ergs/s/cm$^2$. The [SII] correction or calibration factor is

$$\mathrm{[SII]\,cal} = \mathrm{F([SII]}\lambda\lambda 6716, 6731)/1.86 \times 10^{-14}\mathrm{ergs/s-cm^2}\,. \tag{3}$$

In general, we consider using red NLs to calibrate the H$\alpha$ complex superior to the [OIII] cal, due to wavelength dependent calibration errors that can result from uncorrected DAR in slit spectra, for example. A slope, or curvature, of the continuum, induced by calibration errors was evident in a few spectra, as noted in Table 2. For the calibration of the H$\alpha$/[NII] complex we use the following in order of preference:

1. We average the correction factors H$\alpha$ + [NII] cal and [SII] cal if they agree to within 15% (table note f in Table 2).

[2] https://github.com/grzeimann/Panacea

[3] https://github.com/grzeimann/LRS2Multi

**Table 2.** Optical Fitted Spectral Data in QSO Rest Frame

| Date | Telescope | Spectrum reference | 5100 Å Flux Density[a] | Extrapolated from | Hα [a] Broad Line EW (Å) | Hβ [a] Broad Line EW(Å) | [OIII] Cal Factor | Hα+[NII] Cal Factor | [SII] Cal Factor | Hα Correction Factor | Comments |
|---|---|---|---|---|---|---|---|---|---|---|---|
| 3/8/1985 | INT 2m | Marziani et al. (1993) | $1.2 \pm 0.18$ | 4800 Å | $228^{b}$ | $61^{b}$ | ... | .... | ... | ... | H$\beta$ and [SII] very noisy |
| 6/15/1985 | INT 2m | Marziani et al. (1993) | $1.02 \pm 0.10$ | 4800 Å | $392^{b}$ | $84^{b}$ | ... | .... | ... | ... | ... |
| 7/31/1986 | INT 2m | Marziani et al. (1993) | $1.24 \pm 0.12$ | 4800 Å | $254^{b}$ | $71^{b}$ | ... | .... | ... | ... | [SII] low S/N |
| 6/18/1988 | INT 2m | Marziani et al. (1993) | $1.51 \pm 0.15$ | 4800 Å | $318^{b}$ | $75^{b}$ | ... | .... | ... | ... | [SII] low S/N |
| 6/8/1989 | INT 2m | Marziani et al. (1993) | $1.55 \pm 0.16$ | 4800 Å | $433^{b}$ | $76^{b}$ | ... | .... | ... | ... | [SII] low S/N |
| 7/31/1989 | INT 2m | Marziani et al. (1993) | $1.65 \pm 0.17$ | 4800 Å | $370^{b}$ | $70^{b}$ | ... | ... | ... | ... | [SII] low S/N |
| 2/1/1991 | KPNO 4 m | Boroson and Meyers (1992) | $2.0 \pm 0.20^{c}$ | ... | $267^{c}$ | $104^{c}$ | ... | ... | ... | ... | ... |
| 2/19/1991 | KPNO 2.1 m | Marziani et al. (1993) | $2.05 \pm 0.21$ | 4800 Å | ... | $83^{b}$ | ... | ... | ... | ... | ... |
| 5/9/1992 | MMT 4.5 m | Marcha et al. (1996) | $2.0 \pm 0.3$ | ... | ... | ... | ... | ... | ... | ... | ... |
| 6/6/1995 | WHT 4.1 m | Corbett et al. (1998) | $2.05 \pm 0.21^{d,e,f}$ | ∼ 6000 Å | $292^{e,f}$ | ... | ... | 1.13 | 1.06 | 1.09 | (Hα+[NII] cal)/ [SII] cal)=1.07 |
| 2/7/1997 | KPNO 2.1 m | Gezari et al. (2007) | $1.62 \pm 0.16^{e,f}$ | ∼ 5500 Å | $205^{e,f}$ | ... | ... | 1.37 | 1.38 | 1.38 | (Hα+[NII] cal)/([SII] cal)=1.00 |
| 6/9/1997 | KPNO 2.1 m | Gezari et al. (2007) | $1.55 \pm 0.16^{e,f}$ | ∼ 5500 Å | $239^{e,f}$ | ... | ... | 1.15 | 1.34 | 1.25 | (Hα+[NII] cal)/([SII] cal)=0.86 |
| 1/29/1998 | KPNO 2.1 m | Gezari et al. (2007) | $1.43 \pm 0.14^{e,g}$ | ... | $191^{e,f}$ | $53^{e,i}$ | 1.66 | 1.53 | 1.59 | 1.56 | (Hα+[NII] cal)/ [SII] cal)=0.94 |
| 4/8/1998 | MDM 2.4 m | Gezari et al. (2007) | $1.88 \pm 0.28^{e}$ | ... | $235^{e}$ | $73^{e,j}$ | 1.29 | 1.55 | 1.77 | 1.55 | [OIII] and [SII] cal discrepant[j] |
| 6/3/2000 | KPNO 2.1 m | Gezari et al. (2007) | $1.69 \pm 0.17^{e,g}$ | ... | $228^{e,f}$ | $53^{e,g}$ | 1.42 | 1.15 | 1.36 | 1.26 | (Hα+[NII] cal)/([SII] cal)=0.85 |
| 7/21/2001 | KPNO 2.1 m | Gezari et al. (2007) | $1.79 \pm 0.18^{e,h}$ | ∼ 5500 Å | $266^{e,h}$ | ... | ... | 0.47 | 0.57 | 0.57 | (Hα+[NII] cal)/([SII] cal)=0.82 |
| 2/19/2002 | Copernico 1.82 m | This paper | $2.30 \pm 0.23^{e,g}$ | ... | ... | $82^{e,g}$ | 0.85 | ... | ... | ... | ... |
| 4/8/2003 | Copernico 1.82 m | This paper | $2.81 \pm 0.28^{e,g}$ | ... | ... | $83^{e,g}$ | 0.85 | ... | ... | ... | ... |
| 10/18/2005 | OAN/SPM 2.1 m | Torrealba et al. (2012) | $2.72 \pm 0.27^{j}$ | ... | $291^{e,h}$ | $63^{j}$ | ... | 1.26 | 0.98 | 0.98 | (Hα+[NII] cal)/([SII] cal)=1.29 |
| 2/27/2006 | SDSS 2.5 m | Runnoe et al. (2015) | $2.69 \pm 0.27^{k}$ | ... | $329^{e,h}$ | $94^{k}$ | ... | 0.75 | 1.02 | 1.02 | (Hα+[NII] cal)/[SII] cal)=0.73 |
| 3/20/2010 | KPNO 4 m | Runnoe et al. (2015) | $2.62 \pm 0.26^{k}$ | ... | $341^{e,h}$ | $83^{k}$ | ... | 1.97 | 1.44 | 1.44 | (Hα+[NII] cal)/([SII] cal)=1.47 |
| 6/25/2014 | APO 3.5 m | Runnoe et al. (2015) | $2.40 \pm 0.24^{e,f,l}$ | ... | $258^{e,f}$ | ... | ... | 1.43 | 1.66 | 1.54 | (Hα+[NII] cal)/([SII] cal)=0.86 |
| 3/24/2015 | SAO RAS 6 m | Afanasiev et al. (2019) | $2.34 \pm 0.23$ | ... | ... | ... | ... | ... | ... | ... | Used only for 5100 Å flux density |
| 3/15/2022 | Copernico 1.82 m | This paper | $1.52 \pm 0.15^{e,g}$ | ... | ... | $45^{e,g}$ | 0.85 | ... | ... | ... | ... |
| 3/22/2022 | Copernico 1.82 m | This paper | $1.65 \pm 0.17^{e,g}$ | ... | ... | $38^{e,g}$ | 0.85 | ... | ... | ... | ... |
| 2/20/2023 | Copernico 1.82 m | This paper | $1.30 \pm 0.13^{e,g}$ | ... | $268^{e,h}$ | $79^{e,g}$ | 0.89 | 2.23 | 1.49 | 1.49 | (Hα+[NII] cal)/ ([SII] cal)=1.50 |
| 1/03/2024 | HET 10 m | This paper | $1.65 \pm 0.17^{e}$ | ... | ... | $59^{e}$ | ... | ... | ... | ... | LRS-B integral field spectrophotometry |
| 3/30/2024 | HET 10 m | This paper | $1.65 \pm 0.17^{e}$ | ... | $287^{e}$ | ... | ... | ... | ... | ... | LRS2-R integral field spectrophotometry |

[a] Host galaxy plus AGN continuum at 5100 Å $\times 10^{-15}$ ergs/s/cm$^2$-Å; column 5 indicates if this value is extrapolated from another wavelength. Rest frame equivalent widths (EW) measured relative to the host plus AGN continuum at 5100 Å in column 4. The EW uncertainty is 5%.

[b] All values from Marziani et al. (1993).

[c] 5100Å flux density estimated from Fig.1 and EW from Table 2 of Boroson and Meyers (1992).

[e] This paper. Calibration using narrow lines fluxes from HET observations in 2024 with integral field LRS2 spectrograph, last two rows.

[f] Adopted calibration based on H$\alpha$ + [NII] narrow lines averaged with [SII] narrow line calibration.

[g] Adopted calibration based on [OIII]$\lambda$5007.

[h] Adopted calibration based on [SII] doublet.

[i] Discrepant continuum shape at ends of spectra, [OIII] cal and [SII] cal are suspect. We rely on H$\alpha$+[NII] cal =1.55, everywhere, since this is our only calibrator free of these effects. The "out of family" fitted H$\beta$ EW is likely caused by this discrepancy which occurs in the blue wing of the broad line and the FeII blend. Comparing the 4/8/1998 and 1/29/1998 spectra shows that EWs of all lines are negligibly changed, only the continuum at the ends is noticeably different.

[j] H$\beta$ flux and 5100Å flux density from Torrealba et al. (2012). H$\beta$ EW "out of family" (too small). Our calibration yields an EW = 106Å (which is an outlier at the other extreme).

[k] H$\beta$ flux and 5100Å flux density from Runnoe et al. (2015).

[l] The continuum is noisy. All fits are from this paper.

2. If they differ by more than this, we use the correction factor which agrees best with the [OIII] cal (if this line is observed).

3. If 1) is not satisfied and neither agreement with the [OIII] cal factor is within 10% or [OIII] is not observed, we use the [SII] cal correction factor if the S/N $> 5$ (table note h in Table 2).

Figure 15 in Appendix A presents the $\sim$40 years of calibrated spectra in a single plot to illustrate the changes in H$\alpha$ broad line profile over this period.

### 3.2. *Broad Line Equivalent Widths*

This is a partially obscured quasar, so it is more difficult to accurately extract the AGN continuum from the host starlight than it would be for an unobscured quasar. The importance of the galaxy continuum can be judged from the HET spectrum in Fig. 2 that clearly shows stellar photospheric features in the blue, such as the break around Ca II H,K $\lambda$3934,3968. The problem is exacerbated by our efforts to combine new spectra with numerous old spectra that we have recovered from the various archival sources (as referenced in Table 2) in order to track long-term evolution. The first step is to implement the same spectral energy distribution template for the host for all the recent epochs (the new epochs presented in this paper, all epochs after 1991). The template chosen is an evolved stellar population (11 Gyr) from the templates of (Bruzual and Charlot 2003). This template was our fit to the host of the SDSS spectrum, the spectrum with the most contamination from the host. However, the general problem of the relative flux scale of host and AGN continuum is still difficult to solve. An estimate of the host contribution depends on the aperture and the image size ("seeing"), which are not known for most observations. Thus, the host/AGN continuum decomposition has to be determined as part of the multi-component fitting process. This method (the only available option) does not guarantee high precision. For our best spectra, we find a host contribution of 22%- 30%. This compares favorably with the 27% host fraction estimated in observations of 1989 and 1990 (Eracleous and Halpern 1994). Since we cannot make a reliable decomposition in most epochs, as a point of practicality, we simply consider the 5100 Å continuum in the following as the sum of host plus AGN power law in the computation of the EW. This is table note a in Table 2.

Since the quasar is partially obscured, we are not observing an intrinsic quasar property when we compute the EW. However, our scientific interests are trends that involve relative changes in the EW from epoch to epoch. There are two issues hampering this effort, the host/AGN decomposition and outlier data points. The aforementioned $\sim \pm 10\%$ uncertainty in the host/AGN decomposition that applies to most epochs in Table 2. There are also outlier epochs due to unknown factors likely related to wavelength dependent calibration issues. The outliers will be considered as data scatter from the systematic uncertainties of our methods. The philosophy of this paper is to consider data with apparent systematic errors on the same footing as the other data. Since the data set is rather small, and since the object is intrinsically variable, mistakenly removing one of these epochs could skew the statistical analysis away from an actual trend or artificially improve the statistical significance of an hypothesis test. This situation is an unavoidable consequence of the need to use heterogeneous data samples in order to achieve the goal of a 40 year EW light curve. Only the very strongest of statistical trends can be supported by our methods of analysis. Even then we are cautious, and utilize visual evidence such as marked changes to the emission line profile to confirm that the trend is obvious.

We note that OQ208 is rather stable in the optical for a modestly obscured quasar ($\sim$ 75% quasar, $\sim$ 25% host), the B magnitude varies by $\pm 0.6$ magnitudes from 1915 to 2009 (Geffert et al. 2015; Craine and Warner 1973). Thus, a crude test of the validity of our narrow line flux calibrations would be if the 5100 Å light curve derived from the calibrated spectroscopy exhibits a similar range of variation to the B magnitudes. Our 40 year light curve, in Figure 3, also shows $\sim \pm 0.6$ magnitudes of variation. Furthermore, it is noted that the 40 year 5100 Å combined (host + AGN) continuum changes rather smoothly in Figure 3, indicating that there are probably only a few outliers. Thus, our narrow flux calibrations in Table 2 are consistent with the historical optical variability behavior of this source. After performing multiple fits on some of the epochs in Table 2, we determine that the total broad line EW computed relative to the combined continuum has an uncertainty of $\pm 5\%$.

The table of our results shows the time evolution of the total BEL flux. However, the proportion of flux in the red BC, blue BC and VBC is continually evolving. Many papers over the years have toiled with the fundamental cause of such variations in OQ208 (Osterbrock and Cohen 1979; Gaskell 1983; Marziani et al. 1993; Gezari et al. 2007). Many ideas have been suggested, such as elliptical disk distributions, hot spots, binary black holes, warped disks, inflow and outflow. This topic is beyond the scope of this treatment. The implications of the new data in Table 2 and our detailed fits will appear elsewhere (Marziani et al. in preparation).

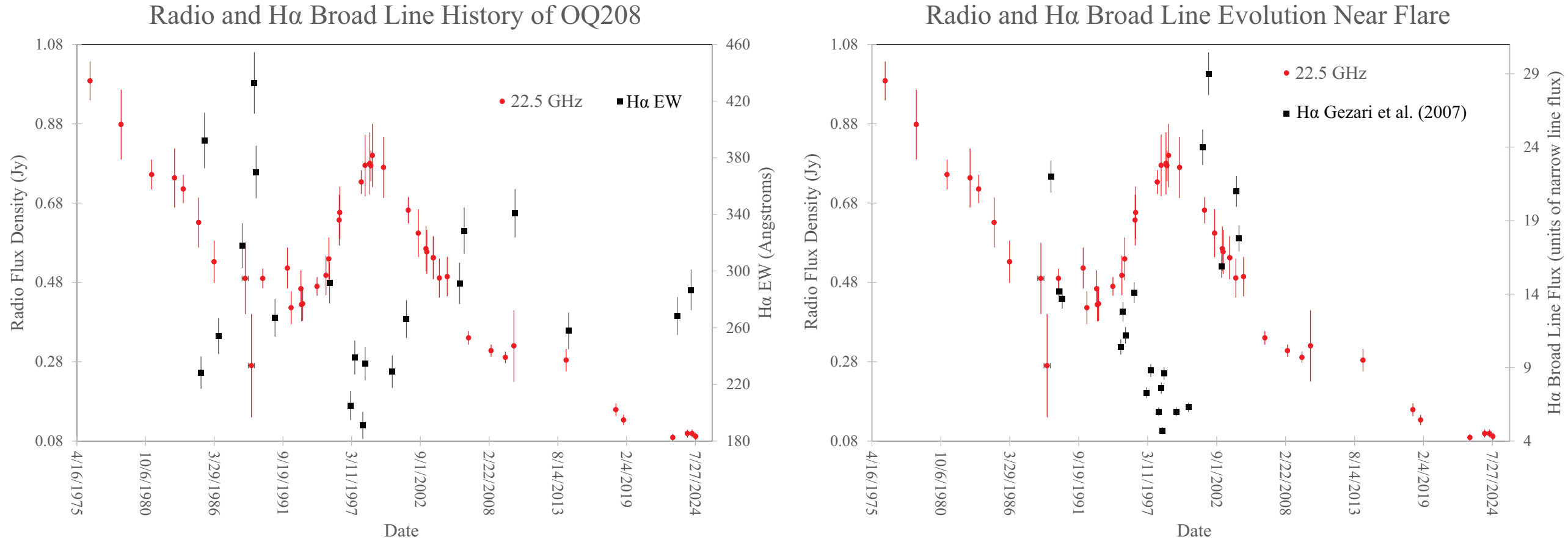

**Figure 4.** The $\sim$ 58 year history of the radio source OQ208 from Figure 1 plotted with the H$\alpha$ broad line EW from Table 2 is displayed in the left panel. It will be shown in this study, due to the steep spectral indices of the components above 20 GHz, that the total 22.5 GHz flux density of OQ208 is an excellent surrogate for the nucleus in all, but the very weakest states. The 1998 flare is the main topic of this study at which time the EW is small. Note that the 15 GHz data were removed from the light curve in this figure in order to reduce clutter. The right panel shows the H$\alpha$ broad line flux in units of the H$\alpha$ narrow flux from Gezari et al. (2007). Note that the relative error assigned by Gezari et al. (2007) for their data and that of our EW data are both 5%.

### 3.3. *Broad Line Emission During the Flare*

Table 2 indicates that the EW of H$\alpha$ is relatively small between 2/7/1997 and 6/3/2000. This is the time frame of the flare peak in Figure 1. The left panel of Figure 4 is a scatter plot of the H$\alpha$ EWs overlayed on the data presented in Figure 1. Figure 4 is very suggestive of a profound connection between the high jet state and the efficiency of the accretion flow to stimulate broad line emission. The primary uncertainty in the plot is that the continuum chosen has a host contribution. Thus, we need other evidence to show that this has not skewed the data, so as to make the plot deceiving to the eye. Firstly, in support of this interpretation is the plot in the right panel of Figure 4 that shows the results of Gezari et al. (2007) superimposed on the light curve. They defined the flux of the H$\alpha$ broad line in units of the H$\alpha$ narrow line flux. So, this is an EW of sorts that does not depend on the absolute flux calibration. This is an independent analysis that utilizes their uniquely determined AGN continuum level (which includes the host/AGN decomposition) for each fitting of the H$\alpha$ complex. The deep drop in the broad line strength during the flare is even more pronounced than we find in Table 2. Most importantly, the strong trend is found independent of the AGN/host decomposition. Thirdly, we plot the H$\alpha$ complex before, during and after the flare in Figure 5 (all of the H$\alpha$ spectra are shown in Appendix A). This figure is an"eyeball test"; there are no aspects of the fitting process involved. There are no estimates of the continuum level or the broad line FWHM which is ambiguous to define due to the three Gaussian component decomposition of the multiply peaked lines and the dependence on the fitting details. The narrowing of the complex during the flare is visually obvious and it is not plausible to construe this an illusion of a time varying host/AGN decomposition. The distinctive bulging of the line complex produced by the BEL has all but disappeared when the flare peaks. The smaller BEL flux and EW during the flare in these three analysis exists independent of the uncertainty in the host/AGN decomposition. It is therefore concluded, that Table 2 accurately describes the strong trend of the BEL change during the flare peak. This study (in particular Section 6) is focused on understanding whether there is a physical connection between the 22.5 GHz flare and the apparent decrease of the EW.

We comment on the reason for our analysis as opposed to just referencing the Gezari et al. (2007) data. They do not determine the optical continuum level. Their data, as presented, does not distinguish if the broad line emission during the flare peak is depressed simply by a low accretion luminosity in this state (which would mean a weak photo-ionizing flux) and the large radio flux is just a coincidence, or if there is a direct connection to jet launching. This motivated our desire to reveal as much of the EW and radio evolution as possible through archives and new observations. The investigation of this phenomenon is the fundamental emphasis of the remainder of this study. Our efforts will utilize the highest resolution VLBA observations that are available in each epoch.

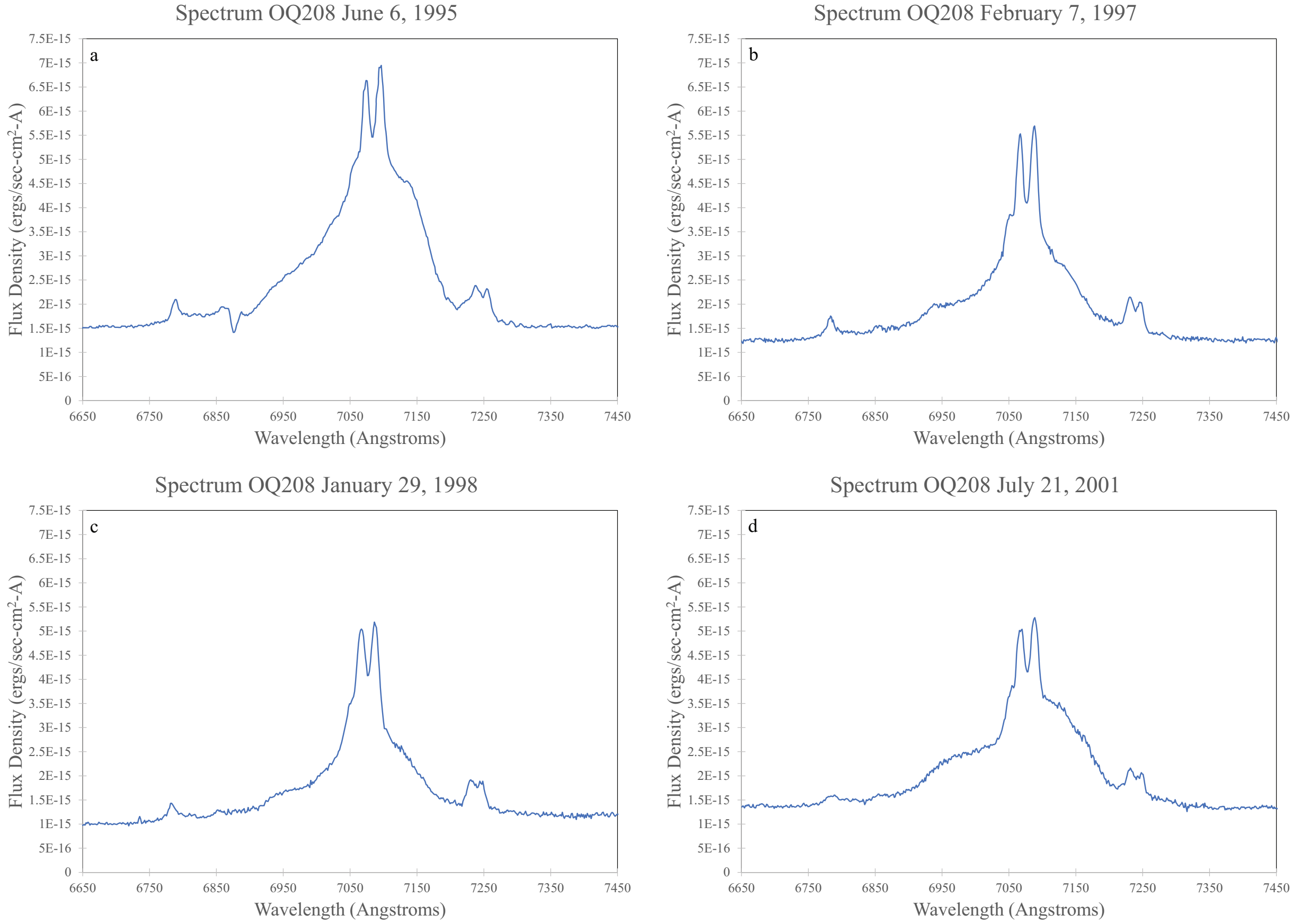

**Figure 5.** The time evolution of the H$\alpha$ complex from just before the flare, 6/6/1995 (frame a), to the flare peak (frames b and c), 2/7/1997 and 1/29/1998, to just after the flare (frame d), 7/21/2001. The data is in the quasar rest frame and has been re-calibrated with our NL method described above and detailed in Table 2. The H$\alpha$ complex is much narrower and weaker at the flare peak. Before and after the flare peak, the red and blue boundaries of the H$\alpha$ complex bulge outward considerably as expected from a stronger broad line component. Clearly, this stark contrast is intrinsic and is not related to an artifact of the host/AGN continuum decomposition.

## 4. LOCATING THE NUCLEUS OF OQ208

The location of the nucleus of OQ208 has been a source of some uncertainty. The most popularized view is that it lies in a region of low surface brightness and is actually invisible in all radio images (Lister 2003; Wu et al. 2013). We discuss the rationale for this scenario and provide kinematic evidence that it is actually located in the compact component where $\sim 80\%$ of the VLBA 15 GHz emission resides during flare states. In order to begin this investigation, we present the 15.4 GHz image from the MOJAVE observation on December 12, 2011 (Figure 6) in order to define the component names (Lister et al. 2018). The choice of component names are those of Wu et al. (2013) that were applied to an observation in 2009. The top panel of Figure 6 presents the positions of the identified components as fit in Appendix B, tagged by epoch, to summarize the kinematic evolution of the system from 1994-2012. Most notably is a narrow channel extending southwest from the bright component NE1. J1 and J2 resides in this channel at the earliest times and are in motion in 1994 and 1995. A careful inspection of the squares, representing the position of J1, shows both inward and outward motion. This is explored in the earliest stages of VLBA high frequency monitoring, in 1994 and 1995 in the lower panel of Figure 6. J1 moves outward rapidly until the end of 1995, stagnates for a few years about 1.5 mas from NE1 and then retreats, relatively slowly in the top panel. All three components, J1, J2 and J3 move away from NE1 during the time frame, 1994-1995. The VLBA Gaussian model fits used in this study are tabulated in Appendix B.

Figure 7 explores this outward motion from NE1 in detail during 1994 -1995. The main uncertainty to resolve is whether the components are moving away from NE1 or if NE1 moving away from these components. The first step is to define a static frame. The diffuse component, NE2, was considered to have the least proper motion of all the components (Wu et al. 2013). We use its location in June 1994 as a static point on the sky plane, and only consider 1.5 years of evolution in order to ensure this is a robust assumption.

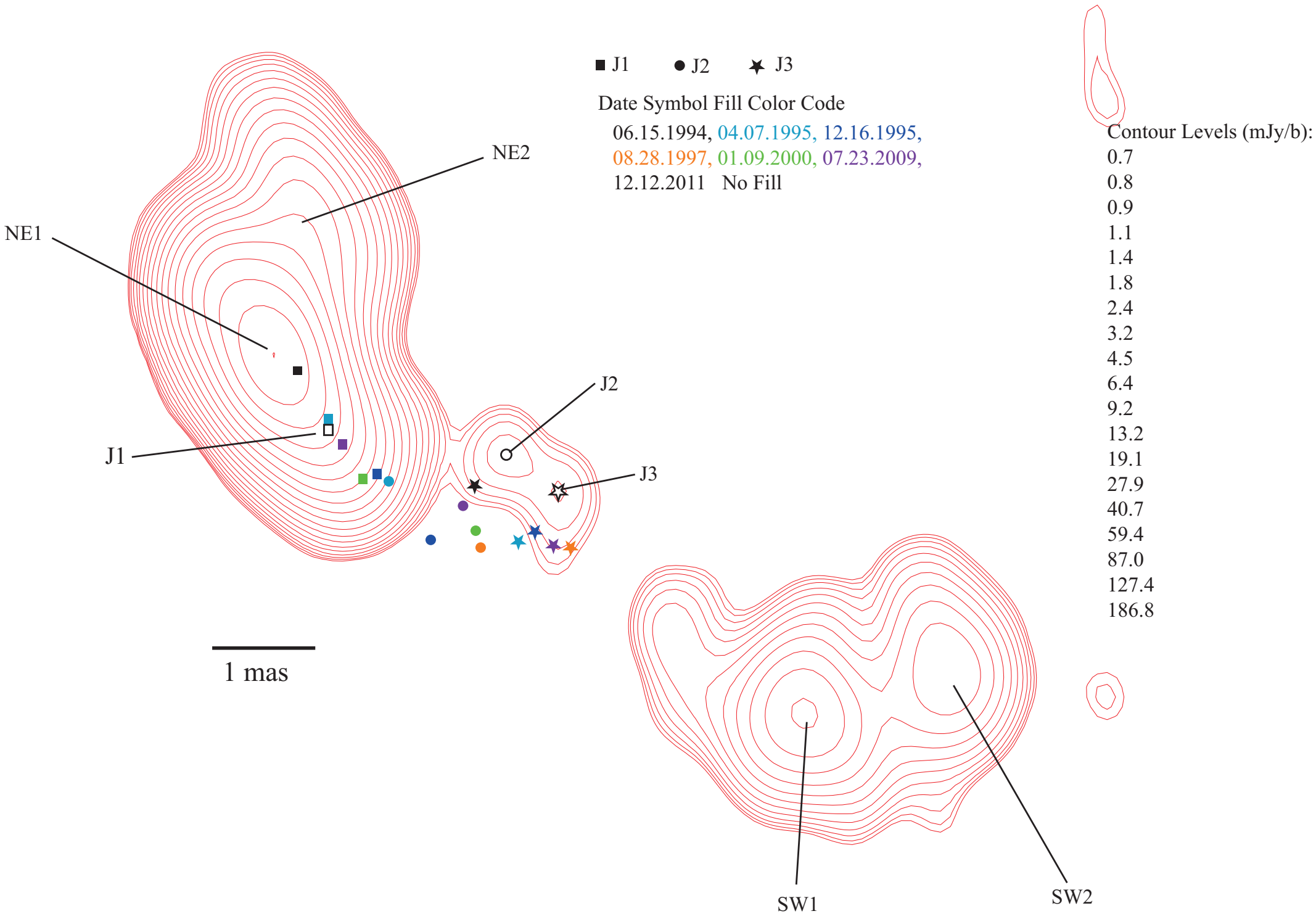

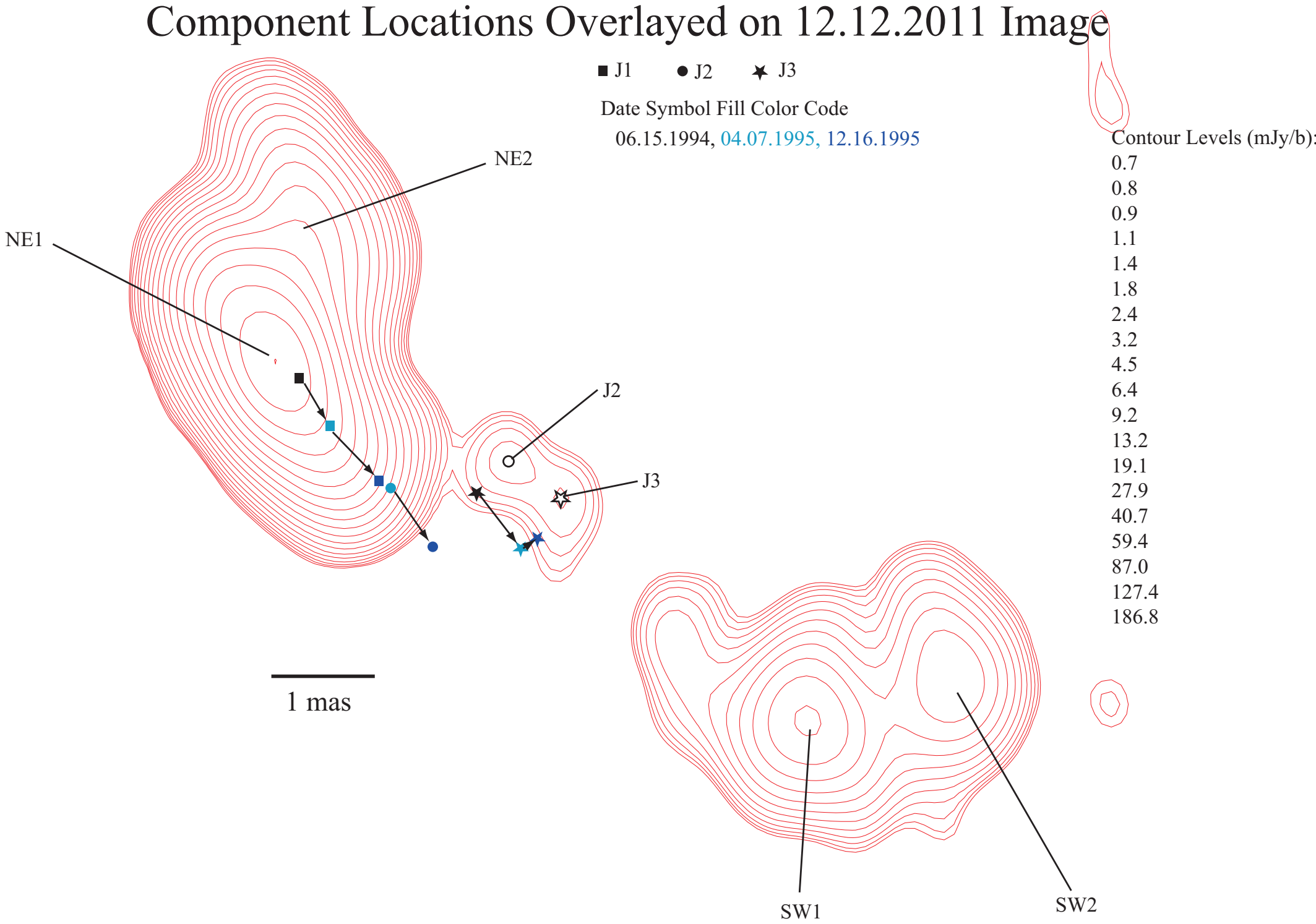

**Figure 6.** This 15.4 GHz VLBA image is used to identify the components defined by Wu et al. (2013). They are NE2, NE1, J1, J2, J3, SW1 and SW2. The vast majority of the flux is from NE1. Using the location of the flux centroid of NE1 as a fixed reference point, the positions of J1, J2 and J3 are shown at different epochs. The bottom frame shows that J1, J2 and J3 are moving away from NE1 in 1994 -1995.

The top panel of Figure 7, shows the centroid of the flux density of NE1 relative to NE2. The origin in this coordinate system is NE1 from June 1994. NE2 is located at (-0.388 mas, 1.202 mas) in all epochs. We added two 15.4 GHz observations that were not made public on the MOJAVE website, November 3 and 27 1995. November 3 is not of the quality of the public observations (4 scans over ∼3hrs, provides low density coverage in the uv plane), but should still be useful for finding peak flux locations. All the centroid positions are located within ($-0.071 \pm 0.071$ mas, $0.023 \pm 0.110$ mas). Notice that the change in position is randomly distributed on a scale $\sim \pm$ one tenth of the restoring beam (∼ 1.00-1.25 mas x 0.55-0.65 mas at PA≈ 10°). The range of uncertainty is approximately 0.1 times the restoring beam. The estimated systematic positional uncertainty of 15.4 GHz VLBA observations (from imperfect u-v coverage and phase noise) is 10% of the restoring beam FWHM (Homan et al. 2002; Lister et al. 2009). It is concluded that the scatter in the top panel of Figure 7 is random and the true position of the nucleus is only known to within 0.1 times the restoring beam FWHM. The displacements shown in the right hand panel are much larger than this, so we can make meaningful measurements of the displacement and the uncertainty. Note that, the largest change in centroid position is between two epochs that are so close in time that any changes in the physical position should be small.

The bottom panel of Figure 7 shows the displacement of the components from NE1 as a function of time in 1994 and 1995. The positional errors follow Homan et al. (2002); Lister et al. (2009) and are computed as follows:

1. Bright compact components, peak flux density > 5 post process rms. Uncertainty is 0.1 beam FWHM for both RA and Dec coordinates (Lister et al. 2009). This applies to the centroid.

2. Faint components, peak flux density < 5post process rms. Uncertainty is 0.2 beam FWHM for both RA and Dec coordinates (Lister et al. 2009).

3. Large components with a low contrast between the component and the surface brightness from the other components (i.e. J1 in early 1995) or low contrast with the rms noise (SW2), we add 0.1 of the FWHM of component in quadrature with the uncertainty in 2) for both RA and DEC.

4. Finally, the uncertainty in the centroid is added in quadrature with the uncertainty in the component.

The results are plotted in the bottom panel of Figure 7. Notice that there is no measurable displacement of the relatively distant radio lobe, SW2, away from NE1 in this 1.5 year interval. Yet, J1 and J3 definitely move away from NE1 (and hence towards SW2). The superluminal, outward speeds in the bottom panel of Figure 7 that are estimated from the slope of the linear fits are poorly constrained, since there are essentially only three distinct epochs: 1994, spring 1995 and the end of 1995. Furthermore, the linear fits are not that good. Figure 6 shows that J1 stagnates ∼ 1.5 mas from NE1 from 1996-2000. After which, it retracts toward NE1 at ≈ 0.23c (Wu et al. 2013). The kinematic analysis of Lister (2003); Wu et al. (2013) did not consider the 1994-1995 observations. Their analysis begins after 1995, so they only track the retraction stage of the motion of J1, not the earlier outward motion relative to NE1. Note that all three components in the top panel of Figure 6 reach a maximum separation then start to retreat toward NE1.

There is more evidence of the nucleus being located within NE1 based on the brightness temperature. First of all, notice that in the top right hand panel of Figure 8 that NE1 is partially resolved into three subcomponents, NE1a, NE1b and NE1c in 2000 at 22 GHz with a uniformly weighted beam (our highest resolution during the flare). The brightness temperature is computed according to the methods of Kellermann & Owen (1988):

$$T_{\rm b}[{\rm K}] = 1.22 \times 10^{12}(1+z)\frac{S_{\nu}}{\theta_1\theta_2\nu^2}\ , \tag{4}$$

where $\nu$ is the observed frequency measured in GHz, $\theta_1$ ($\theta_2$) is the major (minor) axis of the elliptical Gaussian fitted full width at half-maximum (FWHM) measured in mas, and $S_{\nu}$ is the flux density of the component in Jy. On 1/29/2000, the brightness temperature of the subcomponent of NE1bc (NE1b is blurred with NE1c) at 15.4 GHz is $6.00 \times 10^{10}$K compared to that of J1, $1.05 \times 10^{10}$K in Appendix B (Table 5M). Furthermore, matched resolution observations on 1/29/2000 - 1/31/2000 with VLBA at 15.4 GHz and 22.2 GHz and global VLBI at 8.65 GHz are used to define the spectra of the various components in Appendix C. The flaring nuclear component NE1bc has the flattest intrinsic optically thin synchrotron power law (see Table 6).

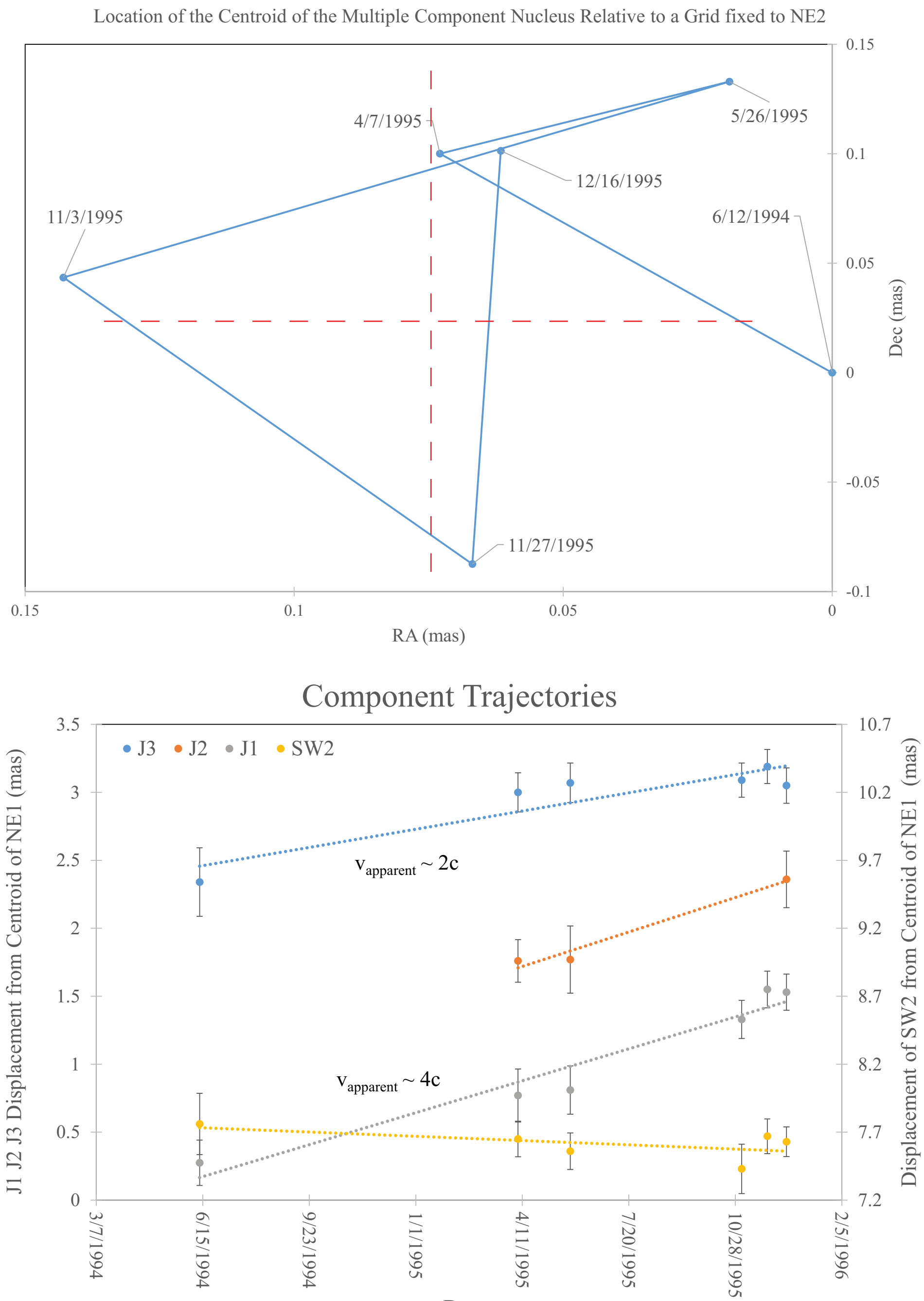

**Figure 7.** The top panel shows the centroid of the flux density of NE1 from June 1994 to December 1995 in a coordinate system fixed to the position of NE2. The figure indicates that the centroid is at a constant position with respect to NE2 from 1994 through 1995 within the systematic measurement uncertainty (Lister et al. 2009). The two November 1995 observations create most of the scatter and their positions are "far" from each other as well the December observation. These two observations are the bulk of the systematic error in the scatter plot. The systematic error bars on each data point are too large to show on this plot without reducing the size, thereby making the details difficult to see. The large red cross is indicative of a typical error associated with the centroid position of the 15.4 GHz VLBA observations, the 1994 error bar is larger. The bottom panel shows the centroid of NE1 is stationary with respect to SW2 from 1994 through 1995 within measurement uncertainty. It also shows that J1, J2 and J3 all move away from NE1 with high velocity. The change in displacement is larger than the measurement uncertainty.

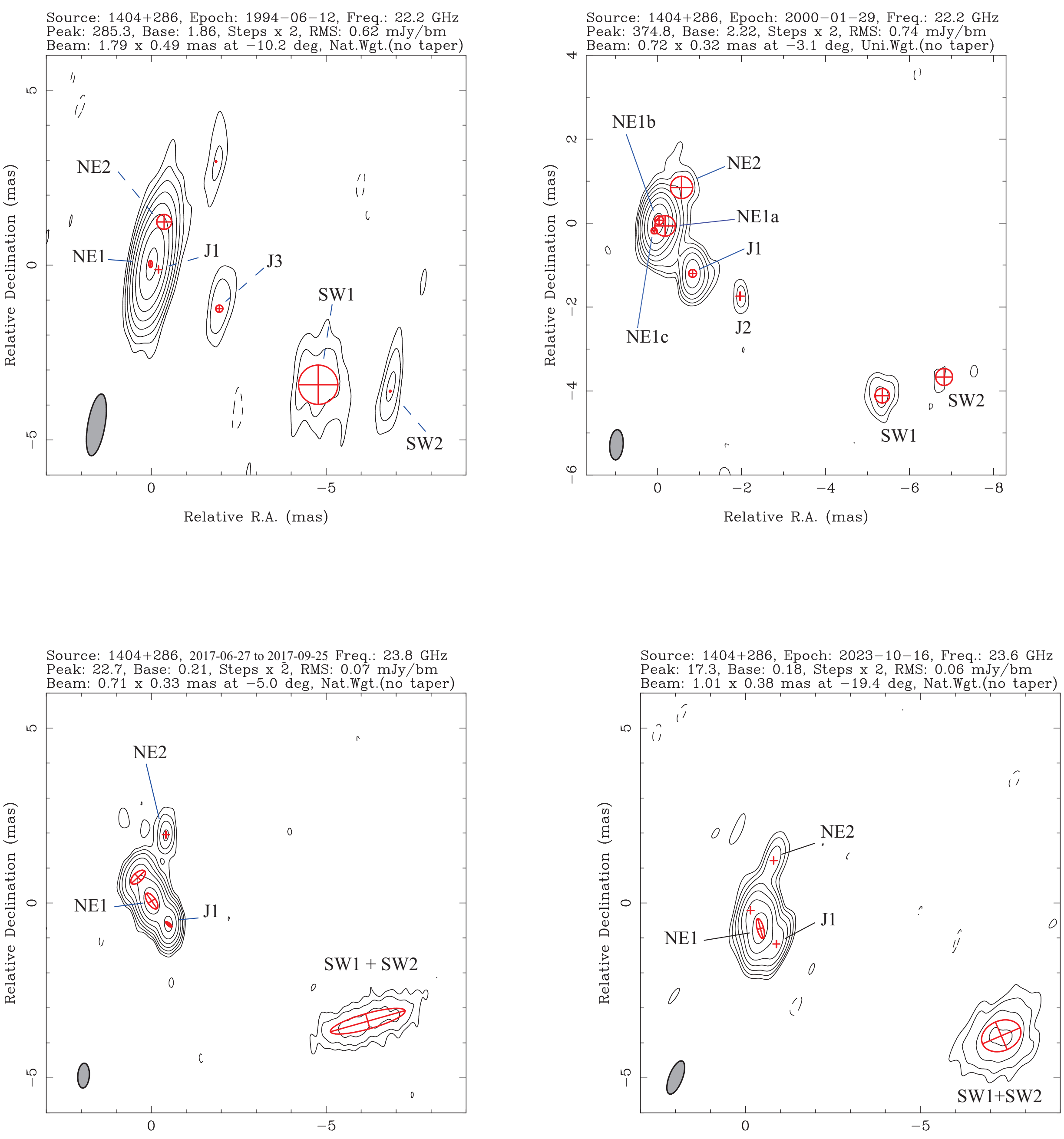

**Figure 8.** The 22 GHz images and fitted components, 1994 (top left), 2000 uniformly weighted (top right), 2017 (bottom left) and 2023 (bottom right). Note the change in the nuclear morphology over time. The observations are self-calibrated, so the origin is approximately the centroid of NE1. Notice that NE1 resolves into 3 subcomponents (NE1a, NE1b and NE1c) in the 2000 observation that is during the strong radio flare. The restoring beam is shown in the lower left of each panel.

## 5. THE FINE SCALE STRUCTURE OF THE FLARING NUCLEUS

The previous section showed that the central engine resides within NE1. In this section, we use high radio frequency VLBI observations to look for an evolving sub-structure within the component, NE1. Figure 8 displays the VLBA 22 GHz images obtained from the NRAO archives and our new observation in 2023. The four panels show the history of the nucleus over a 30-year time span. The first 15 or 22 GHz VLBA observation was project BL007 on 6/12/1994. This is shown in the top left panel. The nucleus is one component that is elongated in an almost north-south direction. A component, apparently J1, is very nearby to the southwest. This image is from an epoch before the 1998 flare started, as indicated in Figure 1 and Table 1. The image in the top right panel is from the BS73 project on 1/29/2000. The observation occurs during the early stages of the flare decay, but the nucleus is still very bright. This image is created with a uniformly weighted restoring beam to maximize the resolution of the nucleus. Notice that the nucleus has three components located within 0.25 mas of each other. This triangular structure was first detected on November 3, 1995 and was last detected in April 2006 with 15.4 GHz VLBA fits. The triangular structure has a large component to the west and two small components. The components overlap on scales smaller than the restoring beam, in general, so it is not obvious if this is an accurate decomposition. However, this same basic configuration exists in the model fits to the VLBA observations for 11 years, indicating that this identification is robust. We defer to Section 5.2 a detailed analysis of the resolution into three components during the flare. Notice that J1 is far away from NE1 in 2000, $\sim 1.45$ mas from the nucleus, more than 1 mas farther than was in 1994 due to the superluminal motion shown in Figure 7. The bottom left panel is from project BS260 in 2017 and has the highest resolution. It combines four observations taken about 1 month apart. This is after the flare shown in Figure 1. The contour levels in Figure 8 indicate a much lower the flux density of the nucleus in 2017 compared to the first two epochs in Figure 8. J1 has now contracted back towards the nucleus, and a new component (with no label) has appeared northeast of the nucleus. The three components form a triple linear structure. This linear triple first appeared in the 15.4 GHz observation on 12/12/2011, shown in Figure 6. The bottom right observations shows that the linear triple has faded in September 2023 and has contracted.

The first issue that we want to study in detail is the time evolution of the VLBA nucleus from 1994 to the flare peak in 1998. This analysis is different from that of Wu et al. (2013) in that they treat NE1 as one component in all epochs. We focus on three details. First, resolving the nucleus, second finding the flux density of each component, and third, locating the positions of each component as a function of time as accurately as possible. The first thing that we tried were 43 GHz VLBA observations. We find a detection on project BL080 on February 15, 2000. However, we were unable to achieve a self-calibration primarily due to the correlator bandwidth available at that time. We place a lower limit of 45 mJy on the source. We could not obtain a successful self-calibration until 2009 on project BS193, well after the flare. However, all we could claim is the detection of a weak ($\sim$30 mJy) point source with a possible 0.5 mas long faint extension with $\sim$ 1-2 mJy pointing towards J1. We imaged OQ208, numerous times after 2009 always settling for a point source at 43 GHz that started to fade in 2016 (see Section 7 for details). Being unsuccessful at 43 GHz, we concentrated on 15 and 22 GHz. We found only one high-quality observation at 22 GHz during the flare, the one in the top right panel of Figure 8.

### 5.1. *Clustering of the Nuclear Components and the Dusty Torus*

Thus, understanding the evolution of the nucleus as the flare grew to its peak in 1998 relies heavily on the 15 GHz observations. There are two main concerns with our attempt to achieve the three objectives in the paragraph, above. One is the overlap of the large diffuse western component, NE1a, with the compact components, NE1b and NE1c, noting that there are few or no baselines long enough to properly resolve this structure. Thus, we require the best uv coverage and lowest phase noise observations in order to extract the trends. This requirement excludes the 11/3/1995 and 11/27/1995 observations, the outliers in Figure 7, in favor of the superior 12/16/1995 observation (also, see the table notes in Tables 5D-5F). The other main issue is locating the positions of the components accurately. To this end, we need a static coordinate system. Again, the best choice is one based on fixing the position of NE2 and minimizing the amount of time elapsed during the analysis in order to minimize the drift from a slow moving NE2 absolute position on the sky plane. The top panel of Figure 9 shows a few things. First the components seem to cluster into three different locations. We label these subcomponents of NE1 as NE1a, NE1b and NE1c as they were labeled in the top right panel of Figure 8. The 6 epochs were chosen for the following reason. 1994 is the first epoch (pre-flare). 4/7/1995 and 5/26/1995 are during the very early stages of the flare. The two epochs are close in time and a consistent pattern in both nuclear decompositions indicates that the feature of a double nucleus (NE1a and NE1b) is robust. Similarly, we choose two observations close in time at the flare peak, 3/3/1998 and 3/27/1998 that are used to verify each other. We also added one intermediate epoch, 12/16/1995. Thus, we see a nucleus that starts as one component and evolves to three in a year and a half.

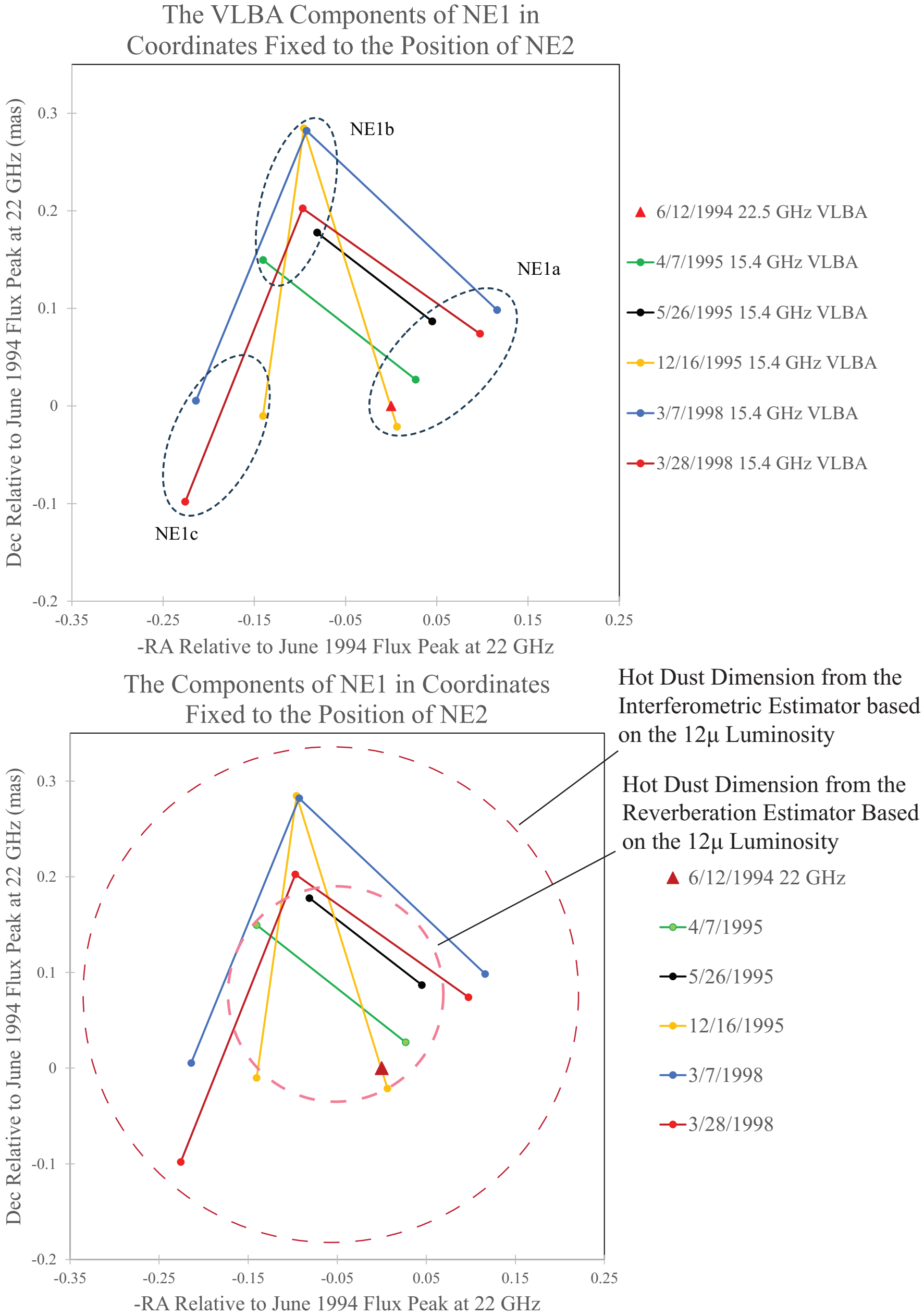

**Figure 9.** Subcomponents of NE1. The markers in the panels indicate the positions of the subcomponents. The subcomponents for each date are connected by lines for visualization purposes. The flare begins as a single component in 1994 (NE1a). A second component (NE1b) appears to the east in the spring of 1995. Then in late 1995 a third component (NE1c) illuminates to the south. The bottom panel shows a spatial connection to the location of the dust sublimation radius and the zones of strong jet dissipation based on the results of the GRAVITY Collaboration as discussed in the text.

The main question this evolving nuclear structure raises, is why there is so much complex behavior on such a small scale? The answer seems to be related to the fact that this is a very luminous mid to far infrared source consistent with a bright AGN obscured by an abundance of molecular gas (Golombek et al. 1988). This is explored in the bottom panel of Figure 9. A meaningful discussion is motivated by the new GRAVITY, infrared interferometry that studies the inner edge of the molecular disk (comprised of hot dust), also called the ring radius. These data are combined with infrared reverberation analysis to quantify the inner regions of the obscuring molecular gas, or the inner edge of the "dusty torus" (Amorim et al. 2024). Relationships are found for both the ring radius and the reverberation radius as a function of the optical luminosity and the 12$\mu$ luminosity for sizeable samples of type I AGN. As others have found, the ring radius measured by interferometry is larger than the reverberation radius (Amorim et al. 2024). This was interpreted as due to a "bowl shape" molecular gas distribution. The 12$\mu$ relationships for these hot dust radii from Amorim et al. (2024) are more relevant than their optical luminosity relationships in the case of OQ208 since the optical flux is obscured by said molecular gas. Spitzer measured 200 mJy at 12$\mu$ on June 24, 2007 (Willett et al. 2010). The CanariCam on the 10.4 m Gran Telescopio CANARIAS (GTC) measured close to 250 mJy at 12$\mu$ on 3/17/2014 (Alonso-Herrero et al. 2016). The bottom panel of Figure 9 uses the relationships for size from Amorim et al. (2024) with the Spitzer 2007 12$\mu$ specific flux. It is hard to overlook the fact that all three components seem to cluster near the inner edge of the molecular gas based on the 12$\mu$ luminosity. The interpretation here is that as the flare begins it creates a strong jet that impacts dense molecular gas clouds that are distributed in an irregular fashion (not a torus). Any preferred axis of the gas does not seem to be aligned with the jet direction. The strong jet, during the flare, propagates through the clouds and is deflected a few times along the way. This creates two or more working surfaces where strong dissipation occurs, that are bright in the radio. None of these three bright spots need to be the exact location of the central engine. However, it seems plausible that the central engine is inside or very near to the central hole in the molecular gas through which the jet emerges. This is the physical motivation for choosing the centroid for the best estimate of the central engine position in the right panel of Figure 7.

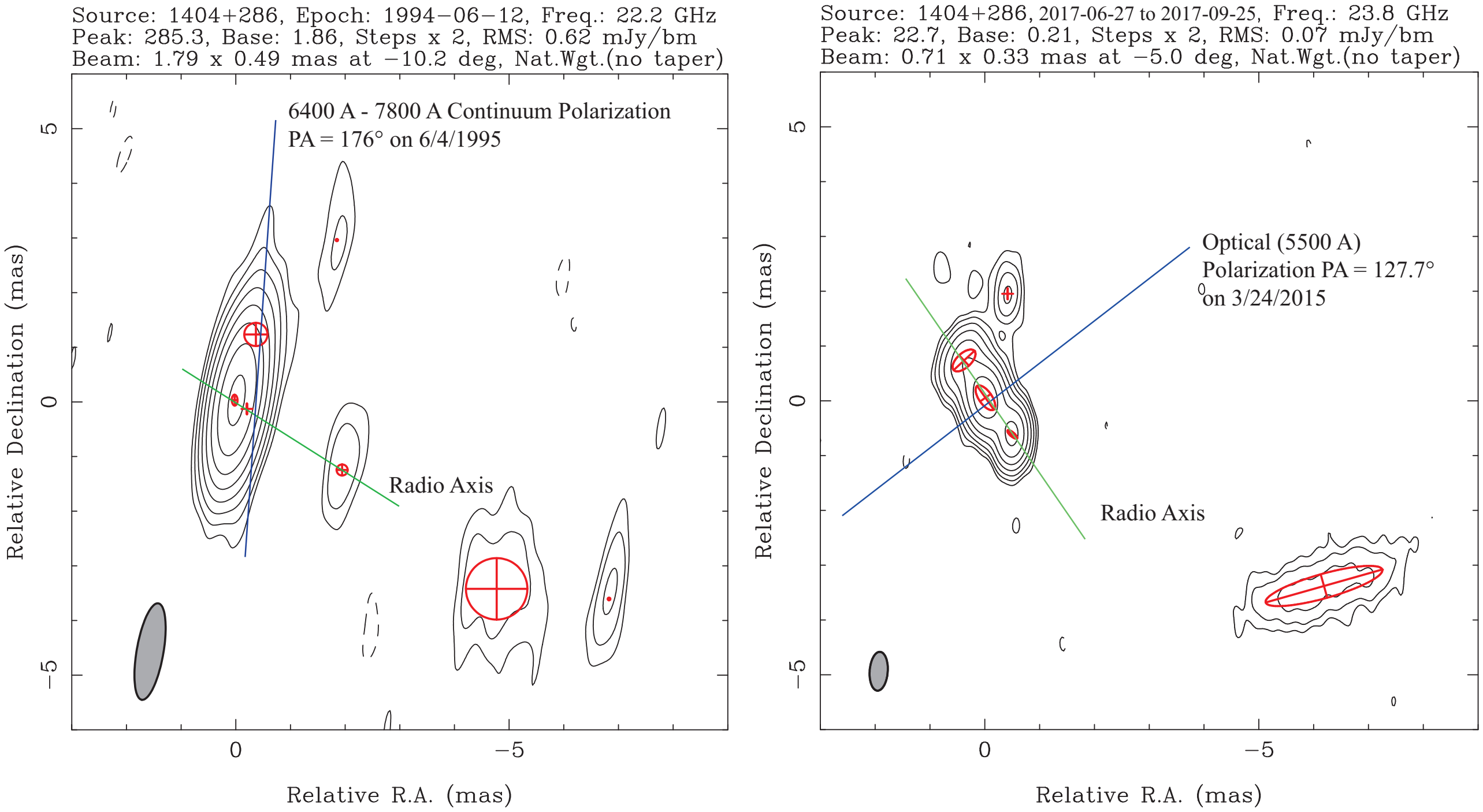

**Figure 10.** In this figure, we show the relationship between polarization in visible light and the radio axis. They rotate between 1994-1995 and 2015-2017. The ionization cone and jet axis both rotate clockwise while maintaining a nearly orthogonal orientation.

Figure 10 highlights the strong interaction between the jet and the molecular gas. In 1994 (left panel), NE1, J1 and J3 form a well-defined radio axis. Comparing this to the line connecting NE1a and NE1b on 4/7/1995 and 5/26/1995 in Figure 9, we see a common alignment direction with the 1994 radio axis. We have overlayed the 6400Å − 7800Å continuum polarization PA measured on 6/4/1995 from spectropolarimetry (Corbett et al. 1998). The radio axis and the polarization PA are nearly orthogonal. The continuum polarization is $\sim 1\%$. From Table 1 and Appendix B, the 15.4 GHz luminosity is $< 1\%$ of the 5100Å luminosity (see also Figure 12) and from Figure 16 in

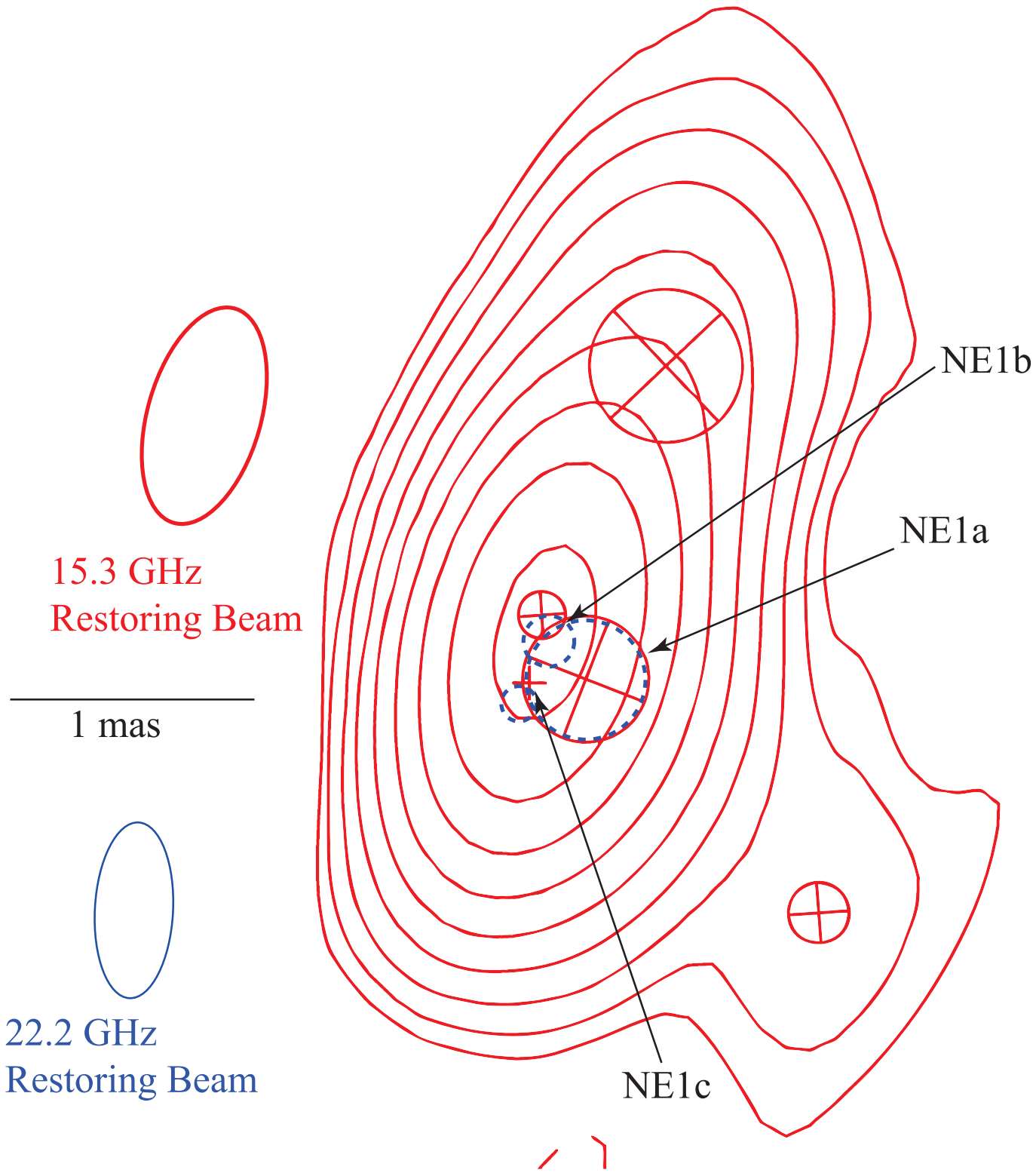

**Figure 11.** The image of NE1 on May 18, 1996 is excised from the image of OQ208 at 15.3 GHz in (Lister 2003). We excised the region in order to magnify the crowded region of interest. We plot the Lister (2003) contours in red as well as his circular Gaussian fits. After registering the fitted model components at 22.2 GHz on January 29, 2000 in the top right panel of Figure 8 to NE1a, we overlay all three circular Gaussian fits as dotted blue circles.

Appendix C, the synchrotron spectrum from the nucleus has a downward spectral break between 22 GHz and 43 GHz. The SED decreases between 15.4 GHz and 5100Å. Thus, we conclude that $\ll 1\%$ of the optical continuum is from jet synchrotron emission. Thus, we have a clear interpretation. The polarized flux is light that is scattered off of electrons in the ionization cone which is formed (loosely collimated) by the "hole" in the molecular gas or dusty torus (Brown and McClean 1977; Goodrich and Miller 1994). Note that in the right frame of Figure 10, both the radio axis of 2017 and the polarization PA of 3/24/2015 from Afanasiev et al. (2019) are plotted. Again, they are nearly orthogonal and we have the same explanation of the polarization as in 1994/1995. The ionization cone (hole in the molecular gas) and the radio axis rotate clockwise in consort after the flaring event. The change in the opening in the molecular gas that is accompanied by a similar change in the jet direction is more evidence of a strong interaction between the jet and molecular gas.

### 5.2. *The Resolution of the Nucleus During the Flare*

We performed model fitting in the spatial frequency domain (not in the image domain), working directly with the visibility data. Moreover, the spatial resolution is not uniform over the image. Traditionally (CLEAN), the restored map is convolved with a fixed beam size determined by the uv-coverage and data weighting scheme. But in fact, the resolution depends on a signal-to-noise ratio (SNR) defined as $S_{\rm peak}/\sigma_{\rm rms}$, where $S_{\rm peak}$ is the peak intensity of the fitted Gaussian component and $\sigma_{\rm rms}$ is the noise on a residual map after the source model is subtracted from data. Thus, it sets a corresponding lower limit on a fitted component size and a proxy of a resolution limit.

Minimal size of a Gaussian component fitted to VLBI data is

$$d = \max\left(\text{component's size};\ \text{resolution limit}\right),$$

where the resolution limit (e.g. Lobanov (2005), Kovalev et al. (2005)) is defined as

$$d_{\rm lim} = 2^{1+\beta/2}\, b_{\varphi} \left[\frac{\ln 2}{\pi} \ln\left(\frac{\rm SNR}{\rm SNR - 1}\right)\right]^{1/2} ,$$

where $\beta = 0$ for natural weighting, $\beta = 2$ for uniform weighting, $b_{\varphi}$ is the half-power beam size measured along a position angle $\varphi$ of the jet component.

In structure model fitting of OQ208, the bright NE1 region components have SNR ranging from ∼100 to ∼500, which implies that the effective resolution achieved in this area is better than 0.1 beam size. Thus, the NE1 features a, b and c are well separated. However, we do not require 0.1 beam size resolution for this analysis.

With that being said, the ability to accurately determine the component separations and flux density assigned to each of the resolved components in interferometric data depends on several factors, e.g., the length of the longest baselines, gaps in the $(u, v)$ coverage, as well as the signal-to-noise ratio of the data. An accurate number is experiment dependent. Recall from Section 4 that the empirically determined (systematic) positional uncertainty is $\sim 10\%$ of the restoring beam width for these particular, 15.4 GHz observations (Homan et al. 2002; Lister et al. 2009). Thus, we expect some inaccuracy in the positional separation and flux density ratio between NE1b and NE1c which are separated by $\sim 25\% - 30\%$ of 15.4 GHz natural beam width. When dealing with a complex region comprising very close features, there is an unavoidable flux density "trading" between them. We address this in the following.

Before performing fitting simulations to assess the magnitude of the flux trading, we can restrict our analysis by motivating a standard 3 component model for NE1. Our highest resolution flare image is in the top right panel of Figure 8, 22.2 GHz on January 29, 2000. The restoring beam is 0.72 mas x 0.32 mas at PA= $-3.1^{\circ}$. A fit with 3 components has smaller residuals than one with 1 or 2 components. Adding additional weak components makes minor changes to the residuals. In a literature search, we found a VLBA image published in a conference proceeding at 15.3 GHz on May 18, 1996 from the same family of MOJAVE observations that we are studying (Lister 2003). This independently generated fit is virtually identical morphologically to our 2000 22.2 GHz image as indicated in Figure 11. The component NE1a is the same size in both the model fits. Over this long of a time frame (1996-2000), we can no longer use NE2 as our fixed reference point. Thus, we register our comparison, arbitrarily to NE1a. So, Figure 11 is a morphology analysis and provides no information on how the triad moved relative to fixed coordinates in the sky plane. NE1b is also the same size at 22.2 GHz and 15.3 GHz, but has shifted closer to NE1a in the 22.2 GHz image. With higher resolution we now have a finite (small) size for NE1c, which was represented as a point source at 15.3 GHz. It is slightly farther from NE1a in the 22 GHz image. The shifts in NE1b and NE1c are less than the 0.1 beam width uncertainty in the component positions and do not necessarily represent a physical difference (Homan et al. 2002; Lister et al. 2009).

Since the 22.2 GHz image has the highest resolution, it is worth analyzing the fidelity of the triad resolution in this observation. Previously run simulations have been generated to determine the limits of the fidelity of VLBA model fits in the visibility domain as a function of resolution and how well this agrees with theoretical estimates (Punsly et al. 2021; Marscher 1985). The main issue is that once the components are closer than the resolution limit of the longest baseline, the fidelity is degraded even though it is still resolved. The simulations were run with typical VLBA (u, v) coverage at 43 GHz of the monthly monitoring by a Boston University based research effort, the VLBA-BU Blazar Monitoring Program[4]. Data simulations found that small bright components can be faithfully resolved to within half of the uniformly weighted beam width without introducing flux sharing effects or positional inaccuracy more than $\sim 10\%$ of the synthesized beam width stated above (Punsly et al. 2021). This is consistent with the resolution of the longest baselines that was determined by an independent analysis to be resolution = $(91{\rm M}\lambda/\text{baseline-length})$ mas (Marscher 1985). The longest baselines for the 15.4 GHz (22.2 GHz) observations (Mauna Kea to St. Croix) can faithfully resolve structures as small as 0.21 mas (0.14 mas), predominantly in the east-west direction, in theory. For our observation, the distances from NE1b to NE1c, NE1b to NE1a and NE1c to NE1a are 0.278 mas (0.48 uniform beam widths), 0.195 mas (0.51 uniform beam widths), 0.286 mas (0.77 uniform beam widths), respectively. Even though this is not our exact observational experiment, the agreement between the Marscher (1985) theoretical estimate and the Punsly et al. (2021) simulations suggests that the half beam width "rule of thumb" is a reasonable estimate for the highest resolution that has high fidelity. The 22.2 GHz model fit would be of high fidelity, if not for the large size of NEla. However, high frequency has an added advantage, NE1a is very steep spectrum (Table 6). Thus, the flux density of NE1a is significantly lower than the two compact components NE1b and NE1c. Since NE1b and NE1c are located near the half maximum radius of NE1a, we expect that NE1a is robbing or giving only a few percent of the flux density of NE1b and NE1c, at most. The converse cannot be stated, since NE1a might have a more significant fraction of its smaller, diffuse total flux density altered by flux sharing with NE1b and NE1c. Thus, we posit that NE1b and NE1c are faithfully represented in the 22.2 GHz model fit.

[4] http://www.bu.edu/blazars/VLBAproject.html Jorstad et al.(2017)

The motivation for analyzing Figure 11 in some detail is to more intelligently argue the universality of the basic triad structure for NE1 during the flare. Our philosophy is to consider all the flare states as a whole and find a consistent morphology. Our starting point is the highest resolution flare image in the top right panel of Figure 8, 22.2 GHz on January 29, 2000. The morphology of the fit seems to be robust based on the resolution of the VLBA at 22.2 GHz and the independent fit of the 15.4 GHz observation on May 18, 1996 (Lister 2003). We argue that a three component decomposition at 15.4 GHz is highly plausible during the flare, if the same configuration that minimizes the residuals results in similar dimensions for the both the components and the overall configuration in all epochs. NE1a, NE1b, NE1c, typically have FWHM of $\sim 0.48$ mas, $\sim 0.2$ mas and $< 0.1$ mas, respectively. NE1b and NE1c are always east of NE1a and NE1c is always south and east of NE1b. Thus, we conclude that the triad morphology of Figure 11 is persistent during the flare. The 22.2 GHz observation is our best case situation for a high fidelity resolution of the triad. The other observations do not satisfy the half beam width "rule of thumb".

The main issue with the exact nature of the 3 component decomposition in the 15.4 GHz observations is the the u-v coverage is not dense enough from the long baselines to reliable compare epochs to the desired accuracy. This is a systematic uncertainty. For example, the 11/27/1995 observations in Table 5E have only 6 scans while the 12/15/1995 observations have 8 scans over a longer period of time and have denser u-v coverage. These observations are quasi-simultaneous. Notice the flux density difference between NE1b and NE1c, 333 mJy (278 mJy) and 142 mJy (184 mJy), respectively on 11/27/1995 (12/15/1995). However, the combined flux density of NE1b and NE1c (NE1bc) is 473 mJy and 462 mJy for 11/27/1995 and 12/15/1995, respectively. Combining the components mitigates the effects of the systematic uncertainty due to u-v coverage. This improved decomposition removes the largest contribution to the systematic uncertainty (induced by the paucity of long north-south baselines). This consideration renders the systematic uncertainty to be governed by the empirical VLBA data analysis of these 15.4 GHz observations (Homan et al. 2002; Lister et al. 2009). These uncertainties are applied after the model fitting. Since more reliable results are obtained for the flux density if we combine NE1b and NE1c as one entity, none of the results of this study depend explicitly on decomposing the 3 components. The following discussion depends only on the overall size of the configuration. The fits to NE1b and NE1c are not used explicitly in our central analysis (of Section 6), only the flux density of the union of the components, NE1bc, whether it was fit as a single elliptical Gaussian function (lower resolution images) or as two circular Gaussian functions. Our results in Section 6 depend on NE1bc.

Thus, we are motivated to quantify the magnitude of the flux sharing between NE1bc and NE1a in the 15.4 GHz model fits. We ran Monte Carlo Markov Chain (MCMC) sampling simulations in order to test the stability of our fits. Based on our the discussion above, we pick 3 subcomponents for NE1, NE2, SW1, SW2 and J1 (see Figure 6) to represent the source for a total of 7 (circular) components. We ignore the weak components J2 and J3 that are far from the nucleus. For the visibilities of a given OQ208 15.4 GHz observation, we are simulating the fitted models that minimize the residuals. We fit complex visibilities in the Fourier domain using using parallel tempering for MCMC sampling to random walk the FWHM, flux density and position of each model component. We used the InterFit tool written by Alexander Plavin in Julia [5]. This procedure is designed to derive not only the best fit model but also Bayesian error estimates on every fitted parameter. There were no systematic uncertainties, it is a test of the ambiguity in the fitting process. Note that systematic uncertainties are assigned to the model fitted components in Appendix B based on an empirical VLBA data analysis of the 15.4 GHz observations (Homan et al. 2002; Lister et al. 2009). There were 11 tempering rounds (2048 equally probable realizations of the model fit) and 20 parallel chains. We find that the flux density of NE1a has a 8.1% uncertainty in the fit and the flux density of the fit to NE1bc has 4.9% uncertainty. The uncertainty of the other bright component, J1, (in a less busy region) has only a 1.5% uncertainty in the model fitted flux density. The larger uncertainty in the NE1 region is likely due to the flux sharing contribution. Our error budget for the model fits in Appendix B (and therefore Figure 12) is 5% for NE1bc and 10% for NE1a. Thus, our budgeted model fit errors are consistent with the MCMC sampling simulation.

## 6. THE NUCLEAR EVOLUTION OF THE FLARE EVENT

[5] https://github.com/aplavin/InterFit.jl

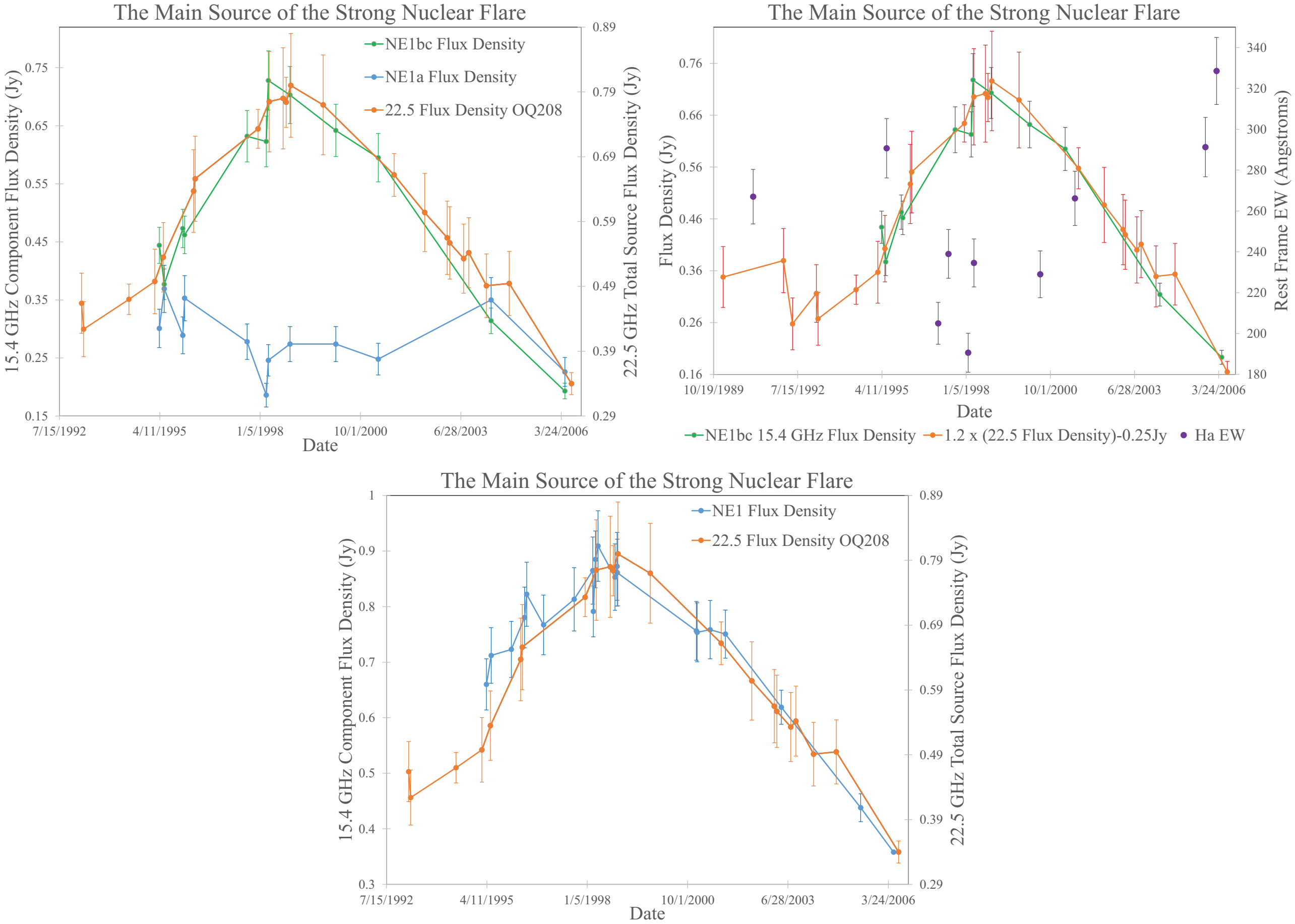

**Figure 12.** Identifying the main source of the strong nuclear flare in radio emission. The evolution of the nuclear subcomponents, NE1a and NE1bc (the union of NE1b and NE1c), before, during and after the flare are plotted in the top left hand panel. The 22.5 GHz flux density is also plotted. The plot shows that 22.5 GHz is a good surrogate for NE1bc in the absence of VLBA data, after a linear re-scaling, during the flare. NE1a is fairly constant during the flare. The flare is actually primarily a flare of NE1bc. The fact that the rise, fall and peak of the flare is very similar whether depicted in terms of NE1bc 15.4 GHz VLBA flux density or 22.5 GHZ VLA total flux density is an independent corroboration that the methodology in Section 5.2 for decomposing NE1 into NE1a and NE1bc is robust during the time period plotted. Notice that fitting the flare with a single elliptical Gaussian component for NE1 in the bottom panel (the fits are from Wu et al. (2013), except March 7 1998 and March 28 1998 were refit by us with the improved 2017 MOJAVE calibration for these epochs) does not match the rise of the 22.5 GHz VLA light curve as well as the NE1bc light curve fits the rise of the VLA data. This is a consequence of the fact that NE1a has a much steeper spectrum than NE1bc (Table 6, Appendix C). Thus, its contribution to NE1 is enhanced at 15.4 GHz, especially before the flare nears its peak. Decomposing NE1 as NE1a plus NE1bc has improved our understanding of the flare. The top right hand side overlays the H$\alpha$ EWs from Table 2 and drops the steady NE1a component for clarity. Since there are three things being plotted, the left axis is needed for both flux densities. The linear re-scaling in the right panel yields a 22.5 GHz light curve that is very similar the one in the left panel. One can see a decline in the EW as NE1bc flares.

In this section, we explore the nuclear sub-component evolution of the radio flare. The first issue that is established in the top left panel of Figure 12 is that the two compact nuclear components NE1b and NE1c are the flaring components, while NE1a is fairly steady until 2006. The notation, NE1bc, means the union of NE1b and NE1c. The reasons for introducing NE1bc are that in some epochs (spring 1995) NE1c is too faint for a confident detection and in some epochs the two components are best fit by one elliptical Gaussian component instead of two circular Gaussian components. Regardless, as discussed in Section 5, there will an ambiguity in the decomposition on such small scales. The NE1bc 15.4 GHz flux density in Figure 12 is the total flux density from NE1b and NE1c. The VLBA flux density uncertainty has two contributions, the absolute flux density and the uncertainty in the component flux density. A 5% absolute flux density uncertainty for 15.4 GHz and 7% for 22.5 GHz has been estimated (Homan et al. 2002). In terms of the component fit uncertainty, from Section 5.2 and Appendix B, we have 5% for NE1bc and 10% for NE1a. We then add this uncertainty in quadrature with the absolute flux uncertainty. These uncertainties are shown in Figure 12. We also plot the 22.5 GHz flux density from Table 1. This plot ends in 2006 because we have no other data until 2008. The plot establishes that NE1bc is the flaring component and the 22.5 GHz total flux density is a useful surrogate for tracking the time variations of NE1 in active states. As discussed at the end of Section 4, NE1bc has the highest brightness temperature of any component in OQ208. Furthermore, it has the flattest intrinsic synchrotron power law of any component in our matched resolution observation on January 29-31, 2000 in Appendix C (see Table 6). Thus, it is likely the farthest upstream towards the launch point of any feature. Being associated with the flaring component of NE1 is consistent with these two circumstances. The resolution is insufficient to locate the position of the launch point within NE1bc (recall, from the beginning of Section 5, that the 43 GHz observation two weeks later, unfortunately, failed to produce an image).

The flaring component being established, we now turn to the connection with broad line equivalent width in the top right panel of Figure 12. In order to add the EW to the right plot we re-plotted the total 22 GHz flux density as 1.2 times the flux density with a 0.25 Jy offset on the left vertical axis. The flare of NE1bc is a factor of 2 increase from the level in the spring of 1995 compared to a factor of 1.5 increase for the total 22 GHz flux density. Measuring the flux density of NE1bc in isolation essentially removes the flux density of the non-flaring components, thereby enhancing the contrast of the flare from the background radio flux.

## 7. CHARACTERIZING THE FLARE PROFILE

In this section, we define a normalized jet luminosity as $L_{15.4}$(NE1bc)/L(5100 Å) in analogy to the definition of the equivalent width. In section 3.2, we estimated that 70%-78% of the observed optical flux is partially obscured AGN emission. It is of potential intreset to compare the jet luminosity to the accretion luminosity in order to explore a possible physical connection in OQ208. In particular, all known models of jet production from black holes require accretion to either power the jet or pin magnetic flux to the event horizon (Blandford and Znajek 1977; Blandford and Payne 1982; Blandford et al. 2019; Punsly 2008). We consider this parameter even though its physical meaning depends on the physical assumptions. For example, under the assumption that changes in $L_{15.4}$(NE1bc) represent a change in jet power, the normalized jet luminosity is a measure that can elucidate a change in the ratio of jet power to accretion power and hence a change in the physical dynamics of the jet launching region that might explain the Balmer EW evolution during the flare in the top right hand panel of Figure 12.

We created Figure 13 to provide an alternative definition of the flare profile. In this figure, the flare is characterized by the normalized jet luminosity. However, this is a challenge due to sparse spectral and VLBA sampling. Technically, the observed jet is emitted earlier than the observed 15.4 GHz luminosity of NE1bc, $L_{15.4}$(NE1bc), by less than the light travel to time to the hot dust (< 6 light-months), based on Figure 9 and assuming a relativistic bulk flow. Similarly, the broad line region has been estimated to be ~1.5 light months from the central engine (Afanasiev et al. 2019). But, this is likely underestimated due to the fact that the observed fluxes are partially obscured. With this consideration, we expect less than 4.5 months of time lag between changes in the 15.4 GHz flux density of NE1bc and changes in the broad line flux. However, either component, NE1b or NE1c, could be very close to the jet launch region and the other at the inner edge of the molecular torus. The existing VLBA data does not preclude this configuration. Thus, the time difference, $\Delta T$, between the jet launch time and the emission time of the ionizing flux for the broad lines can only be constrained as:

$$-4.5\,\mathrm{months} < \Delta \mathrm{T} < 2\,\mathrm{months}\,. \tag{5}$$

We need to compromise on the quality of the individual data points. We use the VLBA archives and Table 2 to try to find optical observations close in time to good VLBA observations at a common frequency (15.4 GHz). We aim for $\mid T \mid < 130$ days based on Equation (5). This is rather long, but it is still very difficult to find time differences this small in practice. The time difference for each "pair" of observations is shown in red in Figure 13 for each point that is located at the VLBA time. The time indicated is the number of days until the optical observations. Based on Equation (5), the larger time differences should be associated with the optical data preceding the radio data. However, it does not work out that way with the available data. This expedience should not be grossly inaccurate as the optical luminosity changes rather slowly over 130 days intervals based on Figure 3 and the top left panel of Figure 13. Furthermore, most of the change is in the radio flare, not the optical continuum.

Notice that in terms of normalized jet luminosity, the flare is much more pronounced than for the flux density of NE1bc plotted in Figure 12. There is now a stark contrast between June 1995 and August 1997. We have three metrics of the flare height in Figures 12 and 13, the 22.5 GHz total flux density, the 15.4 GHz flux density of NE1bc and $L_{15.4}$(NE1bc)/L(5100 Å). These yield a contrast of the flare peak in March 1998 versus the baseline state in May 1995. This contrast is given by the ratio of the flare value of the metric quantity to the baseline value of the metric quantity. For the 22.5 GHz total flux density, the 15.4 GHz flux density of NE1bc and $L_{15.4}$(NE1bc)/L(5100 Å), we get ratios of ≈ 1.5, ≈ 1.9 and ≈ 2.9, respectively. Each of the three light curves has its advantages for defining the flare. The normalized jet luminosity light curve has the most pronounced peak, but the sparsest sampling. The 22.5 GHz VLA light curve has the densest sampling and is the only light curve that extends to epochs before the flare rises. The NE1bc light curve is a compromise between the other two. All three are useful in our analysis.

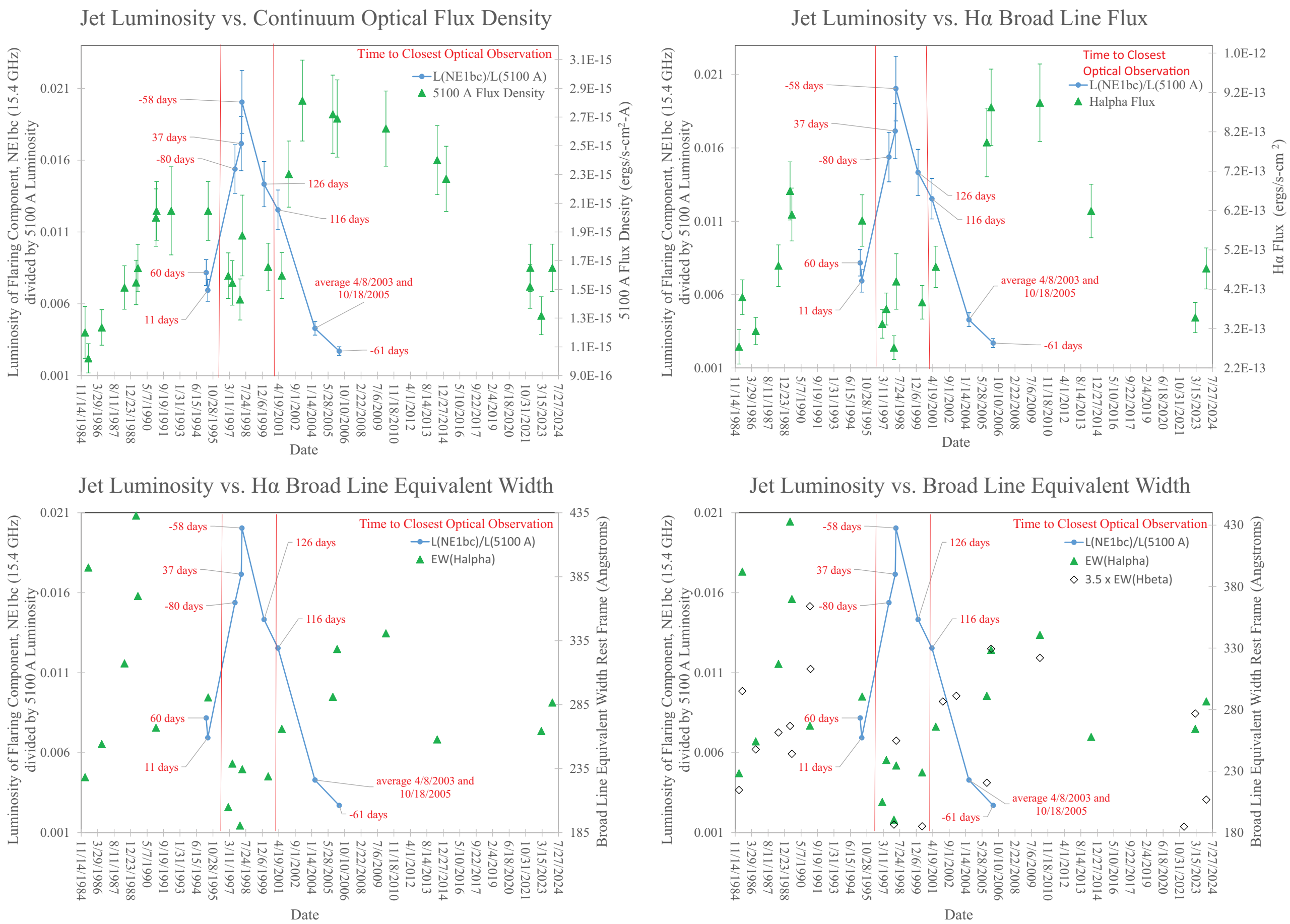

**Figure 13.** The normalized jet luminosity as a function of time indicates a very sharply defined strong flare which is approximated by the time frame demarcated between the red vertical lines. The top two plots are the normalized jet luminosity with the optical continuum and Hα broad line flux overlayed. The bottom left hand panel is the fundamental analysis of this paper. We drop the errors on the EW (already shown twice before) for the sake of visual clarity. The bottom right hand panel includes the H$\beta$ EWs from Table 2. The 3.5 multiplicative factor is ≈ the median Balmer decrement. We have not discussed the H$\beta$ EW previously because there are only 3 measurements during the flare and one has been flagged as an outlier in Table 2. The number of days between the closest optical observation and the VLBA measurement is shown in red.

We proceed to analyze statistical tests of the trends in Figure 13. We use nonparametric statistical tests, the one sided, two sample Komogorov-Smirnov test and the Exact Wicoxon Rank Sum test[6]. There is formally no minimum sample sizes required for these tests. With small samples, like we have here (5 flare epochs and 15 nonflare epochs with Hα data) the tests do not have the ability to distinguish between two modestly different samples with high statistical significance. They are useful for two samples that are different at a high significance level. The results for the Kolmogorov-Smirnov (Wicoxon Rank Sum) tests are as follows:

[6] All analyses were performed using R Statistical Software (v4.5.1; R Core Team 2025)

**Table 3.** Testing the Null Hypothesis, $H_o$: H$\alpha$EW(flare) $\not< H\alpha$EW(nonflare)

| Time Sampling Window (Days) | Flare Window | Flare EW | Non-Flare EW | Number of Flare Epochs | Number of Non-Flare Epochs | K-S Acceptance Probability | Wilcoxon Acceptance Probability | Comments |
|---|---|---|---|---|---|---|---|---|
| 1 | 2/7/1997-6/3/2000 | as fit: Table 2 | as fit: Table 2 | 5 | 15 | 0.0004 | 0.0003 | Fiducial |
| 1 | 2/7/1997-6/3/2000 | as fit + $1\sigma$ | as fit- $1\sigma$ | 5 | 15 | 0.0044 | 0.0027 | $\sigma = 5\%$ |
| 1 | 2/7/1997-7/21/2001 | as fit: Table 2 | as fit: Table 2 | 6 | 14 | 0.0039 | 0.0007 | 2001 possible see Figs.12 and 13 |
| 1 | 6/6/1995-6/3/2000 | as fit: Table 2 | as fit: Table 2 | 6 | 14 | 0.0039 | 0.0032 | 1995 unlikely see Figs.12 and 13 |
| 1 | 6/6/1995-7/21/2001 | as fit: Table 2 | as fit: Table 2 | 7 | 13 | 0.0124 | 0.0042 | 1995 unlikely see Figs.12 and 13 |
| 100 | 2/7/1997-6/3/2000 | as fit: Table 2 | as fit: Table 2 | 4 | 13 | 0.0004 | 0.0004 | |
| 100 | 2/7/1997-6/3/2000 | as fit + $1\sigma$ | as fit- $1\sigma$ | 4 | 13 | 0.0063 | 0.0017 | $\sigma = 5\%$ |
| 100 | 6/6/1995-7/21/2001 | as fit: Table 2 | as fit: Table 2 | 6 | 11 | 0.0149 | 0.0036 | 1995 unlikely see Figs.12 and 13 |

1. The hypothesis that the 5100 Å specific flux is low during the flare compared to the non-flare epochs has a statistical significance of 90.23%(82.78%).

2. The hypothesis that the H$\alpha$ BEL flux is low during the flare compared to the non-flare epochs has a statistical significance of 98.83%(98.36%).

3. The hypothesis that the H$\alpha$ EW (computed relative to the host+AGN continuum) is low during the flare compared to the non-flare epochs has a statistical significance of 99.96%(99.97%).

The main issue with these tests is providing a meaningful value of the statistical significance with the small sample sizes, sparse time sampling and uneven time sampling. These issues are exacerbated by the uncertainty in the flare window shown in Figure 13. Characterizing the flare window benefits from also considering the 22.5 GHz light curve in Figure 12. The EW analysis is the main priority of this study. Thus motivated, Table 3 are the result of our efforts to test the resiliency of the result 3, in the context of the concerns above. The first line is our fiducial test, result 3. We treat all the data on equal footing, we do not consider the EW uncertainties and we adopt the flare window from Figure 13. The dates 2/7/1997-6/3/2000 represent the range of optical observations that are within the flare window.

Below this we begin to investigate the possible shortcomings of the fiducial analysis. In the second row, we address the measurement uncertainty in EW from Table 2 and Section 3. In order to investigate the possible effects on result 3, we add the error to *all* the flare EWs in order to make them larger (in practice, it seems implausible that all are underestimated). Similarly, we subtract the error from *all* the non-flare EWs (similarly implausible). The motivation for these implausible blanket data perturbations is to stress the statistical analysis to check if it is even remotely possible that the these two data sets are actually similar within the uncertainty of observational and fitting errors. In row 3, we expand the flare window towards the present by adding 7/21/2001. Looking at Figure 13, it was a judgement call not to add 2001 to the window. The normalized jet luminosity was dropping rapidly at this time, but it was still elevated. This is deemed a reasonable choice for the flare window. The fourth row extends the flare window towards the past adding 6/6/1995. The normalized jet luminosity is not elevated on this date and the 22 GHz total flux is just starting to rise on 5/17/1995 after years of being constant within the error bars (see the top right hand panel of Figure 12). Adding 20 months to the window towards the past is marginally possible since the flare might be on the rise. In row 5, we get more aggressive and expand the window in both directions. This is as wide of a window that can be justified by Figures 12 and 13 with the available optical spectra. In row 6, we consider the possibility that the statistical analysis is skewed by the uneven time sampling. In particular, two measurements close in time could be sampling the same slowly evolving state providing twice the weight to this epoch. We pick an averaging (we average the two EW values) window of 100 days, long enough to capture a few pairs and short enough not to expect major changes in double peaked broadline profiles (Gezari et al. 2007). We skew the data by assigning asymmetric errors in row 7. In row 8 when we skew the data with the asymmetric errors and expand the sampling window to a maximum size. Table 3 indicates that the notion that the EWs are depressed during the flare relative to the rest of the time history is robust independent of the uncertainty in the flare window, the uncertainty in the EW measurements and the available flexibility in time sampling. This verifies that which seems visually apparent in Figure 4.

The observations provide one more piece of evidence that defines the EW behavior during the flare. The EW drop is not due to increased absorption in the broad line region. Figure 14 shows this is not the case as there is no measurable increase in the Balmer decrement during the NE1bc flare. Figure 14 provides motivation for the next two sections that consider the role of the central engine of the quasar in the EW drop during the flare.

Figure 14. The Balmer decrement in OQ208 over a 39 year period. The red vertical lines indicate the time frame of the flare estimated in Figure 13. There is no indication that there is an enhanced decrement (more obscuration of the BLR) during the flare. The Balmer decrement has a median of 3.77 and mean of 4.04. The linear fit to the data is shown as a dotted line.

## 8. THE NATURE OF THE NE1BC FLARE EMISSION

The next two sections are more speculative physical interpretations of the flare and the coinciding drop in the EW. This section is predominantly observation based and Section 9 is a theoretical discussion. In these two sections, we develop a possible consistent physical explanation of Figures 1, 12 and 13. NE1bc is apparently located on or near the hot dust region or dust sublimation radius based on Figure 9. NE1bc is $\sim 0.4$ mas x 0.2 mas in size (see Tables 5L-5N), or approximately 1.9 x 1.0 light years projected on the sky plane. Yet in Figure 12, the NE1bc flux density increases by $\sim 50\%$ from December, 15 1995 to August, 28 1997 (20 months). If this were due to a change in the environment (the inner regions of the dusty torus), this would require the molecular gas to redistribute itself with a velocity that were a significant fraction of the speed of light. However, this is a slow moving non-relativistic medium (molecular gas). For a $5 \times 10^8$ solar mass central black hole, the virial mass estimate of Wu et al. (2009), it would take a virialized gas cloud at the hot dust reverberation (interferometry) radius, 0.5 lt-yrs. (1.25 lt-yrs.), $\sim 7$ yrs.($\sim 16$ yrs.) to traverse 0.1 lt-yrs.

The working surfaces between the jet and the molecular torus (NE1b and/or NE1c based on the top right panel of Figure 8) likely includes shocks. From a geometric standpoint, the fitted data is consistent with a stationary configuration (as noted in Section 5, see also Appendix B), even though the angular resolution is not sufficient to verify this with high certainty. The large changes from December, 15 1995 to August, 28 1997 is unlikely to involve a change of the piston (the dusty gas), as noted above this region is not changing dramatically in 20 months. It would seem to require either large changes in Doppler enhancement that are associated with changes in orientation or large changes in the jet power. This discussion involves the former, the latter is discussed below. Using Equation 4 for $T_b$, we can check for evidence of relativistic motion. Appendix C is a discussion of a matched resolution analysis on 1/29/2000 to 1/31/2000 based on 15.4 GHz and 22.2 GHz VLBA and 8.6 GHz global VLBI (near the flare peak). For NE1bc, from Tables 5L-5N, $T_b$ is $1.35 \times 10^{10}$K, $6.00 \times 10^{10}$K, $7.17 \times 10^{10}$K at 22.2 GHz, 15.4 GHz and 8.6 GHZ (near the peak of the NE1bc spectrum, see Figure 18), respectively. The 15.4 GHz value is much lower than the $10^{11}\text{K} - 5 \times 10^{13}$K brightness temperatures found for the VLBI cores of Doppler boosted blazars in a large 15 GHz VLBA sample of bright compact radio sources (Kovalev et al. 2005). OQ208 ($T_b = 4.57 \times 10^{10}$K in their Table 2), is located at the very low end of the $T_b$ distribution in their Figure 9. The measured $T_b$ values are consistent with the equipartition value associated with the inverse Compton limit for stationary synchrotron sources (Readhead 1994). Furthermore, the fits in Appendix B and our Figure 9 do not indicate any superluminal motion of NE1b, NE1c or NE1bc within the measurement uncertainty. Superluminal motion equates to $> 0.2$ mas/year change in separation.

This would create a clear resolvable separation (with 15.4 GHz VLBA) that occurs during the flare. This is not evident in Figure 9, nor in the fits of Appendix B. Thus, there is no evidence of enhanced Doppler beaming creating the jet flare. The observations do not support the notion that flux density increase in NE1bc from December, 15 1995 to August, 28 1997 is a consequence of a change in the shock orientation or geometry that deflects the downstream flow toward the line of sight so as to create significant Doppler enhancement.

Thus, changes in the luminosity of NE1bc is more likely related to jet power as opposed to changes in the jet direction or the configuration of the working surface. For most bright high frequency compact sources, the 15.4 GHz flux from NE1bc would not be a reasonable surrogate for jet power. However, by contrast to the vast majority of such sources, the large NE1bc flux density is not a consequence of Doppler enhancement and the 15.4 GHz flux density is optically thin as well as steep spectrum at least up to 150 GHz (Appendix C: Table 6 and Figure 18).

The decay side of the flare in Figure 12 and Figure 1 contains valuable information when considered with the 22.2-23.8 GHz VLBA observations in Appendix B and Table 4. Since the observations in Table 4 cannot resolve NE1, we consider the total flux density of NE1, the flux density of NE1a + NE1bc. On January 29, 2000, the flux density of NE1 is $558 \pm 50$ mJy and on $\sim$10/16/2009 it is $86 \pm 14$ mJy. The 22 GHz flux density drops by a factor of $\sim 6.5$ in $\sim 9.5$ years. Since the flare peak does not appear to be Doppler enhanced, it is difficult to conceive of a local process (relative to NE1bc) that can make such a large change in such a short period of time. Furthermore, this decay continues until 10/16/2023 (see also Figure 1), at which time, the NE1 flux density is $\sim$1/21.5 of its value in 2000. However, rapid and extreme changes in jet power from the small jet launching region can be propagated along the jet to effect the observed rapid and extreme changes in jet luminosity $\sim$ 0.5-1.0 light years away.

Furthermore, the decline of the 22.5 GHz flux density in Figure 1 can be used in conjunction with the 22 GHz VLBA observations to show that the decline is not a local event, but a large scale event. Consider 22 GHz VLA and 22.5 GHz VLBA observations that are made quasi-simultaneously. One can take the difference of the VLA flux density and the NE1 VLBA flux density to find the excess flux density above that of the nucleus, NE1. The excess flux density was $\approx 214$ mJy at the time of the January 29, 2000 VLBA observation. For the VLBA observation on 4/5/2009, 4/19/2018 and 10/16/2023 in Table 4, we have VLA observations close in time in Table 1. The excess above nuclear flux density in 4/5/2009, 4/19/2018 and 10/16/2023 decays from the peak value during the flare of $\approx 214$ mJy to $\approx 187$ mJy, $\approx 103$ mJy and $\approx 73$ mJy, respectively. The decay from the flare peak is a a non-localized phenomenon. The decay of the excess flux from 2000 to 2009 is accounted for by the fading of J1 from 81 mJy to 53 mJy. From 2009 to 2023 there is continual fading of J1, but the larger effect is fading of more extended flux that is of low surface brightness and not detected by VLBA. From a causality standpoint, independent local physical processes distributed over an extended region is not a viable explanation of the non-local nature of the 22.5 GHz flux decay in OQ208. The observational data, taken in totality, indicate a single causative agent that is predominantly responsible for the flux decay in OQ208. The data is compatible with a decrease in jet power that is transmitted to increasing distances over time.

If the decline of the flare is driven by a decline in jet power, it would seem to follow that the even faster flare rise (the NE1bc flux density increases by $\sim 50\%$ from December, 15 1995 to August, 28 1997: 20 months) is also due to a rise in jet power given the apparent absence of significant Doppler enhancement. This suggests that the anti-correlation between the 15.4 GHz flux density of NE1bc with the H$\alpha$ EW during the 1998 flare is a manifestation of an anti-correlation of jet power with the H$\alpha$ EW. This is motivation to investigate a possible physical interpretation of the anti-correaltion in the next section.

## 9. INTERPRETING THE FLARE IN TERMS OF MAGNETICALLY ARRESTED ACCRETION AND THE EUV DEFICIT

In this theoretical section, we describe the relevant dynamics associated with magnetically arrested accretion and the EUV emission region. This section is largely a review of previously published material aimed at providing context for magnetically arrested accretion in the case of OQ208. We propose a possible physical explanation of Table 3 and Figures 4, 12 and 13 in terms of magnetically arrested accretion.

### 9.1. *The EUV Deficit of Radio Loud Quasars*

The ionizing luminosity, $L_{\rm ion}$, for hydrogen in a quasar is primarily extreme ultraviolet (EUV) emission, $\sim 912\text{Å} - 100\text{Å}$, $L_{\rm EUV}$ (Laor et al. 1997; Telfer et al. 2002). A straightforward explanation of the depressed EW of H$\alpha$ broad line is that this is a manifestation of the tendency for radio loud quasars to have a depressed EUV continuum relative to the EUV continuum of radio quiet quasars at matched UV luminosity, "the EUV deficit" of radio loud quasars (Laor et al. 1997; Telfer et al. 2002; Shang et al. 2011; Punsly 2014). For powerful radio jets, the magnitude of the EUV deficit tends to increase with the increase of the ratio of the long term time averaged jet power to the bolometric luminosity of the accretion flow (Punsly 2015). The existence of this correlation was part of the motivation to introduce the normalized jet luminosity. The only known source with quasi-simultaneous evidence of an EUV flux that is anti-correlated with jet power is the high redshift, compact, steep spectrum quasar 1442+101. This source is unresolved with the VLA and is very core dominated in 8.6 GHz VLBI images. Over a $\sim 40$ year period, the 38.5 GHz rest frame flux density increases (decreases) as the EUV flux decreases (increases)(Punsly et al. 2015). That study considered the 38.5 GHz flux density variation as a manifestation of a variation of the jet power since this emission is not Doppler boosted and is optically thin in 1442+101. There is also no evidence of superluminal motion. As a corollary, there is no "optical deficit" for loud quasars, the optical and UV continua are indistinguishable from those of radio quiet quasars (Shang et al. 2011). In particular, see the insert in Figure 7 of (Shang et al. 2011).

### 9.2. *The Location of the EUV Emitting Gas in Quasars*

Being both obscured and low redshift, the EUV continuum cannot be observed directly for OQ208. However, a detailed study of quasar EUV variability of quasars was able to approximate the location of the EUV emitting region in SDSS quasars exhibiting EUV variability (Punsly et al. 2016). Variability time scales were married with virial black hole mass estimates based on the FWHM of broad lines and the General Relativistic magnetohydrodynamic simulations of slim disks (Sadowski et al. 2011). Assuming a high spin black hole, the EUV emitting region was found to be within $\sim 7M - 12\mathrm{M}$ from the inner edge of the accretion disk, where $M$ is the geometrized mass of the black hole (Punsly et al. 2016). $M$ is the approximately the event horizon radius in the high spin case. Even though this result depends on EUV variability time scales and virial black hole estimates, it utilizes the infall times from the EUV region of magnetized slim disk simulations and hence is model dependent. Another independent attempt to find the location of the EUV emitting region in an AGN comes from the Extreme Ultraviolet Explorer observations of NGC 5548. Simultaneous UV and EUV monitoring indicated that the EUV emission was consistent with the Wien tail of the optically thick thermal emission from the innermost regions of the accretion disk (and not consistent with an optically thin region) (Marshall et al. 1997).

### 9.3. *The Jet-Accretion Flow Interaction*

The jet is likely launched in large scale magnetic flux tubes that are located in the inner accretion disk or black hole ergosphere (Blandford and Znajek 1977; Blandford and Payne 1982; Punsly 2008; Blandford et al. 2019). The only region of the disk or corona that is near the jet launching region is the innermost region. If the EUV continuum does not have a large contribution from the inner regions of the disk or its atmosphere (perhaps an inner corona) it would raise the question: why is the only difference in the continuum of radio loud quasars and radio quiet quasars in the EUV? The jet base and the EUV emission region are likely either co-spatial and/or adjacent to each other. One would reasonably expect that there is a significant interaction between the base of a powerful quasar jet and the inner accretion disk and/or inner corona. This is plausibly the causative agent for the EUV deficit of radio loud quasars. Simple analytic models of this interaction can reproduce the relationship found in the radio loud quasar population between the magnitude of the EUV deficit and the ratio of jet power to the bolometric luminosity of the accretion flow (Punsly 2015). In particular, the stronger the jet, the larger volume of the inner accretion flow that is displaced by the magnetic flux tubes comprising the jet base. This same dynamic might be occurring when NE1bc flares in OQ208, decreasing $L_{\mathrm{ion}}$ and therefore the Balmer EW. It is worth noting that, perhaps the most powerful jet in the Universe, the jet in the CSS quasar, 3C298, (the largest 178 MHz luminosity in the 3C catalog) has broad lines with roughly 50% of the EW of a typical quasar of its UV luminosity and it has a very steep EUV spectrum (Punsly et al. 2022).

The simple analytic models mentioned above are a manifestation of magnetically arrested accretion (Igumenshchev et al. 2003; Igumenshchev 2008; Punsly 2015). There are currently no general relativistic numerical simulations of black hole accretion systems that radiate a high fidelity quasar accretion disk spectrum with a powerful jet. However, the aforementioned analytic models capture the relevant detail: the inner regions of the accretion flow is disrupted by large scale magnetic fields (as depicted in Figure 15) in all magnetically arrested numerical simulations (Igumenshchev 2008; Tchekhovskoy et al. 2011).

## 10. THE VLBA ANALYSIS OF THE POST FLARE CENTRAL ENGINE

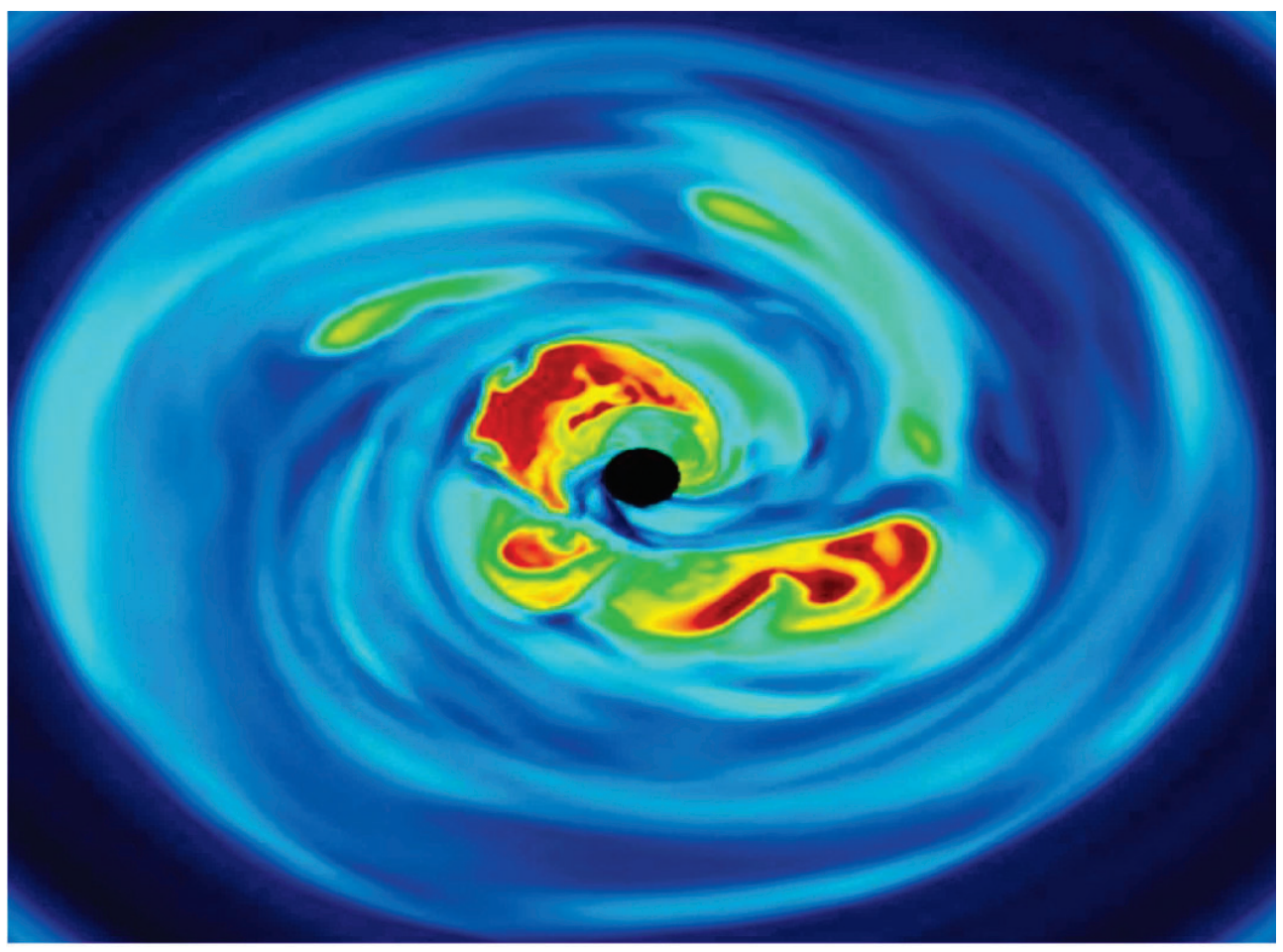

**Figure 15.** The disruption of the inner accretion disk in a magnetically arrested numerical simulation is shown in this figure. The logarithmic false color contour plot of the vertical magnetic field pressure from a magnetically arrested simulation run from Igumenshchev (2008) is used to schematically illustrate the concepts posited in this section. One can see Figure 2 of Punsly et al. (2009) for the false color scale, but this is not necessary for this discussion. The image was previously used in Punsly (2015) for the same purposes. The false color accentuates the two-fluid nature of the accretion flow. The red to yellow regions are magnetic islands (low density), regions of weak turbulence (from magneto-rotational instabilities) and therefore weak disk emission. The dark blue (the densest plasma) and green are turbulent strong disk emission regions with weak large scale vertical magnetic fields.

Notice the drastic difference in the nuclear morphology between the top right panel and the bottom left panel in Figure 8. During the flare the nucleus, NE1, is resolved as a bright triangular triple. After the flare in 2017, it is a symmetric linear triple that is an order of magnitude fainter on a much larger scale ($\sim$ 1.5 mas) and NE1 is unresolved. The southern component is J1 after it retracts back toward the nucleus. It looks like a different radio source. The linear triple was first detected in the 12/12/2011 observation shown in Figure 6. The northern component (the component that is not labeled in the bottom panels of Figure 8) seems to appear out of nowhere and we cannot find a counterpart in any previous observation. The triple persists in the 2018 observation, project code B263, at 23.8 GHz and in the 10/16/2023 observation in the bottom right hand panel of Figure 8. Table 4 shows the VLBA data reductions that we used to study the evolution of the nucleus after the flare had subsided. The best observation is clearly the 2017 observation in the lower left panel of Figure 8. The 2017 image is in fact a composite of 4 (project code BS260) epochs observed over 3 months that were combined. Thus, the u,v coverage is quite good in spite of the fact that OQ208 is observed as a calibrator. Combining observations is potentially fraught with issues (sky changing, instrumental oddities), so we performed numerous careful comparisons between the 4 input epochs and the composite result. We did not find that any problems or odd artifacts had been introduced. We also produced various other composites, combining pairs of epochs (e.g. just epochs a+b, a+d and c+d) and they all produced a similar image, with the 4 combined producing an image with higher dynamic range. All of our 43 GHz images were point sources.

Our study was initiated based on Jiang et al. (2022) who reported the fading of the continuum and the H$\beta$ broad line in 2022. We noticed that the radio flux was also fading. Thus, we began a study of the recent VLBA images. Figure 16 shows the radio data from Tables 1 and 4 (suitably re-scaled for comparison purposes) and the optical continuum from Table 2. All the quantities appear to be decaying simultaneously. Even though our data are very sparse, the fading of the H$\alpha$ broad lines in the right hand panel seems to be associated with the fading of the continuum. The fading of the NE1 luminosity occurs when the broad lines are becoming weak, the opposite behavior that is observed during the flare. This indicates a state change in the jet launching region.

## 11. SUMMARY AND CONCLUSION

In this study, we analyzed the long term (58 year) radio behavior of OQ208 with a special emphasis on the time frame after the advent of the VLBA. We also investigated the broad line evolution during the last 40 years. In particular, we have numerous archival VLBA and optical spectroscopic observations during a strong 15.4 GHz nuclear (NE1bc) flare that peaked in 1998. In Figure 4 and Section 7, we found that the H$\alpha$ broad line EW (computed relative to the host plus AGN continuum) is depressed during the flare rise and peak.

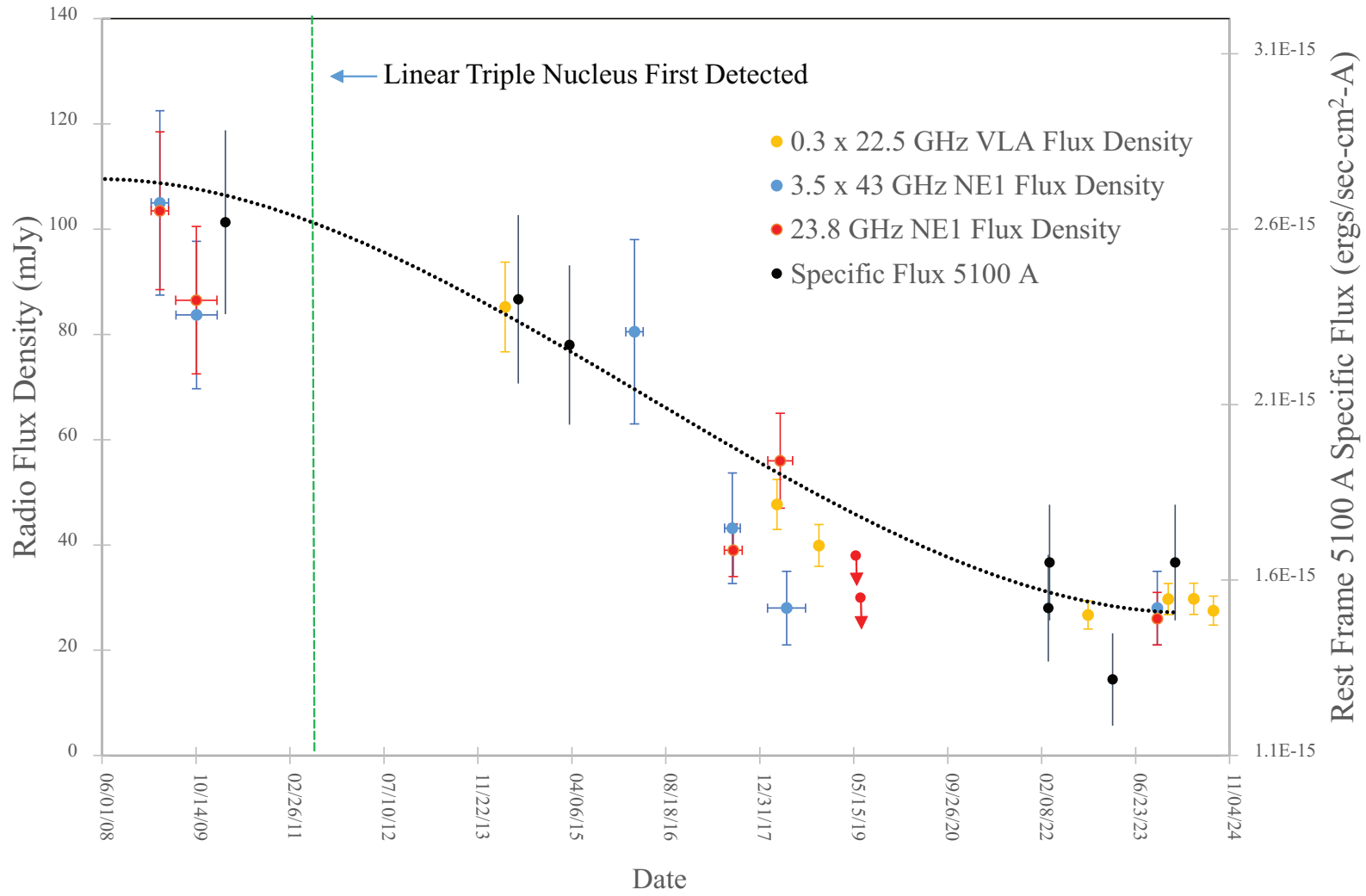

**Figure 16.** The data in Tables 1, 2 and 4 are used to plot the time evolution of OQ208 after the 1998 flare. In the top panel, we see the 5100 Å flux and the radio data are plotted. We added a black dashed third order polynomial fit to the optical continuum that goes back to 2006. This is simply for visualization purposes, but was chosen because there seems to be two inflection points. In an approximate sense, from 2009 - 2010, it is a rather flat light curve, then there is a decent until it hits a local minimum in 2022, the second inflection point. The bottom panel is the same, but with the H$\alpha$ flux. A similar pattern is seen.

The discussion of this result was developed as follows. The light curve constructed in Section 2 showed two large high frequency radio flares, one in the mid 1970's and a smaller one that peaked in 1998. It was shown in Section 3 that the EW of broad lines during the radio flare $\approx$1997-2001 were considerably smaller than during the entire 40 year broad line history (see Figures 4 and 5, as well as the related discussions). We then explored the connection to the central engine of the radio jet. In Section 4, we argued that the so-called northeast lobe, NE1, that is an extremely dominant feature at frequencies above 15 GHz is actually the host of the jet central engine. Firstly, six VLBA epochs from 1994 and 1995 showed three components moving away from NE1 (two were superluminal), while NE1 was static with respect to the two diffuse components (lobes) NE2 and SW1. Corroborating evidence of this identification is presented in Section 4. It was noted that subcomponents of NE1 have the highest brightness temperatures in the radio source. Furthermore, the subcomponent NE1bc has the flattest intrinsic power law of any feature in the SSA powerlaw decomposition of the northeast complex that was derived from matched resolution observations in January 2000 (Appendix C). Section 5 developed the discussion of the subcomponents of NE1. The three subcomponents NE1a, NE1b and NE1c appear to be distributed near the inner edge of the dusty torus. There is the possibility that one of the components is located inside of the dust sublimation radius, much closer to the jet launch point. The data are not sufficient to isolate which component (NE1a or NE1b), if any, is much closer to the jet launch point. Due to resolution concerns, we combined NE1b and NE1c as NE1bc. In Section 6, it was determined that the flare in total 22 GHz flux density was essentially a flare in NE1bc. This is consistent with NE1bc having the highest $T_b$ and flattest intrinsic power law spectra of any component. Section 7 is a more in depth characterization of the flare.

**Table 4.** The Data from VLBA Observations of NE1 Post Flare

| Date | Project | Frequency (GHz) | Flux Density (mJY) | Comments |
|---|---|---|---|---|
| 4/5/2009±46days | BS193 | 43.3 | 30 ± 5 | Calibrator (Schinzel et al. 2012). Average of approximate point source in 4 epochs |
| 4/5/2009±46days | BS193 | 23.8 | 104 ± 15 | Calibrator (Schinzel et al. 2012). Average of 4 epochs |
| 10/16/2009±110days | BS194 | 43.3 | 24 ± 4 | Calibrator (Schinzel et al. 2012). Average of approximate point source in 8 epochs |
| 10/16/2009±110days | BS194 | 23.8 | 86 ± 14 | Calibrator (Schinzel et al. 2012). Average of 8 epochs |
| 3/4/2016±46days | BW115 | 43.3 | 23 ± 5 | Calibrator. Average of approximate point source in 3 epochs |
| 8/7/2017±41days | BS260 | 43.3 | 12 ± 3 | Calibrator(Roder et al. 2024). Average of approximate point source: 3 epochs, BS260ABC |
| 8/17/2017±44days | BS260 | 23.8 | 39 ± 5 | Calibrator. All 4 epochs stacked for a low noise image. Individual images are consistent |
| 5/28/2018±101days | BS263 | 43.3 | 8 ± 2 | Calibrator(Roder et al. 2024). All epochs 6 are an approximate point source, average |
| 4/19/2018±66days | BS263 | 23.8 | 56 ± 9 | Calibrator(Roder et al. 2024). Nuclear triple detected in first 4 epochs, average |
| 5/26/2019 | BS281A1 | 23.8 | $\lesssim 38$ | Calibrator. Upper limit, secondary cannot be resolved from nucleus |
| 6/10/2019 | BS281A2 | 23.8 | $\lesssim 30$ | Calibrator. Upper limit, secondary cannot be resolved from nucleus |
| 10/16/2023 | BR254a | 43.3 | 8 ± 2 | Pointed observation, only 8 antennas, no SC |
| 10/16/2023 | BR254a | 23.6 | 26 ± 5 | Pointed observation, only 8 antennas, no SC |

We developed a possible physical explanation of the Balmer EW behavior during the radio flare in Sections 8 and 9 based on magnetically arrested accretion. During the 39 year history of our combined radio and optical observations in Tables 1 and 2, we identify 1997-2001 as a possible time frame with magnetically arrested states in OQ208. The paper presents an interpretation consistent with a transient magnetically arrested accretion scenario, rather than an actual measurement of magnetically arrested accretion parameters or timescales.

In section 10, we note another state change in the jet accretion system after 2011. The jet is relatively weak and the nuclear region is a 2.5 pc long linear triple radio source. Unlike the possible magnetically arrested state, the jet luminosity, the optical luminosity and the broad line flux all decrease in unison after 2014. It is a system that is governed primarily by the accretion rate - the magnetic field in the jet base does not provide any measurable feedback.

In regards to the history of the source, recall the discussion in Section 4 of the top frame of Figure 6. J1, J2 and J3 all reach a maximum distance from the nucleus, then eventually contract towards NE1. J1 reaches a maximum probably in early 1996, then stagnates for about 4 years then starts retracting. It is difficult to pinpoint the exact date of maximum extension from the data in Appendix B for J2 and J3, since they are not always detected. J3 is at the maximum measured distance from NE1 in August 1997 and J2 is at maximum in October 1998. In order to understand this phenomenon in a broader context, note that in Section 5, during the flare, the 3 components of the nucleus seem to cluster at or within the hot dust at the inner edge of the molecular torus in Figure 8. The irregular, non-colinear, pattern seems to imply a jet experiencing multiple deflections from interactions with the molecular gas. These strong interactions are supported by the approximate co-rotation of the radio jet axis and the polar scattering (ionization) cone, shaped by an aperture in the molecular gas, from before the flare to after the flare in Figure 10. In this context, the implication is that the retracting trajectories of J1, J2 and J3 are in a sense failed jets that are thwarted by the molecular clouds. They might be thought of as precursors to ejections, MHD discontinuities that signal an increase in jet power to the downstream flow. They are patterns that will travel the fastest in highly magnetized corridors in the jet (in a highly magnetized region the magneto-sonic wave speed is near the speed of light). Overall, in the past 30 years, jet propagation is highly obstructed on all scales from 0.1 mas to 4 mas. There must be a large bend in the northward counter-jet going from the linear triple in the lower left panel of Figure 8 to NE2. The approaching jet must also bend dramatically from J1 to reach SW1 and SW2. The only region that seems to escape the obstructions are the “distant” ejecta, SW1 amd SW2, that seem to follow a linear outgoing path.

A special thanks to Michael Eracleous for finding the 6 spectra from 1997-2001 and sharing these with us. These were critical to our results. He also supplied the 2010 and 2014 spectra. We are grateful to the referee for constructive comments and guidance that improved the manuscript. 22 GHz VLA data was graciously shared by Tanmoy Laskar, Deanne Coppejans, Eli Pattie and Joe Michail. Margo Aller generously provide the University of Michigan data. We thank the staff of the Hobby Eberly Telescope for support of our LRS2 observations. This research was partially funded through the McDonald Observatory. The Hobby-Eberly Telescope (HET) is a joint project of the University of Texas at Austin, the Pennsylvania State University, Ludwig-Maximilians-Universität München, and Georg-August-Universität Göttingen. The HET is named in honor of its principal benefactors, William P. Hobby and Robert E. Eberly. The Low Resolution Spectrograph 2 (LRS2) was developed and funded by the University of Texas at Austin McDonald Observatory and Department of Astronomy and by Pennsylvania State University. We thank the Leibniz-Institut für Astrophysik Potsdam (AIP) and the Institut für Astrophysik Göttingen (IAG) for their contributions to the construction of the integral field units. We acknowledge the Texas Advanced Computing Center (TACC) at The University of Texas at Austin for providing high performance computing, visualization, and storage resources that have contributed to the results reported within this paper. This work is based in part on observations collected at Copernico 1.82m telescope (Asiago Mount Ekar, Italy) INAF - Osservatorio Astronomico di Padova. PM acknowledges financial support from the Spanish MCIU through project PID2022-140871NB-C21 by “ERDF A way of making Europe”, and from the Severo Ochoa grant CEX2021-515001131-S funded by MCIN/AEI/10.13039/501100011033. CO acknowledges support from the Natural Sciences and Engineering Research Council (NSERC) of Canada. This work makes use of the Sloan Digital Sky Survey IV, with funding provided by the Alfred P. Sloan Foundation, the U.S. Department of Energy Office of Science, and the Participating Institutions. SDSS-IV acknowledges support and resources from the Center for High-Performance Computing at the University of Utah. The SDSS web site is www.sdss.org. This research has made use of data from the MOJAVE database that is maintained by the MOJAVE team (Lister et al. 2018) AF was funded by the European Union ERC-2022-STG - BOOTES - 101076343. The National Radio Astronomy Observatory and Green Bank Observatory are facilities of the U.S. National Science Foundation operated under cooperative agreement by Associated Universities, Inc. The work of ABP was carried out within the state assignment of the Federal State Budget Scientific Institution “Crimean Astrophysical Observatory of RAS”.

# APPENDIX

## A. A COMPENDIUM OF SPECTRA

This Appendix shows all of the spectra in the Hα region. They are arranged in Figure 17, vertically in chronological order, and offset in flux density by an arbitrary amount, in an attempt to minimize the overlap of the lines from different epochs. The plot illustrates the difference in data quality (i.e., resolution and S/N ratio) and shows gross changes in the broad-line shape. Details of the different line shapes and model fits will be compared in a future paper (Marziani et al., in preparation).

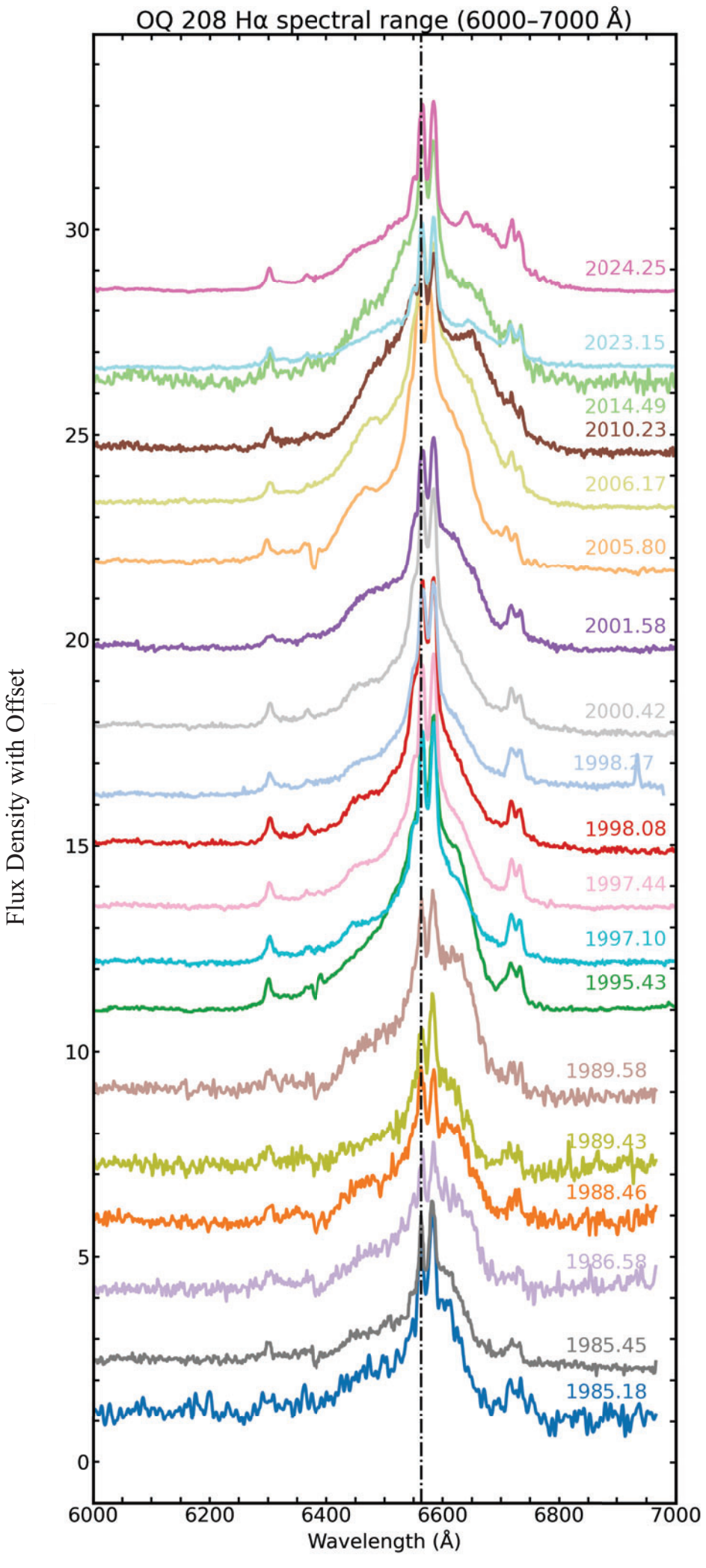

**Figure 17.** The compilation of spectra around Hα spanning nearly 40 years, arranged in chronological order on the same flux density scale. An offset is applied between spectra to separate them and reveal the evolution of the broad line profile shape.

## B. VLBI MODELS

We do not implement the statistical uncertainties based on noise in our fits (Fomalont 1999; Lee et al. 2008). These tend to underestimate the errors. Thus, these outputs from AIPS are not listed in the fits below. We use more conservative empirical, systematic positional (flux density) uncertainties as discussed in Section 5 (Section 7) which we repeat here for ease of reference. For positional errors we utilize the following philosophy.

1. Bright compact components, peak flux density $> 5$ post process rms. Uncertainty is 0.1 beam FWHM for both RA and Dec coordinates (Lister et al. 2009). This applies to the centroid of the nuclear flux density.
2. Faint components, peak flux density $< 5$post process rms. Uncertainty is 0.2 beam FWHM for both RA and Dec coordinates (Lister et al. 2009)
3. Large components with a low contrast between the component and the surface brightness from the other components (i.e. J1 in early 1995) or low contrast with the rms noise (SW2), we add 0.1 of the FWHM of component in quadrature with the uncertainty in 2) for both RA and DEC
4. Finally, the uncertainty in the centroid is added in quadrature with the uncertainty in the component.

Our VLBA observations have an absolute flux density uncertainty of 5%. To this we need to add an uncertainty due to the fitting process in quadrature. In terms of the flux density uncertainty of the component fits, multiple fitting of the same epoch suggest using 5% for bright compact components (NE1c, NE1b, NE1bc and J1 in epochs in which it is pronounced such as 1/29/2000) and 10% for large diffuse components (NE1a and NE2). These rules are applied as indicated in the development of the text.

The absolute flux density uncertainty for the global VLBI observations is 10% (Pushkarev and Kovalev 2012).

**Table 5.** VLBI Gaussian Model Fits

| A | 6/12/1994 | 22.2 GHz | VLBA | Project BL007 | |
|---|---|---|---|---|---|
| Component | Flux Density (mJy) | radius (mas) | PA deg | Major Axis FWHM (mas) | aspect ratio/PA (deg) |
| NE2 | 19.5 | 1.26 | -17.9 | 0.425 | 1/... |
| NE1 | 256.1 | 0 | ... | 0.200 | 0.527/1.7 |
| SW2 | 8.8 | 7.75 | -118.1 | 0.05 | 1/... |
| SW1 | 19.1 | 5.91 | -125.9 | 1.12 | 1/... |
| J3 | 6.9 | 2.34 | -123.1 | 0.203 | 1/... |
| J1 | 59.8 | 0.28 | -125.3 | 0 | ... |
| *a* | 4.0 | 3.47 | -32.5 | 0.04 | 1:. |
| B | 4/7/1995 | 15.4 GHz | VLBA | Project BK016 | |
| NE2 | 91.8 | 1.19 | -24.3 | 0.492 | 1/... |
| NE1a | 300.9 | 0.16 | -124.6 | 0.378 | 1/... |
| NE1b | 443.6 | 0.05 | 48.8 | 0.344 | 1/... |
| SW2 | 16.1 | 7.68 | -116.8 | 0.503 | 1/... |
| SW1 | 41.2 | 6.50 | -126.1 | 0.899 | 1/... |
| J3 | 14.5 | 3.03 | -126.3 | 0.411 | 1/... |
| J2 | 9.6 | 1.79 | -137.2 | 0 | ... |
| J1 | 61.0 | 0.80 | -136.9 | 1.14 | 1/... |
| C | 5/26/1995 | 15.4 GHz | VLBA | Project BP17 | *b* |
| NE2 | 106.4 | 1.13 | -22.0 | 0.587 | 1/... |
| NE1a | 371.1 | 0.11 | -129.4 | 0.379 | 1/... |
| NE1b | 383.0 | 0.05 | 61.7 | 0 | ... |
| SW2 | 18.8 | 7.58 | -117.3 | 0.575 | 1/... |
| SW1 | 7.8 | 5.55 | -127.6 | 0.455 | 1/... |
| J3 | 19.0 | 3.10 | -127.0 | 0.425 | 1/... |
| J2 | 4.4 | 1.79 | -137.4 | 0 | ... |
| J1 | 36.8 | 0.83 | -142.0 | 0.697 | 1/... |
| D | 11/03/1995 | 15.4 GHz | VLBA | Project BA12 | *c* |
| NE2 | 62.2 | 1.29 | -25.4 | 0.487 | 1/... |
| NE1a | 284.6 | 0.25 | -164.1 | 0.471 | 1/... |
| NE1b | 167.9 | 0.26 | -23.2 | 0.220 | 1/... |
| NE1c | 357.6 | 0.05 | 72.4 | 0 | ... |
| SW2 | 27.0 | 7.44 | -118.4 | 1.42 | 1/... |
| SW1 | 28.8 | 6.35 | -127.2 | 0.839 | 1/... |
| J3 | 6.7 | 3.10 | -123.8 | 0.339 | 1/... |
| J1 | 13.5 | 1.34 | -136.0 | 0 | ... |
| E | 11/27/1995 | 15.4 GHz | VLBA | Project BA11 | *d* |
| NE2 | 75.6 | 1.22 | −24.8 | 0.596 | 1/... |
| NE1a | 289.1 | 0.12 | -83.6 | 0.456 | 1/... |
| NE1b | 332.6 | 0.06 | -2.4 | 0 | ... |
| NE1c | 141.5 | 0.31 | 172.4 | 0.18 | 1/... |
| SW2 | 17.1 | 7.72 | -116.9 | 0.595 | 1/... |
| SW1 | 29.5 | 6.62 | -126.2 | 0.704 | 1/... |
| J3 | 12.0 | 3.23 | -126.4 | 0 | ... |
| J1 | 7.4 | 1.59 | -133.3 | 0 | ... |
| | 4.0 | 5.74 | -129.4 | 0 | ... |
| *e* | 32.2 | 0.74 | -142.2 | 0.665 | 1/... |
| F | 12/15/1995 | 15.4 GHz | VLBA | Project BK37A | *f* |
| NE2 | 81.5 | 1.19 | -23.0 | 0.614 | 1/... |
| NE1a | 352.6 | 0.17 | -144.0 | 0.533 | 1/... |
| NE1b | 277.8 | 0.17 | 0.6 | 0.22 | 1/... |
| NE1c | 184.1 | 0.14 | 160.0 | 0 | ... |
| SW2 | 14.9 | 7.66 | -117.4 | 0.406 | 1/... |
| SW1 | 37.4 | 6.52 | -126.3 | 1.06 | 1/... |
| J3 | 13.9 | 3.08 | -124.6 | 0.680 | 1/... |
| J2 | 2.4 | 2.40 | -139.7 | 0 | ... |
| J1 | 14.4 | 1.56 | -138.4 | 0.323 | 1/... |

**Table 5.** VLBI Gaussian Model Fits -continued

| G | 8/28/1997 | 15.4 GHz | VLBA | Project BK052A | |
|---|---|---|---|---|---|
| Component | Flux Density (mJy) | radius (mas) | PA deg | Major Axis FWHM (mas) | aspect ratio/PA (deg) |
| NE2 | 69.8 | 1.18 | -25.0 | 0.640 | 1/... |
| NE1a | 277.7 | 0.15 | -121.8 | 0.532 | 1/... |
| NE1b | 415.4 | 0.11 | -9.4 | 0.18 | 1/... |
| NE1c | 217.1 | 0.21 | 158.4 | 0 | ... |
| SW2 | 14.9 | 7.71 | -117.4 | 0.623 | 1/... |
| SW1 | 34.6 | 6.56 | -127.2 | 0.766 | 1/... |
| J3 | 5.6 | 3.49 | -123.1 | 0.478 | 1/... |
| J2 | 3.2 | 2.76 | -132.8 | 0 | ... |
| J1 | 41.9 | 1.47 | -141.8 | 0.338 | 1/... |
| H | 3/07/1998 | 15.4 GHz | VLBA | Project BK052B | $^b$ |
| NE2 | 57.9 | 1.04 | -27.6 | 0.530 | 1/... |
| NE1a | 186.1 | 0.28 | -131.4 | 0.498 | 1/... |
| NE1b | 436.7 | 0 | ... | 0.198 | 1/... |
| NE1c | 186.0 | 0.30 | 156.4 | 0 | ... |
| SW2 | 12.2 | 7.69 | -118.1 | 0.482 | 1/... |
| SW1 | 27.2 | 6.68 | -127.1 | 0.636 | 1/... |
| J1 | 52.1 | 1.55 | -143.3 | 0.198 | 1/... |
| I | 3/28/1998 | 15.4 GHz | VLBA | Project BC077B | $^b$ |
| NE2 | 64.0 | 1.11 | -25.9 | 0.611 | 1/... |
| NE1a | 246.0 | 0.23 | -123.5 | 0.542 | 1/... |
| NE1b | 509.5 | 0 | ... | 0.198 | 1/... |
| NE1c | 218.0 | 0.33 | 156.9 | 0 | ... |
| SW2 | 19.0 | 7.76 | -117.5 | 0.875 | 1/... |
| SW1 | 33.5 | 6.68 | -127.2 | 0.662 | 1/... |
| J2 | 4.9 | 2.88 | -133.3 | 0 | ... |
| J1 | 69.2 | 1.54 | -143.1 | 0.266 | 1/... |
| J | 10/30/1998 | 15.4 GHz | VLBA | Project BK052D | |
| NE2 | 67.3 | 1.13 | -28.2 | 0.604 | 1/... |
| NE1a | 273.6 | 0.19 | -139.7 | 0.557 | 1/... |
| NE1b | 467.9 | 0.11 | -14.5 | 0.210 | 1/... |
| NE1c | 235.2 | 0.18 | 152.2 | 0 | ... |
| SW2 | 17.0 | 7.74 | -117.9 | 0.681 | 1/... |
| SW1 | 34.4 | 6.69 | -127.7 | 0.595 | 1/... |
| J2 | 4.9 | 2.88 | -133.3 | 0 | ... |
| J1 | 94.0 | 1.48 | -144.8 | 0.190 | 1/... |
| K | 1/29/2000 | 22.2 GHz | VLBA | Project BS73 | $^g$ |
| Component | Flux Density (mJy) | radius (mas) | PA deg | Major Axis FWHM (mas) | aspect ratio/PA (deg) |
| NE2 | 18.3 | 1.02 | -33.9 | 0.529 | 1/... |
| NE1a | 110.7 | 0.19 | -110.5 | 0.489 | 1/... |
| NE1b | 296.9 | 0.07 | -27.6 | 0.208 | 1/... |
| NE1c | 149.5 | 0.20 | 155.8 | 0.145 | 1/... |
| SW2 | 6.8 | 7.75 | -118.3 | 0.403 | 1/... |
| SW1 | 19.7 | 6.75 | -127.6 | 0.348 | 1/... |
| J2 | 5.2 | 2.62 | -131.6 | 0 | ... |
| J1 | 81.1 | 1.46 | -145.2 | 0.19 | 1/... |

**Table 5.** VLBI Gaussian Model Fits -continued

| L<br>Component | 1/29/2000<br>Flux Density<br>(mJy) | 22.2 GHz<br>radius<br>(mas) | VLBA<br>PA<br>deg | Project BS73<br>Major Axis FWHM<br>(mas) | $h$<br>aspect ratio/PA<br>(deg) |
|---|---|---|---|---|---|
| NE2 | 13.7 | 1.10 | -33.6 | 0.394 | 1/... |
| NE1a | 123.5 | 0.19 | -114.5 | 0.474 | 1/... |
| NE1bc | 414.5 | 0.05 | 158.3 | 0.411 | 0.483/-22.8 |
| SW2 | 6.7 | 7.78 | -118.7 | 0.423 | 1/... |
| SW1 | 18.5 | 6.75 | -127.9 | 0.335 | 1/... |
| J2 | 7.5 | 2.39 | -133.5 | 0 | ... |
| J1 | 79.5 | 1.48 | -146.0 | 0.218 | 1/... |
| M | 1/29/2000 | 15.4 GHz | VLBA | Project BS73 | |
| NE2 | 58.7 | 1.10 | -28.4 | 0.523 | 1/... |
| NE1a | 274.2 | 0.17 | -116.0 | 0.485 | 1/... |
| NE1bc | 641.7 | 0.03 | 79.5 | 0.373 | 0.426/-28.8 |
| SW2 | 14.9 | 7.77 | -118.3 | 0.650 | 1/... |
| SW1 | 36.4 | 6.71 | -127.4 | 0.513 | 1/... |
| J2 | 6.1 | 2.70 | -132.9 | 0 | ... |
| J1 | 112.3 | 1.42 | -145.7 | 0.243 | 1/... |
| N | 1/31/2000 | 8.6 GHz | Global VLBI | | $h$ |
| NE2 | 158.1 | 1.21 | -25.6 | 0.598 | 1/... |
| NE1a | 456.0 | 0.20 | -66.6 | 0.459 | 1/... |
| NE1bc | 633.6 | 0.10 | 137.2 | 0.516 | 0.583/-7.1 |
| SW2 | 39.7 | 7.66 | -118.2 | 0.727 | 1/... |
| SW1 | 57.3 | 6.70 | -128.1 | 0.522 | 1/... |
| J2 | 6.1 | 2.70 | -132.9 | 0 | ... |
| J1 | 121.8 | 1.26 | -145.1 | 0.414 | 1/... |
| O | 3/13/2000 | 2.3 GHz | Global VLBI | | |
| NE2 | 1.311 | 0.28 | -150.3 | 1.86 | 1/... |
| NE1a | 200.6 | 1.55 | 168.9 | 0.414 | 1/... |
| SW2 | 69.7 | 7.38 | -125.9 | 1.68 | 1/... |
| P | 3/27/2001 | 15.4 GHz | VLBA | Project BS85 | |
| NE2 | 53.8 | 2.77 | -102.2 | 0.514 | 1/... |
| NE1a | 247.7 | 2.87 | -124.7 | 0.518 | 1/... |
| NE1bc | 595.0 | 2.65 | 125.5 | 0.401 | 0.415/-28.9 |
| SW2 | 14.6 | 10.38 | -119.5 | 0.718 | 1/... |
| SW1 | 40.8 | 9.42 | -126.9 | 0.518 | 1/... |
| J2 | 3.9 | 5.43 | -129.9 | 0.221 | 1/... |
| J1 | 99.9 | 4.01 | -132.4 | 0.239 | 1/... |
| Q | 3/6/2002 | 8.6 GHz | Global VLBI | | $i$ |
| NE2 | 225.6 | 1.07 | -241.5 | 0.695 | 1/... |
| NE1a+NE1b | 947.0 | 0.03 | -110.2 | 0.526 | 1/... |
| NE1c | 339.2 | 0.27 | 139.2 | 0.093 | 1/... |
| SW2 | 52.2 | 7.67 | -118.7 | 0.728 | 1/... |
| SW1 | 82.4 | 6.70 | -128.2 | 0.516 | 1/... |
| J2+J3 | 6.7 | 2.91 | -139.9 | 0.910 | 1/.. |
| J1 | 180.1 | 1.27 | -150.9 | 0.406 | 1/... |
| R | 4/4/2004 | 15.4 GHz | VLBA | Project BH118C | |
| NE2 | 51.6 | 1.15 | -29.7 | 0.602 | 1/... |
| NE1a | 350.0 | 0.17 | -74.5 | 0.532 | 1/... |
| NE1b | 231.9 | 0.05 | 78.2 | 0.172 | 1/... |
| NE1c | 81.0 | 0.26 | 141.9 | 0 | ... |
| SW2 | 17.4 | 7.81 | -117.2 | 0.703 | 1/... |
| SW1 | 42.8 | 6.81 | -126.7 | 0.479 | 1/... |
| J1 | 128.1 | 1.26 | -145.2 | 0.274 | 1/... |

**Table 5.** VLBI Gaussian Model Fits -continued

| S | 4/28/2006 | 15.4 GHz | VLBA | Project BL137D | |
|---|---|---|---|---|---|
| Component | Flux Density | radius | PA | Major Axis FWHM | aspect ratio/PA |
| | (mJy) | (mas) | deg | (mas) | (deg) |
| NE2 | 96.4 | 0.74 | -43.9 | 1.36 | 0.359/-12.7 |
| NE1a | 226.2 | 0.11 | -91.5 | 0.532 | 1/... |
| NE1b | 161.9 | 0.08 | 66.2 | 0 | ... |
| NE1c | 30.8 | 0.29 | 137.7 | 0 | ... |
| SW2 | 20.2 | 7.94 | -117.9 | 0.963 | 1/... |
| SW1 | 33.9 | 6.85 | -126.2 | 0.523 | 1/... |
| J1 | 127.2 | 1.15 | -143.9 | 0.314 | 1/.... |
| T | 12/12/2011 | 15.4 GHz | VLBA | Project BL178AC | |
| NE2 | 82.4 | 1.01 | -10.9 | 0.981 | 1/... |
| NE1 | 204.0 | 0.05 | -178.1 | 0.39 | 0.81/33.1 |
| New[j] | 70.1 | 0.77 | 40.9 | 0.348 | 1/... |
| [a] | 18.0 | 1.29 | -141.8 | 0.361 | 1/... |
| SW2 | 21.0 | 7.48 | -114.5 | 1.38 | 0.59/-2.9 |
| SW1 | 29.4 | 6.46 | -124.3 | 0.900 | 0.84/-82.4 |
| J2 | 3.8 | 2.36 | -129.1 | 0 | ... |
| J1 | 52.6 | 0.75 | -145.8 | 0.132 | 1/... |
| U | 8/12/2017 | 23.8 GHz | VLBA | Project BS260A-D | [k] |
| NE2 | 1.8 | 2.00 | -12.1 | 0 | ... |
| NE1 | 39.5 | 0.06 | -11.7 | 0.526 | 0.45/35.2 |
| New | 9.9 | 0.83 | 27.2 | 0.515 | 0.50/-47.2 |
| SW1+SW2 | 8.6 | 7.06 | -118.7 | 2.21 | 0.20/-74.2 |
| J1 | 19.4 | 0.79 | -140.8 | 0.222 | 0.34/49.8 |
| V | 10/16/2023 | 23.6 GHz | VLBA | Project BR254A | |
| NE2 | 1.2 | 1.46 | -33.7 | 0 | ... |
| NE1 | 25.9 | 0.85 | -149.8 | 0.589 | 0.31/15.2 |
| New | 5.3 | 0.26 | -145.1 | 0 | ... |
| SW1+SW2 | 6.6 | 8.24 | -117.5 | 1.17 | 0.73/-66.9 |
| J1 | 4.5 | 1.47 | -143.0 | 0 | ... |

[a]This component only occurs in this image

[b]Absolute flux scale re-calibrated by MOJAVE collaboration in 2017

[c]4 scans within ∼3hrs. The uv coverage not as dense as other epochs. This is included because it is the earliest detection of the triple nucleus in Figure 9.

[d]6 scans within ∼7hrs

[e]A jet component, detected only in this epoch, that is closer to the nucleus than J1 was on 4/7/1995

[f]8 scans within ∼7hrs

[g]Based on image restored with uniformly weighted beam

[h]Based on image restored with beam matched to 15.4 GHz resolution.

[i]Best global VLBI. Data used for best NE2 flux density measurement.

[j]The north component of the linear triple nucleus first appears.

[k]Four epochs stacked, 6/27/2017, 7/31/2017, 8/29/2017, 9/25/2017

## C. THE NORTHEAST COMPLEX

The five components, NE1a, NE1b, NE1c, NE2 and J1 are crowded together on the sky plane within a region $< 1.5$mas radius, which we will call the NE Complex.. Each of these components has a high frequency peak and a steep spectral tail. The difference in these (presumably synchrotron self-absorbed, SSA, induced) peaks leads to the source having a different morphology at different frequencies. For example, Figure 8 shows that at 15 and 22 GHz, the NE complex is dominated by NE1. A 5 GHz global VLBI observation in 1993 shows the NE complex as a double radio source with NE2 and NE1 of similar brightness (Stanghellini et al. 1997). At 2.3 GHz, NE2 dominates the NE complex (see below). We explore the confusing situation in this appendix.

The project BS73, not only had the 22.2 GHz observation used in the top right panel of Figure 8, but also had a 15.4 GHz observation. Thus, we can produce a matched resolution fit for the more variable region of the NE Complex, NE1a, NE1b, NE1c and J1. NE1b and NE1c are not resolved at 15.4 GHz. They are fit as one elliptical Gaussian, designated as NE1bc. So, the matched resolution process was to

1. match the UV-ranges, setting an upper limit of 442 M$\lambda$ (the maximum in 15 GHz observations) for the 22 GHz data (for model fitting and further map construction).

2. restore LL-map at each frequency with the 15 GHz beam and a pixel size of 0.05 mas.

The 22 GHz image is then refit with one elliptical Gaussian for NE1b+NE1c. There is a resolution matched 8.6 GHz global VLBI image two days later on 1/31/2000. The 2.4 GHz image is not that good on this date, so we use a more reliable image from 3/13/2000 (Table 5O) since it is dominated by NE2 which is slowly variable. One thing to note is that the highest quality 8 GHz image is from 3/6/2002 (Table 5Q). The data of that experiment (RDV32) had in addition to 10 VLBA antennas, 9 IVS stations participated in that session. The full list is: 1:BR, 2:FD, 3:GC, 4:GG, 5:HN, 6:KP, 7:LA, 8:MC, 9:MK, 10:NL, 11:NT, 12:ON, 13:OV, 14:PT, 15:SC, 16:TS, 17:WF, 18:KK, 19:WZ. We flagged data from the small GGAO dish (4, above), these were too noisy. Gilcreek (3 above) took part in this run, which is a rare case for Astrogeo observations. The large number of radio telescopes formed four sub-arrays and provided quite dense (for the snap-shot mode) uv-coverage. Westford (19, above) was a unique station forming both the shortest (with Hancock, 5 above) and longest (with Tsucuba, 16 above) baseline projections. This image will also be used in the following.

Our analysis is not intended to be an exact fit to all the data in Figure 18. It is intended to qualitatively divide NE Complex into simple SSA power law component spectra in order to explain the intricate broadband spectrum of the subcomponents and the entire complex. The fit is not unique, but one that minimizes the residuals with the high confidence data. The source is too complex to do more than this. We assume an underlying unabsorbed power law in each component. The flux density is defined as $S_\nu(\nu) = S\nu^{-\alpha}$, where $S$ is a constant. A simple solution to the radiative transfer equation occurs in the homogeneous approximation (Ginzburg and Syrovatskii 1965; van der Laan 1966). Using the SSA absorption coefficient from Ginzburg and Syrovatskii (1969), $\mu$, the SSA power law is parameterized as

$$S_\nu = \frac{S_\mathrm{o}\nu^{-\alpha}}{\tau(\nu)} \times \left(1 - e^{-\tau(\nu)}\right) \, , \; \tau(\nu) \equiv \mu(\nu)L \, , \; \tau(\nu) = \overline{\tau}\nu^{(-2.5-\alpha)} \, , \qquad \text{(C1)}$$

where $\tau(\nu)$ is the SSA opacity, $L$ is the path length in the rest frame of the plasma, $S_\mathrm{o}$ is a normalization factor and $\overline{\tau}$ is a constant.

One thing to note is that there is an excess of emission at the lowest frequency that is not accounted for in the compact component model. This is likely from diffuse, fossil, or lobe emission. Otherwise, the fit is rather good considering that some of the data (i.e, 5 GHz and 1.7 GHz) is not contemporaneous. There is also a spectral break at 22 GHz that seems very robust. In particular, during the 43 GHz VLA observation on March 3 1999, the weather seems to have been excellent and both 3C 286 and OQ208 were observed at elevations >70 degrees. The most difficult component to model is NE1a. It is a large component adjacent to, or overlapping, two bright components. Not only is it difficult to accurately model a 0.5 mas component with sparse, imperfect long baseline UV coverage, but the lower surface brightness of NE1 makes it difficult to extract uniquely from the bright NE1bc emission overlapping it. In addition, there is an inherent difficulty in modeling a surface brightness distribution with ad hoc discrete elements (Gaussian functions). Regardless of these qualifying comments, we present the parameters for each SSA fit used to make Figure 18 in Table 6.

At first glance these results seem to be in conflict with the two-point spectral index map made from the matched resolution observations in Figure 20. Such a task is nontrivial. There are inevitable issues like flux trade between neighboring components in model fitting and uncertainty in their positions. Thus, we tried another approach to measure the shift between the images - 2D cross-correlation over all optically thin emission since its position is achromatic.

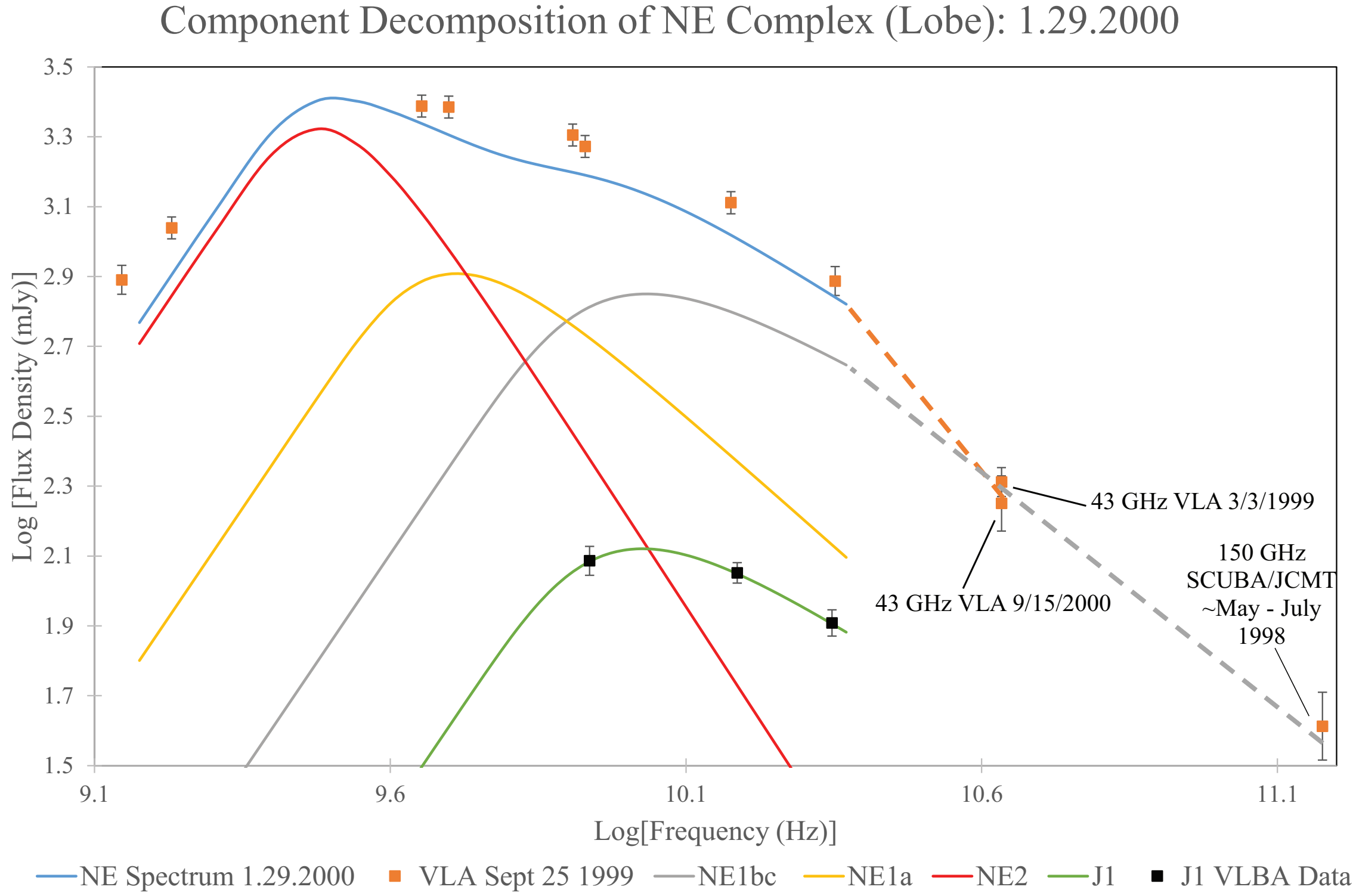

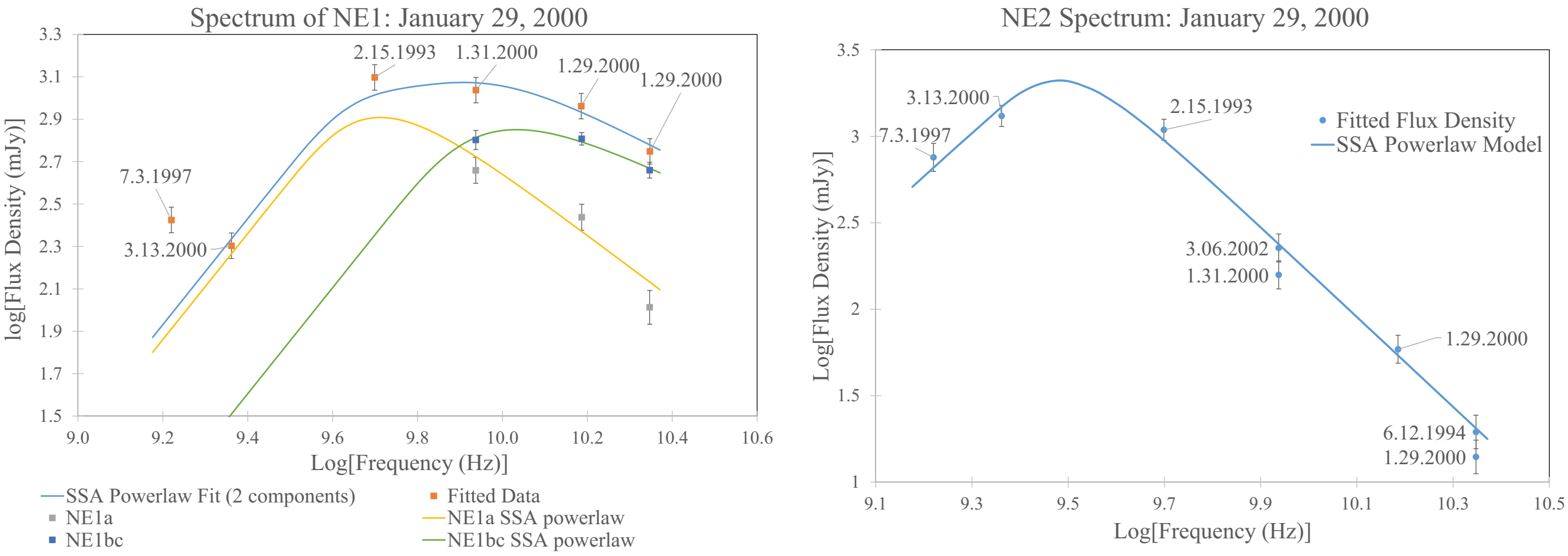

**Figure 18.** The SSA powerlaw fits to components of OQ208 during the matched resolution observations on 1/29/2000. Some of the NE2 data are not simultaneous since this is considered a slowly varying feature and diffuse. Being low surface brightness and large, it is a challenge to image with high frequency VLBI. So maximum sensitivity is considered the highest priority, not simultaneity. The largest flux density measurement is considered the most relevant. The VLA data are from (Dallacasa et al. 2000) except for the 22.5 GHz data which is from our Table 1. The 1997 VSOP data at 1.66 GHz is from Kameno et al. (2000).

Interestingly, image registration results in a zero shift between the 22 and 15 GHz maps. So, we did not shift the images. Figure 20 is what we have in the case that the maps are aligned, applying no shift. Ostensibly, it seems like J1 might be the core, as $\alpha$ progressively decreases along the jet as one moves away from it. However, from the right panel of Figure 7, we know that J1 (as well as J2 and J3) seems to emerge from NE1. Also, near the spectral peaks (see Figure 18), the brightness temperature at 8.6 GHz of NE1bc ($7.17 \times 10^{10}$K) 5.7 times higher than that of J1 (see Table 5N and Section 5). So there is more going on here. The first place to look is the large restoring beam of the VLBA relative to the component separations in the NE complex. This is a potential issue as indicated by the flattest spectrum in Figure 19 occurs in regions along the last contour where there is probably not even a source. The use of a component fit provides higher resolution than that of the surface brightness of the large Gaussian restoring beam. There are two things to consider. First, the larger SSA absorption at the lower frequency makes a two-point spectral index smaller than the intrinsic spectral index. But, more importantly, the overlapping steep spectrum component NE1a, dilutes the flat spectrum of the nearby NE1b. Table 6, explains the discrepancy as primarily caused by insufficient resolution in the spectral index map.

**Table 6.** SSA Composite Model of NE Complex on 1/29/2000

| Component | $\alpha$ | $\tau_{\rm SSA}$(10 GHz) | Flux Density 15.0 GHz (mJy) | Flux Density 22.0 GHz (mJy) |
|---|---|---|---|---|
| NE2 | 2.6 | 0.0028 | 57 | 21 |
| NE1a | 1.5 | 0.062 | 243 | 138 |
| NE1bc | 0.95 | 0.80 | 630 | 470 |
| J1 | 1.07 | 0.84 | 114 | 81 |

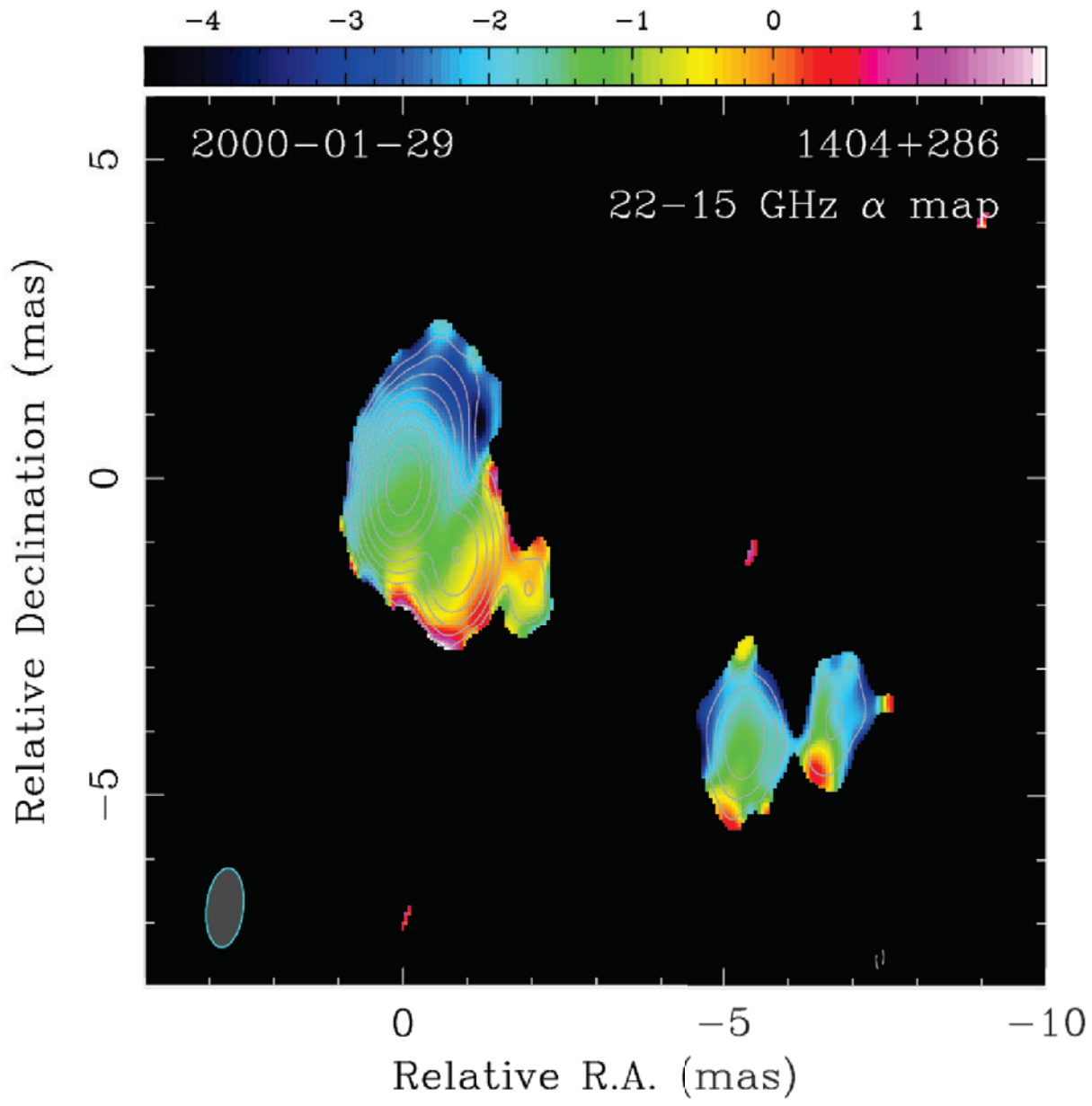

**Figure 19.** The spectral index map from the matched resolution observation on 1/29/2000.